%% file: main.tex
\documentclass{amsart}
\usepackage[english]{babel}
\usepackage{graphicx}
\usepackage{amssymb}
\usepackage{amsmath}
\usepackage[foot]{amsaddr}
\usepackage{amsfonts}
\usepackage{bbm}
\usepackage{xcolor}
\usepackage{float}
\usepackage{algorithm}
\usepackage{algpseudocode}
\usepackage{rotating}

\input{packages.tex}

\input{commands.tex}

\usepackage{tikz}
\usepackage[top=1in, bottom=1.25in, left=1.25in, right=1.25in]{geometry}

\usepackage{xcolor}
\usepackage{hyperref}

\newcommand{\bb}{\mathbf b}

\newcommand{\bk}{\mathbf k}

\newcommand{\bu}{\mathbf u}

\def\cl {\nonumber \\}
\def\el {\nonumber }

\begin{document}

\title[]{{A comparative computational study of different formulations of the compressible Euler equations for mesoscale atmospheric flows in a finite volume framework}} 

\author{M.~ Girfoglio$^1$, A.~Quaini$^2$, G.~Rozza$^1$}
\address{$^1$ mathLab, Mathematics Area, SISSA, via Bonomea 265, I-34136 Trieste, Italy; mgirfogl@sissa.it, grozza@sissa.it 
}
\address{$^2$ Department of Mathematics, University of Houston, 3551 Cullen Blvd, Houston TX 77204, USA; aquaini@uh.edu}

\begin{abstract}
We consider three conservative forms of the mildly compressible Euler equations, called CE1, CE2 and CE3, with the goal of understanding which leads to the most accurate and robust pressure-based solver in a finite volume environment.  
Forms CE1 and CE2 are both written in density, momentum, and specific enthalpy, but employ two different treatments of the buoyancy and pressure gradient terms: for CE1 it is the standard pressure splitting implemented in open-source finite volume solvers (e.g., OpenFOAM\textsuperscript{\textregistered}),
while for CE2 it is the typical pressure splitting found in computational atmospheric studies.
Form CE3 is written in density, momentum, and
potential temperature, with the buoyancy and pressure terms addressed as in CE2.  
For each formulation, we adopt a computationally efficient splitting approach. 
The three formulations are thoroughly 
assessed and compared through six benchmark tests involving dry air flow over a flat terrain or orography. We found that
all three models are able to provide accurate results for the tests with a flat terrain, although the solvers based on the CE2 and CE3 forms are more robust. As for the mountain tests, CE1 solutions become unstable, while 
the CE2 and CE3 models provide results in very good agreement with data in the literature, the CE3 model being the most accurate. Hence, 
the CE3 model is the most accurate, reliable, and robust for the simulation of mesoscale atmospheric flows when using a pressure-based approach and space
discretization by a finite volume method.

\end{abstract}

\maketitle

\section{Introduction}
\label{sec:intro}

Hydrostatic equilibrium is an atmospheric state in which the upward 
directed pressure gradient force is balanced by the downward-directed 
gravitational pull of the Earth. On average, the Earth’s atmosphere is 
very close to hydrostatic equilibrium. Because of this, for decades 
weather prediction models have relied on approximated hydrostatic equations. 
However, non-hydrostatic effects become important when the horizontal 
and vertical scales of motion are similar, which typically happens 
when horizontal scales of the order of 10 km are resolved with a mesh size of order 2 km. 
For motions of larger scale that are resolved with mesh sizes of order 10 km, 
the hydrostatic approximation is well satisfied. In the last decade, 
increasing computational power has made it possible to predict the weather with 
increasingly finer meshes (i.e., with mesh size below 10 km), thereby requiring 
the use of non-hydrostatic models able to capture small-scale mesoscale 
circulations such as cumulus convection and sea-breeze.

Almost all the non-hydrostatic mesoscale codes currently available
employ Finite Difference methods. See, e.g., \cite{WRF, Janjic1994, Benoit1997, Hodur1997, Room2001, Doms2002, Steppeler2002, Wood2014, Saito2006, Yang2008, Pielke1992, Dudhia1993, Xue2000, Arakawa2009, Heus2010}. 
More recently, solvers based on other discretization techniques have been appeared. 
Some adopted a Finite Element (FE) method \cite{marrasEtAl2013a, marrasEtAl2013b, Cote1998} 
including Discontinous Galerkin (DG) methods \cite{Giraldo2013, Kelly2012, Brdar2012}, others
a Spectral Element (SE) method \cite{clima-doc, Giraldo2013, Kelly2012, Brdar2012} 
or a Finite Volume (FV) method \cite{atlas, Gassmann2008, Wan2013, Zhao2009, Bacon2000, Walko2008, Satoh2008, Lee2009, Weller2014, Skamarock2012, Ullrich2012}. Each discretization comes with its
advantages and disadvantages. We choose a FV method for the following reasons: 
i) it enforces conservation of quantities (e.g., mass, momentum, energy) at 
the discrete level, i.e., these quantities  
remain conserved also at a local scale; ii) it can take 
full advantage of arbitrary meshes to approximate complex geometries
(e.g., orography \cite{Tissaoui2023}); iii) many fluid dynamics codes are 
based on FV methods, hence a FV-based code can appeal to a larger workforce.

All the solvers mentioned in the previous paragraph 
are sophisticated enough to reproduce complex atmospheric
processes and some are open-source. However, if one wanted to test and
assess cutting-edge numerical approaches within these open-source solvers, 
their complexity translates into a steep learning curve that could be demotivating.
In order to provide a FV-based open-source package specifically designed for the quick
assessment of new computational approaches for the simulation of 
mesoscale atmospheric flows, we created 
GEA (Geophysical and Environmental Applications) \cite{GEA}. 
Currently, GEA features a pressure-based solver for the mildly compressible Euler equations written in conservative form using density, momentum, and specific enthalpy as variables \cite{GQR_OF_clima, GirfoglioFVCA10} 
and it includes a Large Eddy Simulation techniques based on nonlinear
spatial filtering \cite{CGQR}. It is worth noticing that 
the vast majority of the software for weather prediction 
adopts density-based approaches, which were originally designed 
for high-speed compressible flows. Since atmospheric flows are
mildly compressible, it seemed more appropriate to us to 
adopt a pressure-based approach, historically devised for mildly compressible flows. 


A fundamental question that arose in the development of GEA is: what formulation 
of the compressible Euler equations leads to the most accurate and robust FV pressure-based solver?
This paper addresses this question by comparing three different conservative formulations: form CE1 in density, momentum and specific enthalpy with the standard pressure splitting implemented in relevant open-source FV solvers, e.g.,  OpenFOAM\textsuperscript{\textregistered} \cite{Weller1998}, 
form CE2, which differs from CE1 only in the fact that it uses the 
typical pressure splitting adopted in atmospheric studies, and 
form CE3 in density, momentum, and potential temperature with 
the buoyancy and the pressure terms treated as in CE2.
For each formulation, we adopt a computationally efficient splitting approach. 
The three formulations are thoroughly 
assessed and compared through six benchmark tests. We divide them into two categories: benchmarks with a flat terrain (rising thermal bubble, density current, and inertia-gravity waves) and benchmarks with orography (hydrostatic equilibrium of an initially resting atmosphere, 
steady state solution of linear hydrostatic flow, and steady state solution of linear non-hydrostatic flow). 

This paper is inspired by \cite{giraldo_2008}, 
where many of the above mentioned tests are used to compare three formulations of the Euler equations 
for a SE/DG density-based solver. Despite this inspiration, our work has some important differences with respect to \cite{giraldo_2008}. The first lies in the approach: ours is pressure-based, while it is density-based
in \cite{giraldo_2008}. The second difference is the space discretization: we use a FV method, while
\cite{giraldo_2008} focuses on Se and DG methods. 
Finally, due to our choice for space discretization we consider only conservative formulations of the Euler equations, while a non-conservative form is considered in 
\cite{giraldo_2008}. The only common formulation is CE3. 



The rest of the paper is organized as follows. Sec.~\ref{sec:pbd} describes the three forms of the Euler equations under consideration. Secs.~\ref{sec:splitting} and \ref{sec:space} discuss the time and space discretization, respectively. Then numerical results for the three benchmark tests on flat terrain are presented in Sec.~\ref{sec:flatBench}, while we report the results
related to the three benchmark tests on orography in Sec.~\ref{sec:orography}.
Conclusions are drawn in Sec.~\ref{sec:conc}.


\section{Problem definition}
\label{sec:pbd}

We consider the dynamics of dry atmosphere in a spatial domain of interest $\Omega$ by neglecting the effects of moisture, solar radiation, and heat flux from the ground.
We assume that dry air behaves like an ideal gas.
To state the equations that govern the motion of dry air, let $\rho$ be the air density, $\bu = (u, v, w)$  
the wind velocity, and $e$ the total energy density. Note that 
$e = c_v T + |\bu|^2/2 + g z$, where $c_{v}$ is the specific heat capacity at constant volume, $T$ is the absolute temperature, $g$ is the gravitational constant, and $z$ is the vertical coordinate. 

Conservation of mass, momentum, and energy for the dry atmosphere written in terms of $\rho$, $\bu$, and $e$ (i.e., in primitive variables) over a time interval of interest $(0,t_f]$ are given by: 
\begin{align}
&\frac{\partial \rho}{\partial t} + \nabla \cdot (\rho \bu) = 0 &&\text{in } \Omega \times (0,t_f], \label{eq:mass}  \\
&\frac{\partial (\rho \bu)}{\partial t} +  \nabla \cdot (\rho \bu \otimes \bu) + \nabla p   + \rho g \widehat{\bk} = \boldsymbol{f_u} &&\text{in } \Omega \times (0,t_f],  \label{eq:mome} \\
&\frac{\partial (\rho e)}{\partial t} +  \nabla \cdot (\rho \bu e) + \nabla \cdot (p \bu) = f_e &&\text{in } \Omega \times (0,t_f],
\label{eq:energy}
\end{align}
where $\widehat{\bk}$ is the unit vector aligned with the vertical axis $z$ and $p$ is pressure. In \eqref{eq:mome} and \eqref{eq:energy}, $\boldsymbol{f_u}$ and $f_e$ are possible forcing terms. To close system \eqref{eq:mass}-\eqref{eq:energy}, we need a 
thermodynamics equation of state for $p$. Following the assumption that dry air behaves like an ideal gas, we have: 
\begin{align}
p = \rho R T, 
\label{eq:p}
\end{align}
where $R$ is the specific gas constant of dry air.

For numerical stability, we add an artificial diffusion term to the momentum and energy equations:
\begin{align}
&\frac{\partial (\rho \bu)}{\partial t} +  \nabla \cdot (\rho \bu \otimes \bu) + \nabla p + \rho g \widehat{\bk} - \mu_a \Delta \bu = \boldsymbol{f_u}  &&\text{in } \Omega \times (0,t_f], \label{eq:momentum_stab}\\
&\frac{\partial (\rho e)}{\partial t} +  \nabla \cdot (\rho \bu e) +   \nabla \cdot (p \bu) - c_p \dfrac{\mu_a}{Pr} \Delta T = f_e &&\text{in } \Omega \times (0,t_f], \label{eq:energy_stab}
\end{align}
where $\mu_a$ is a constant (artificial) diffusivity coefficient and $Pr$ is the Prandtl number.
We note that this is equivalent to using a basic eddy viscosity model in Large Eddy Simulation (LES) \cite{BIL05}. 
More sophisticated LES models can be found in, e.g., \cite{CGQR, marrasNazarovGiraldo2015}.

Starting from the compressible Euler equations in formulation \eqref{eq:mass}, \eqref{eq:p}-\eqref{eq:energy_stab}, below we derive three other forms called CE1, CE2, and CE3. 
\subsection{The Euler equations in CE1 form}
Equation set CE1 states conservation of mass, momentum, and energy written in terms of $\rho$, $\bu$, and specific enthalpy $h = c_v T + p/\rho = c_p T$. 


To rewrite \eqref{eq:energy_stab} in terms of $h$, let 
\begin{align}
    K = |\bu|^2/2 \label{eq:kinetic}
\end{align} 
be the kinetic energy density. The total energy density can be written as $e = h - p/\rho + K + gz$. Then, we can rewrite eq.~\eqref{eq:energy_stab} as:
\begin{align}
\frac{\partial (\rho h)}{\partial t} +  \nabla \cdot (\rho \bu h) + 
\frac{\partial (\rho K)}{\partial t} +  \nabla \cdot (\rho \bu K) - \dfrac{\partial p}{\partial t}  +  
\rho g \bu \cdot \widehat{\bk} - \dfrac{\mu_a}{Pr} \Delta h = f_h, \label{eq:energyCE1}
\end{align}
where we have used eq.~\eqref{eq:mass} for further simplification. In \eqref{eq:energyCE1}, $f_h$ is a forcing term.

For CE1, we write the pressure $p$ as the sum of a fluctuation $p'$ with respect to a background state: 
\begin{align}
p = p_{g} + \rho g z + p', \label{eq:p_split}
\end{align}
where $p_{g}= 10^5$ Pa is the atmospheric pressure at the ground.
By plugging \eqref{eq:p_split} into \eqref{eq:momentum_stab}, we obtain:
\begin{align}
\frac{\partial (\rho \bu)}{\partial t} +  \nabla \cdot (\rho \bu \otimes \bu) + \nabla p' + gz \nabla \rho - \mu_a \Delta \bu = \boldsymbol{f_u} \quad \text{in } \Omega \times (0,t_f]. \label{eq:momentumCE1}
\end{align}
We remark that $p' \neq 0$ also in hydrostatic balance conditions because 
$\rho$ does not remain constant in the compressible regime.







In summary, the Euler equations in CE1 form are given by \eqref {eq:mass}, \eqref{eq:p}, \eqref{eq:kinetic}-\eqref{eq:momentumCE1}.
To the best of our knowledge, form CE1 has been investigated in the context of atmospheric flows only in our previous works \cite{GQR_OF_clima, CGQR, GirfoglioFVCA10}. Form CE1 was the first to be implemented in our open-source package GEA \cite{GEA}, which will be used for all the results in this paper.



\begin{rem}
Let $p_0$ and $\rho_0$ be the pressure and density in  hydrostatic balance. 
From \eqref{eq:p_split}, we have 
\begin{align}
p_0 = p_{g} + \rho_0 g z + p'_0, 
\label{eq:p_split_hyd}
\end{align} 
and from \eqref{eq:p} we have 
\begin{align}
\rho_0 = \dfrac{p_0}{R T_0}.
\label{eq:rho_hyd}
\end{align} 
The fluctuation $p'_0$ could be computed by solving:
\begin{align}
\nabla p'_0 + gz \nabla \rho_0 = \boldsymbol{0}. 
\label{eq:hydroCE1}
\end{align}
\end{rem}

\subsection{The Euler equations in CE2 form}
As mentioned in a commit by Henry Weller \cite{Weller} to open-source library OpenFOAM\textsuperscript{\textregistered} \cite{Weller1998}, pressure splitting \eqref{eq:p_split} might lead to numerical issues associated with the interaction of large gradients of the buoyancy force with mesh non-orthogonality. 
Indeed, while splitting \eqref{eq:p_split} does reduce rather 
significantly the errors due to non-orthogonality in many cases, large errors cannot be ruled out. 
This is a potentially serious issue in presence of orography or non-flat terrains in general. 

To address it, we introduce a different pressure splitting:
\begin{align}
p = p_0 + p', \label{eq:p_splitCE2}
\end{align}
where we recall that subindex 0 refers a hydrostatic variable. Note that by replacing the background state $p_g + \rho g z$ 
with $p_0$ we do get $p' = 0$ in hydrostatic balance conditions.
In fact, \eqref{eq:p_splitCE2} is the standard pressure splitting in computational atmospheric studies.
Let 
\begin{align}
\rho = \rho_0 + \rho', \label{eq:rhoPrime} 
\end{align}
with $\rho_0$ given by \eqref{eq:rho_hyd}.
By plugging \eqref{eq:p_splitCE2} and \eqref{eq:rhoPrime} into \eqref{eq:momentum_stab}, we obtain:
\begin{align}
&\frac{\partial (\rho \bu)}{\partial t} +  \nabla \cdot (\rho \bu \otimes \bu) + \nabla p' + \rho' g \widehat{\bk} - \mu_a \Delta \bu = \boldsymbol{f_u}  &&\text{in } \Omega \times (0,t_f]. \label{eq:momeCE2}
\end{align}

To compute $p_0$, 
we solve the following equation
\begin{align}
\nabla p_0 + \rho_0 g \widehat{\bk} = \boldsymbol{0}, 
\label{eq:p0_hydro}
\end{align}
with $\rho_0$ given by \eqref{eq:rho_hyd}.

Form CE2 is given by \eqref {eq:mass}, \eqref{eq:p}, \eqref{eq:kinetic}, \eqref{eq:energyCE1},  \eqref{eq:p_splitCE2}-\eqref{eq:momeCE2}. 
Note that the only difference between CE2 and CE1 is in the treatment of the buoyancy and pressure terms. 
To the best of our knowledge, no solver for the Euler equations in CE2 form has been shared with the large OpenFOAM community. This is not surprising because OpenFOAM was not originally designed for atmospheric applications
and it is the reason why we make it available via GEA \cite{GEA}.



\subsection{The Euler equations in CE3 form}

A quantity of interest for atmospheric problems is the potential temperature
\begin{align}
\theta = \frac{T}{\pi}, \quad \pi = \left( \frac{p}{p_g} \right)^{\frac{R}{c_{p}}}, \label{eq:theta}
\end{align}
i.e., the temperature that a parcel of dry air would have if it were expanded or compressed
adiabatically to standard pressure $p_g = 10^5$ Pa. 
In \eqref{eq:theta}, $\pi$ is the so-called Exner pressure.

If one solves the Euler equations in forms CE1 or CE2, $\theta$ is computed in a post-processing phase. To avoid that, we replace eq.~\eqref{eq:energy_stab} with:
\begin{align}
&\frac{\partial (\rho \theta)}{\partial t} +  \nabla \cdot (\rho \bu \theta) - \dfrac{\mu_a}{Pr} \Delta \theta = f_\theta &&\text{in } \Omega \times (0,t_f], \label{eq:energyCE3}
\end{align}
instead of \eqref{eq:energyCE1}.
In \eqref{eq:energyCE3}, $f_\theta$ accounts for possible forcing terms. As a consequence of replacing \eqref{eq:energyCE1} 
with \eqref{eq:energyCE3}, the thermodynamics equations of state \eqref{eq:p} is recast into:
\begin{align}
p = p_g\left(\dfrac{\rho R \theta}{p_g}\right)^{\dfrac{c_p}{c_v}}.
\label{eq:p2}
\end{align}

Form CE3, given by \eqref {eq:mass}, \eqref{eq:p_splitCE2}-\eqref{eq:momeCE2}, \eqref{eq:energyCE3}, \eqref{eq:p2},
is the most used formulation of the Euler equations in the literature of atmospheric simulations. 

\section{Time discretization}\label{sec:splitting}

This section presents the time discretization for the CE1, CE2 and CE3 equation sets. For this,
let $\Delta t$ be a time step, $t^n = n \Delta t$, with $n = 0, ..., N_f$ and $t_f = N_f \Delta t$. Moreover, we denote by $y^n$ the approximation of a generic quantity $y$ at the time $t^n$. We adopt the Backward Euler scheme for the discretization of the time derivatives, although other schemes are possible (e.g., Backward Differentiation Formula of order 2).

\subsection{The time-discrete CE1 form}\label{sec:CE1_td}
The CE1 equation set 
discretized in time reads: given $\rho^0$, $\bu^0$, $h^0$,  $p^0$, and $T^0$, set $K^0 = |\bu^0|^2/2$ and for $n \geq 0$ find $\rho^{n+1}, \bu^{n+1},h^{n+1},K^{n+1},p^{n+1}$, $p'^{,n+1}, T^{n+1}$ such that: 
\begin{align}
& \frac{\rho^{n+1}}{\Delta t} + \nabla \cdot (\rho^{n+1} \bu^{n+1}) = b^{n+1}_\rho, \label{eq:mass_td}  \\
&\frac{\rho^{n+1} \bu^{n+1}}{\Delta t} +  \nabla \cdot (\rho^{n+1} \bu^{n+1} \otimes \bu^{n+1}) + \nabla p'^{,n+1} + gz \nabla \rho^{n+1} - \mu_a \Delta \bu^{n+1} 
=\bb^{n+1}_\bu,  \label{eq:mom_td} \\
& \frac{\rho^{n+1} h^{n+1}}{\Delta t} +  \nabla \cdot (\rho^{n+1} \bu^{n+1} h^{n+1}) + \frac{\rho^{n+1} K^{n+1}}{\Delta t} +  \nabla \cdot (\rho^{n+1} \bu^{n+1} K^{n+1})  \cl
&\quad- \frac{p^{n+1}}{\Delta t} + \rho^{n+1} g \bu^{n+1}  \cdot \widehat{\bk}  - \dfrac{\mu_a}{Pr} \Delta h^{n+1} = b_e^{n+1}, 
\label{eq:ent_td} \\
&p^{n+1} = p_g + p'^{,n+1} + \rho^{n+1} g z, \label{eq:p_td} \\
& p^{n+1} = \rho^{n+1} R T^{n+1},\label{eq:p_td2}  \\
&h^{n+1} - h^{n} = c_p (T^{n+1} - T^n) \label{eq:T_td}, \\
& K^{n+1} = \frac{|\bu^{n+1}|^2}{2}, \label{eq:K_td}
\end{align}
where $b^{n+1}_\rho = \rho^n/\Delta t$, $\bb^{n+1}_\bu = \rho^{n}\bu^n/\Delta t + \boldsymbol{f_u}^{n+1}$, and
$b_e^{n+1} = (\rho^nh^n + \rho^n K^n - p^n)/\Delta t + f_e^{n+1}$. Notice that in \eqref{eq:T_td} we have chosen to update the value of the specific enthalpy in an incremental fashion. 

A monolithic solver for  coupled problem  \eqref{eq:mass_td}-\eqref{eq:K_td} would have an exorbitant computational cost. For computational efficiency, we adopt a splitting approach  thoroughly described in \cite{GQR_OF_clima}.


\subsection{The time-discrete CE2 form}\label{sec:CE2_td}
The CE2 form differs from the CE1 form in the fact that it introduces pressure splitting \eqref{eq:p_splitCE2}, instead of 
\eqref{eq:p_split}, and density splitting \eqref{eq:rhoPrime}, which leads to a change in the momentum equation, i.e., we take eq.~\eqref{eq:momeCE2} instead of eq.~\eqref{eq:momentumCE1}.

At time $t^{n+1}$, pressure splitting \eqref{eq:p_splitCE2} becomes
\begin{align}
&p^{n+1} = p'^{,n+1} + p_0, \label{eq:p_td22}
\end{align}
which replaces \eqref{eq:p_td}
in the time-discrete CE2 form, 
and density splitting 
\eqref{eq:rhoPrime} becomes:
\begin{align}
&\rho'^{,n+1} = \rho^{n+1} - \rho_0.\label{eq:rhoPrime2} 
\end{align}
Time discrete momentum eq.~\eqref{eq:mom_td} is replaced by:
\begin{align}
&\frac{\rho^{n+1} \bu^{n+1}}{\Delta t} +  \nabla \cdot (\rho^{n+1} \bu^{n+1} \otimes \bu^{n+1}) + \nabla p'^{,n+1} + \rho'^{,n+1} g - \mu_a \Delta \bu^{n+1} =\bb^{n+1}_\bu. \label{eq:mom_td_2} 
\end{align}
Thus, the time discrete CE2 equation set reads: given $\rho^0$, $\bu^0$, $h^0$,  $p^0$, and $T^0$, set $K^0 = |\bu^0|^2/2$ and for $n \geq 0$ find $\rho^{n+1}, \rho'^{,n+1}, \bu^{n+1},h^{n+1},K^{n+1},p^{n+1}, p'^{,n+1}, T^{n+1}$ such that \eqref{eq:mass_td}, \eqref{eq:ent_td}, \eqref{eq:p_td2}-\eqref{eq:K_td}, 
\eqref{eq:p_td22}-\eqref{eq:mom_td_2}
hold.



Since the splitting approach for 
the time discrete CE1 form in \cite{GQR_OF_clima} can be easily adapted for the time discrete CE2 equations, we opt to not report it.

\subsection{The time-discrete CE3 form}\label{sec:CE3_td}
The CE3 form differs from the CE2 form in the following: i) eq.~\eqref{eq:energyCE3} replaces eq.~\eqref{eq:energyCE1}
and ii) eq.~\eqref{eq:p2} takes the place of eq.~\eqref{eq:p}.

The time discretization of \eqref{eq:energyCE3} by the Backward Euler scheme yields:
\begin{align}
&\dfrac{\rho^{n+1} \theta^{n+1}}{\Delta t} + \nabla \cdot (\rho^{n+1} \bu^{n+1} \theta^{n+1}) - \dfrac{\mu_a}{Pr}\Delta \theta^{n+1} = b_\theta^{n+1},
\label{eq:energyCE3_n}
\end{align}
where $b_\theta^{n+1} = \rho^n \theta^n / \Delta t + f_\theta^{n+1}$. As for state equation \eqref{eq:p2}, at time $t^{n+1}$ it becomes
\begin{align}
p^{n+1} = p_g\left(\dfrac{\rho^{n+1} R \theta^{n+1}}{p_g}\right)^{\dfrac{c_p}{c_v}}.
\label{eq:p2_n}
\end{align}
So, the time discrete CE3 equation set reads: given $\rho^0$, $\bu^0$, $\theta^0$ and $p^0$, for $n \geq 0$ find $\rho^{n+1}, \rho'^{,n+1}, \bu^{n+1}$, $\theta^{n+1}, p^{n+1}, p'^{,n+1}$ such that \eqref{eq:mass_td}, 
\eqref{eq:p_td22}-\eqref{eq:mom_td_2},
\eqref{eq:energyCE3_n},
\eqref{eq:p2_n}
hold.

The adaptation of the splitting approach in \cite{GQR_OF_clima}
for the time discrete 
CE3 equations is slightly more complex, hence we describe it below.

Given $\rho^0$, $\bu^0$, $\theta^0$, and $p^0$, for $n \geq 0$ perform
\begin{itemize}
\item[-] Step 1: find first intermediate density ${\rho}^{n+\frac{1}{3}}$, density fluctuation ${\rho}'^{,n+\frac{1}{3}}$ and intermediate velocity ${\bu}^{n+\frac{1}{3}}$ such that 
\begin{align}
& \frac{{\rho}^{n+\frac{1}{3}}}{\Delta t} + \nabla \cdot (\rho^{n} \bu^{n}) = b^{n+1}_\rho,\label{eq:step1} \\
& \rho'^{,n+\frac{1}{3}} = \rho^{n+\frac{1}{3}} - \rho_0,\label{eq:step1_1} \\
& \frac{{\rho}^{n+\frac{1}{3}} {\bu}^{n+\frac{1}{3}}}{\Delta t} +  \nabla \cdot (\rho^{n} \bu^{n}\otimes {\bu}^{n+
\frac{1}{3}}) + \nabla p'^{,n} + \rho'^{,n+\frac{1}{3}} g -  \Delta {\bu}^{n+\frac{1}{3}} =\bb^{n+1}_\bu. \label{eq:step1_2} 
\end{align}
\item[-] Step 2: find potential temperature $\theta^{n+1}$ and second intermediate density ${\rho}^{n+\frac{2}{3}}$ such that
\begin{align}
& \frac{{\rho}^{n+\frac{1}{3}} \theta^{n+1}}{\Delta t} +  \nabla \cdot (\rho^{n} \bu^{n} \theta^{n+1}) - \frac{\mu_a}{Pr }\Delta \theta^{n+1} = b_\theta^{n+1}
, \label{eq:step2_1} \\
& p^{n} = p_g\left(\dfrac{\rho^{n+\frac{2}{3}} R \theta^{n+1}}{p_g}\right)^{\dfrac{c_p}{c_v}}. \label{eq:step2_3} 
\end{align}
\item[-] Step 3: find end-of-step velocity $\bu^{n+1}$, pressure $p^{n+1}$ and pressure fluctuation $p'^{,n+1}$, and end-of-step density $\rho^{n+1}$ 
such that 
\begin{align}
& \frac{{\rho}^{n+\frac{1}{3}} {\bu}^{n+1}}{\Delta t} +  \nabla \cdot (\rho^{n} \bu^n \otimes {\bu}^{n+1}) + \nabla p'^{,n+1} + \rho'^{,n+\frac{1}{3}} g -  \Delta {\bu}^{n+\frac{1}{3}} =\bb^{n+1}_\bu, \label{eq:step3} \\  
& p^{n+1} = p_0 + p'^{,n+1}, 
\label{eq:step3_2} \\
& p^{n+1} = p_0\left(\dfrac{\rho^{n+1} R \theta^{n+1}}{p_0}\right)^{\dfrac{c_p}{c_v}}, \label{eq:step3_3} \\
& \frac{\rho^{n+1}}{\Delta t} + \nabla \cdot ({\rho}^{n+\frac{2}{3}} {\bu}^{n+1}) = b^{n+1}_\rho \label{eq:step3_4}. 
\end{align}
\end{itemize}



\section{Space discretization}\label{sec:space}
We choose a finite volume method for the space discretization of the time-discrete models in Sec.~\ref{sec:CE1_td}-\ref{sec:CE3_td}.
To this end, we consider a partition of the computational domain $\Omega$ into cells or control volumes $\Omega_i$, with $i = 1, \dots, N_{c}$, where $N_{c}$ is the total number of cells in the mesh. Furthermore, we denote by
\textbf{A}$_j$ the surface vector of each face of the control volume, 
with $j = 1, \dots, M$.

This section focuses on the details of space discretization for the hydrostatic balances \eqref{eq:hydroCE1} and \eqref{eq:p0_hydro}, the buoyancy and pressure terms appearing in eq.~\eqref{eq:step1_2} and \eqref{eq:step3}, the state equations \eqref{eq:step2_3} and \eqref {eq:step3_3}, the density splitting \eqref{eq:step1_1}, pressure splitting \eqref{eq:step3_2}, and the conservation equation for the potential temperature \eqref{eq:step2_1}. Details on the space discretization for all the other equations and terms can be found in \cite{GQR_OF_clima}. 



Let us start with the hydrostatic balances.
Following the framework introduced in \cite{bottaKlein2004}, we determine the local hydrostatic reconstructions within each control volume $\Omega_i$ in order to obtain a well-balanced approximation, as opposed to
strategies working with reference states defined globally and used for entire vertical columns in case the mesh is structure vertically (as is usually the case). 
Then, the integral form of the eq.~\eqref{eq:hydroCE1} for each volume $\Omega_i$, after multiplication by $\rho_0$ and the application of the divergence operator $\nabla \cdot$ and the Gauss-divergence theorem, is given by: 
\begin{equation}
 \int_{\partial \Omega_i} \left(\rho_0 \nabla p'_0 +  \rho_0 g z \nabla \rho_0 \right) \cdot  d\textbf{A} = {0}. \label{eq:hydro1}
\end{equation}
Note that there is no time index in \eqref{eq:hydro1} because the hydrostatic balance is related to the initial condition. We approximate \eqref{eq:hydro1} as follows:
\begin{equation}
 \sum_j \left({\rho_0}_j \nabla {p'_0}_{i,j}  +  {\rho_0}_j g z_j \nabla {\rho_0}_{i,j} \right) \cdot  \textbf{A}_j  = {0}, \label{eq:hydro1_1}
\end{equation}
 where ${\rho_0}_j$ is the value of $\rho_0$ associated to the centroid of
face $j$ and $\nabla {p'_0}_{i,j}$ and $\nabla {\rho_0}_{i,j}$ are the
gradients of ${p'_0}_{i}$ and ${\rho_0}_{i}$ at faces $j$. We choose to approximate the gradients 
with second order accuracy. 
Following the same procedure for \eqref{eq:p0_hydro} leads us to:
\begin{equation}
 \sum_j \left({\rho_0}_j\nabla {p_0}_{i,j}  +  \rho_{0_j}^2 g \widehat{\bk} \right) \cdot  \textbf{A}_j  = {0}.\label{eq:hydro2_2}
\end{equation}






Concerning the buoyancy and pressure terms appearing in eq.~\eqref{eq:step1_2}, i.e.,  $\nabla p'^{,n} + \rho'^{,n+\frac{1}{3}} g$, after integration over the control volume $\Omega_i$ and division by the control volume,
we get 
\begin{align}
\nabla p'^{,n}_i + \rho'^{,n+\frac{1}{3}}_i g,
\label{eq:buoyancy_1}
\end{align}
where $p'^{,n}_i$ and $\rho'^{,n+\frac{1}{3}}_i$ are the average pressure and intermediate density fluctuations in control volume $\Omega_i$. In the rest of this section, we will use subindex $i$ to refer to average quantities in control volume $\Omega_i$. 
The gradient term is treated with a
second-order face flux reconstruction in order to suppress spurious oscillations \cite{gradrho}. This is the same scheme used for the buoyancy contribution $   g z_i \nabla \rho_i$ in the CE1 form  \cite{GQR_OF_clima}.


The space-discrete \eqref{eq:step2_3} and \eqref {eq:step3_3} are given by 
\begin{align}
p_i^{n} = p_g\left(\dfrac{\rho_i^{n+\frac{2}{3}} R \theta_i^{n+1}}{p_g}\right)^{\dfrac{c_p}{c_v}}, \quad
p_i^{n+1} = {p_0}_i\left(\dfrac{\rho_i^{n+1} R \theta_i^{n+1}}{{p_0}_i}\right)^{\dfrac{c_p}{c_v}}. \el
\end{align}
Similarly, once discretized in space eq.~\eqref{eq:step1_1} and \eqref{eq:step3_2} become:
\begin{align}
\rho'^{,n+\frac{1}{3}}_i = \rho_i^{n+\frac{1}{3}} - {\rho_0}_i, \quad
p_i^{n+1} = {p_0}_i + p'^{,n+1}_i.
\end{align}


We conclude with the fully discretized form of eq.~\eqref{eq:step2_1}. After applying the Gauss-divergence theorem,
the integral form of eq.~\eqref{eq:step2_1} for each volume $\Omega_i$ is given by:
\begin{align}
& \frac{1}{\Delta t}\int_{\Omega_i} \rho^{n+\frac{1}{3}} \theta^{n+1} d\Omega + \int_{\partial \Omega_i} (\rho^{n} \bu^{n} \theta^{n+1}) \cdot d\textbf{A} - \frac{\mu_a}{Pr} \int_{\partial \Omega_i} \nabla \theta^{n+1}  \cdot d\textbf{A}  = \int_{\Omega_i} {b}_\theta^{n} d\Omega.
\label{eq:energyCE3_disc}
\end{align}
This is approximated as follows:
\begin{align}
\frac{1}{\Delta t} {\rho}^{n+\frac{1}{3}}_i \theta^{n+1}_i + \sum_j  \varphi^n_j \theta_{i,j}^{n+1} - \dfrac{\mu_a}{Pr} \sum_j \nabla \theta_{i,j}^{n+1} \cdot \textbf{A}_j= {b}_{\theta,i}^{n+1}, \quad \varphi^n_j =(\rho^n\bu^n)_{i,j}  \cdot  \textbf{A}_j,\label{eq:e1_d} 
\end{align}
where $ \varphi^n_j$ denotes 
the convective
flux
through face $j$ of $\Omega_i$, which is computed by a linear interpolation of the values from
the adjacent cells. 

\section{Numerical results for tests with flat terrain}\label{sec:flatBench}
In this section, we compare the results obtained with the three forms of the Euler equations for benchmark tests featuring a flat terrain.
We consider three classical benchmarks that
have been widely used to assess atmospheric dynamical cores: the rising thermal bubble (Sec.~\ref{sec:bubble}), the density current (sec.~\ref{sec:Straka}), and  the non-hydrostatic inertia-gravity waves (Sec.~\ref{sec:gravity_waves}).
All test cases involve a perturbation of a neutrally stratified atmosphere. However, for the rising bubble and the density current tests the atmosphere is initially at rest and the background potential temperature is uniform, while for the inertia-gravity waves test there is an initial horizontal flow and
the background potential temperature is stratified with a Brunt-V\"ais\"al\"a frequency. 
We use the setting from \cite{ahmadLindeman2007, Feng2021} for the thermal bubble benchmark
and the setting provided in \cite{carpenterDroegemeier1990,strakaWilhelmson1993}
for the density current benchmark. 
The setting for the inertia-gravity waves test case is taken from \cite{giraldo_2008, bonaventura2023}.

None of the benchmarks in this section has an exact solution. 
In \cite{Skamarock1994}, one finds an analytic solution for the inertia-gravity waves test that holds for the
Boussinesq equations. Such solution does not hold for
the fully compressible Euler equations considered here.  
Hence, for all the tests  we can only have a 
comparison with other numerical data available in the literature.  

Finally, we would like to point out that the results obtained with the CE1 model for the rising bubble and the density current 
have been already presented in \cite{GQR_OF_clima}.
We report them here as well to facilitate the comparison with the results given by the CE2 and CE3 models.

\subsection{Rising thermal bubble} \label{sec:bubble}

The computational domain in the $xz$-plane is $\Omega=[0, 5000]\times[0, 10000]$ m$^2$ and the time interval of interest is $(0, 1020]$ s. Impenetrable, free-slip boundary conditions are imposed on all walls. The initial temperature for the CE1 and CE2 models is defined by 
\begin{equation}
T^0 = 300 - \dfrac{gz}{C_p} + 2\left[ 1 - \frac{r}{r_0} \right] ~ \textrm{if $r\leq r_0=2000~\mathrm{m}$}, \quad T^0 = 300 - \dfrac{gz}{c_p}
~ \textrm{otherwise},
\label{warmEqn01}
\end{equation}
with $r = \sqrt[]{(x-x_{c})^{2} + (z-z_{c})^{2}}$, $(x_c,z_c) = (5000,2000)~\mathrm{m}$ \cite{ahmadLindeman2007,ahmad2018} and with 
$c_p = R + c_v$, $c_v = 715.5$ J/(Kg K), $R = 287$ J/(Kg K).
This is equivalent to the following initial potential temperature perturbation: 
\begin{equation}
\theta^0 = 300 + 2\left[ 1 - \frac{r}{r_0} \right] ~ \textrm{if $r\leq r_0=2000~\mathrm{m}$}, \quad\theta^0 = 300 ~ \textrm{otherwise},
\label{warmEqn01_bis}
\end{equation}
for the CE3 model. The CE1 and CE2 models  
define the initial specific enthalpy as:
\begin{equation}\label{eq:h0}
h^{0} = c_p T^0.
\end{equation}
These initial conditions refer to a neutrally stratified atmosphere with uniform background potential temperature of $ 300~\mathrm{K}$ perturbed by a circular bubble of warmer air. The initial density $\rho^0$ is computed by solving the hydrostatic balance, i.e., eq. \eqref{eq:hydroCE1} for CE1 and eq. \eqref{eq:p0_hydro} for CE2 and CE3. 
The initial velocity field is zero everywhere.

We consider five different meshes with uniform resolution $h = \Delta x = \Delta z = 250, 125, 62.5$, $31.25, 15.625$~m.  
Following \cite{ahmadLindeman2007, GQR_OF_clima}, we set $\mu_a = 15$ and $Pr = 1$.
Both of these are ah-hoc values provided in \cite{ahmadLindeman2007} to stabilize the numerical simulations. As mentioned in Sec.~\ref{sec:pbd}, more sophisticated LES models can be used without affecting the conclusions of this study.

\begin{figure}[htb]
\centering
 \begin{overpic}[width=0.2\textwidth]{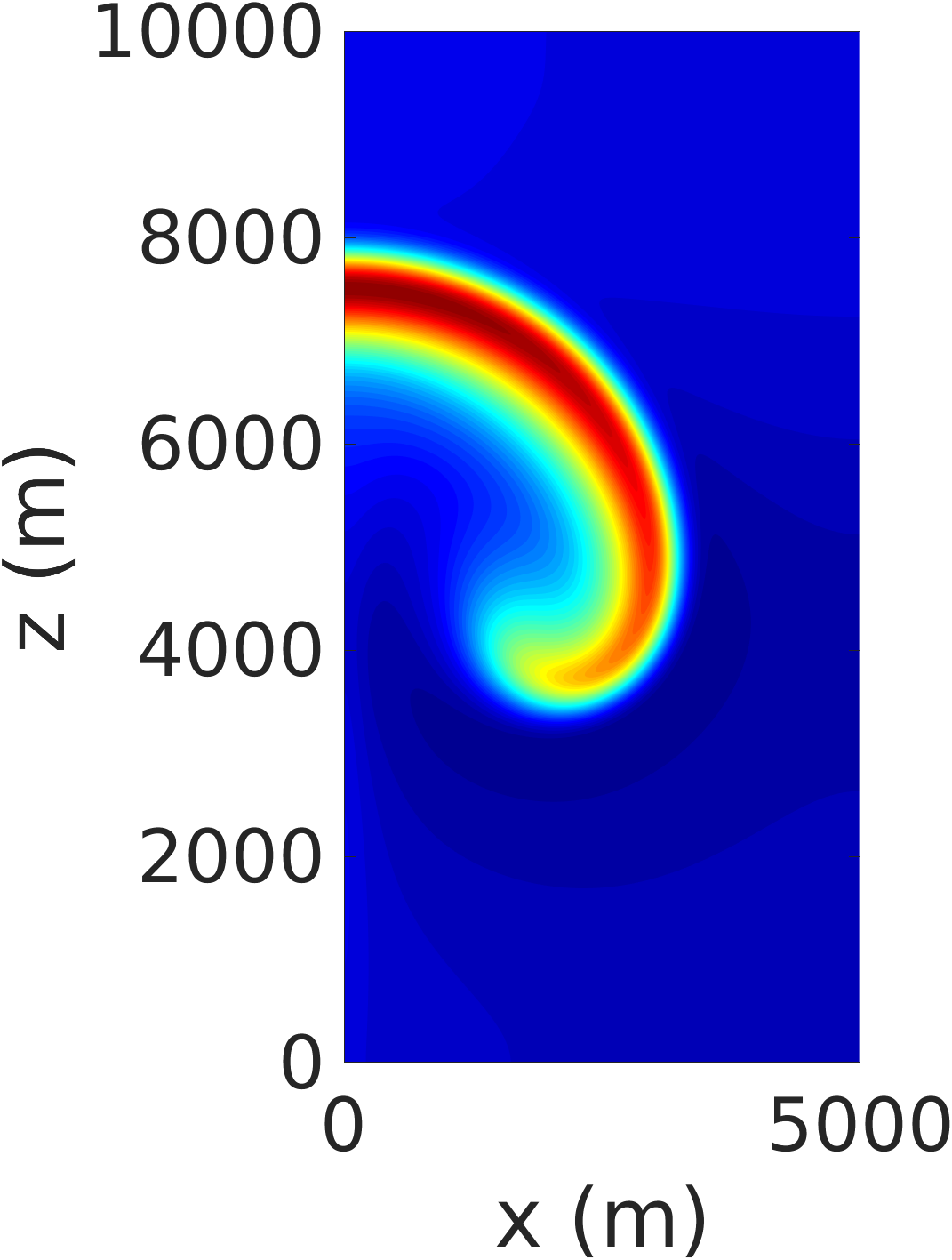}
        \put(40,87){\textcolor{white}{CE1}}
      \end{overpic}
 \begin{overpic}[width=0.2\textwidth]{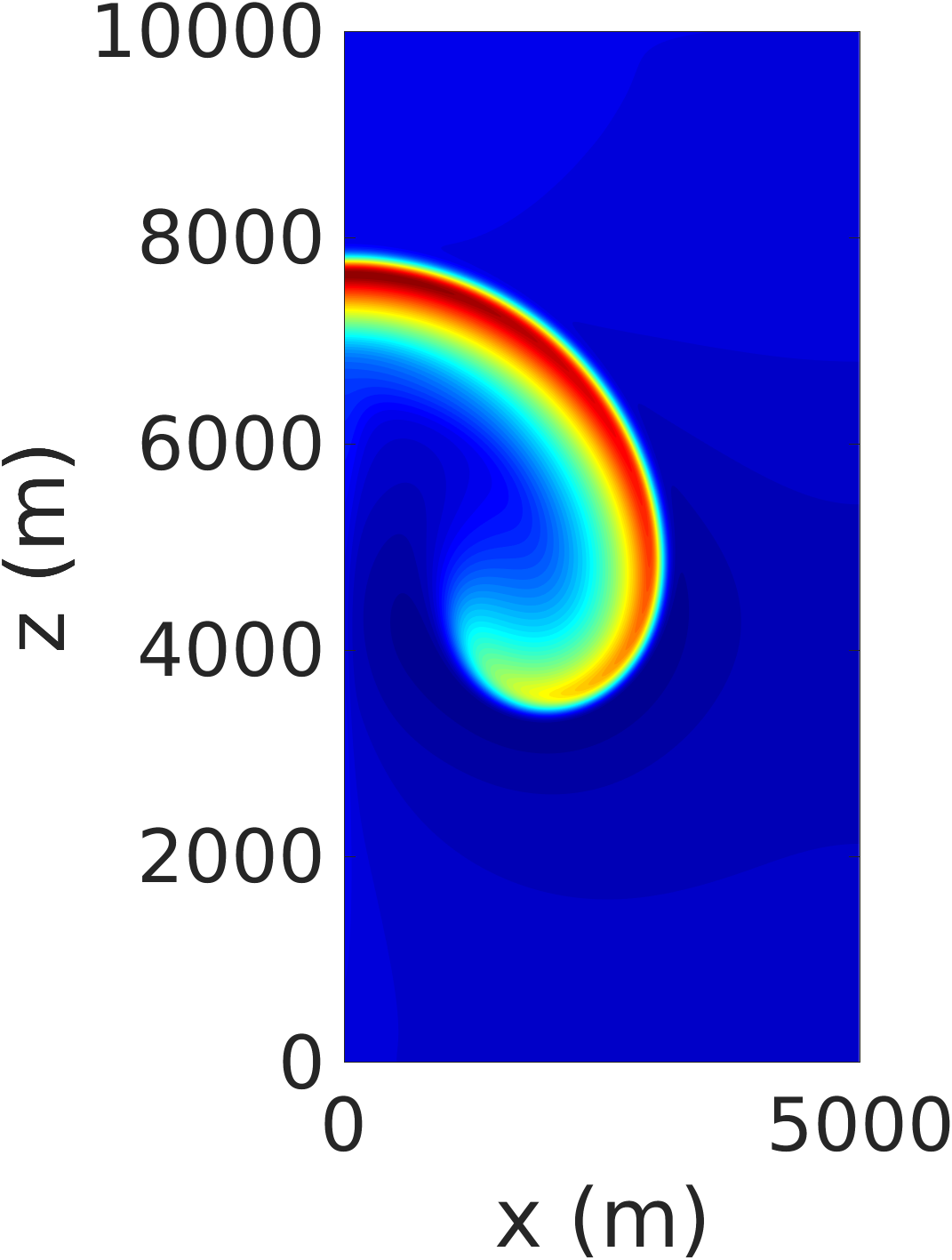}
        \put(40,87){\textcolor{white}{CE2}}
      \end{overpic}
      \begin{overpic}[width=0.235\textwidth]{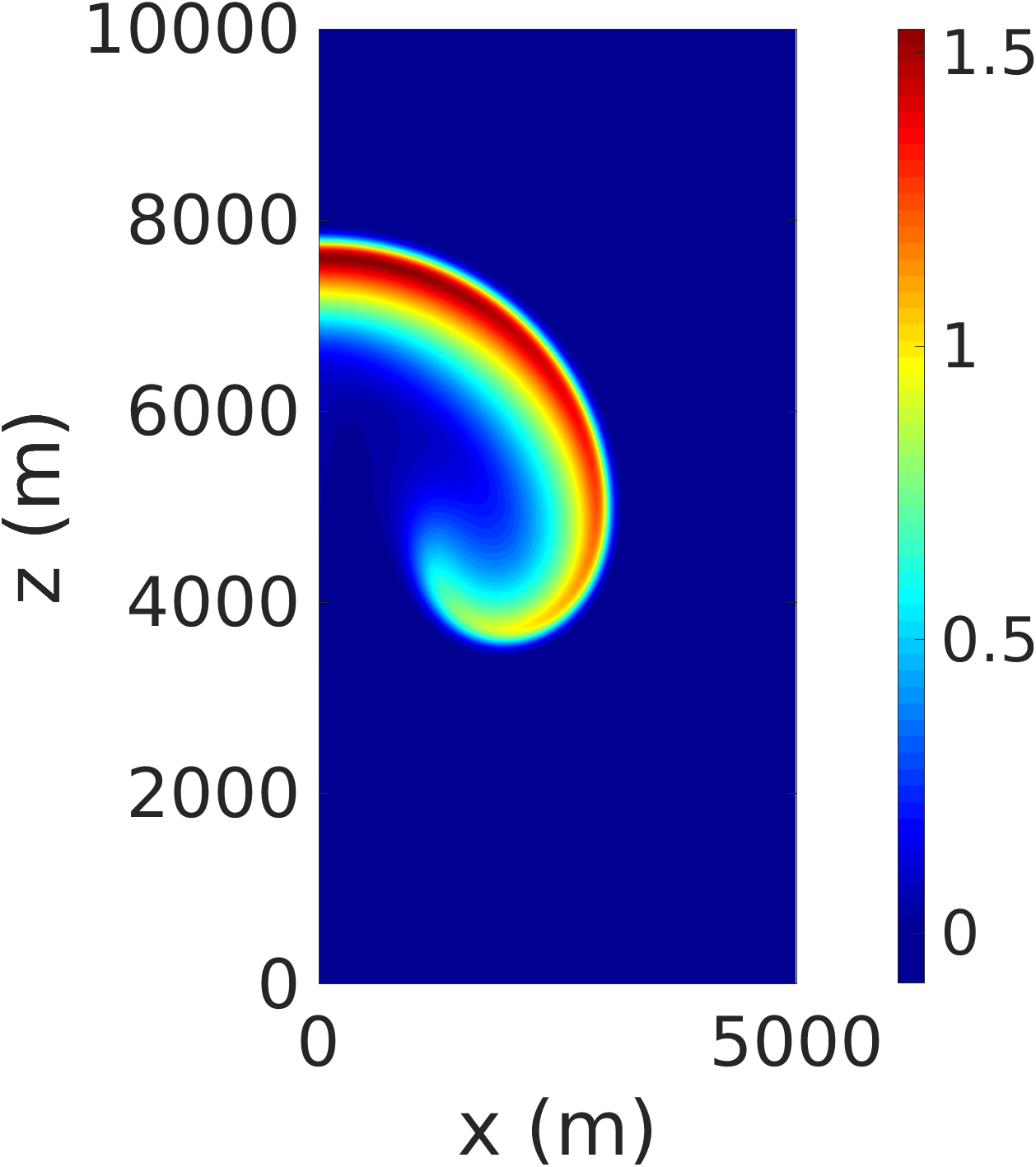}
        \put(40,87){\textcolor{white}{CE3}}
      \end{overpic}
\caption{Rising thermal bubble: perturbation of potential temperature at $t = 1020$ s computed with mesh $h = 15.625$ and 
time step $\Delta t = 0.1$ s.}
\label{fig:RTB3}
\end{figure}

Fig.~\ref{fig:RTB3} reports the perturbation of potential temperature $\theta'$ at $t = 1020$ s computed by CE1, CE2 and CE3 models with mesh $h = 15.625$ m and $\Delta t = 0.1$ s.  
By $t = 1020$ s, the air warmer than the ambient has risen due to buoyancy and deformed into a mushroom shape due to shearing motion. 
The results in Fig.~\ref{fig:RTB3} agree well
with those reported in the literature. See, e.g., \cite{ahmadLindeman2007,ahmad2018,marrasNazarovGiraldo2015}. 
We remark that, to facilitate the comparison of the panels, we have forced the colorbar to range from -0.084 to 1.54 coinciding with the minimum and maximum $\theta'$ over the three models (see Table \ref{tab:2}).

\begin{figure}[htb]
\centering
 \begin{overpic}[width=0.45\textwidth]{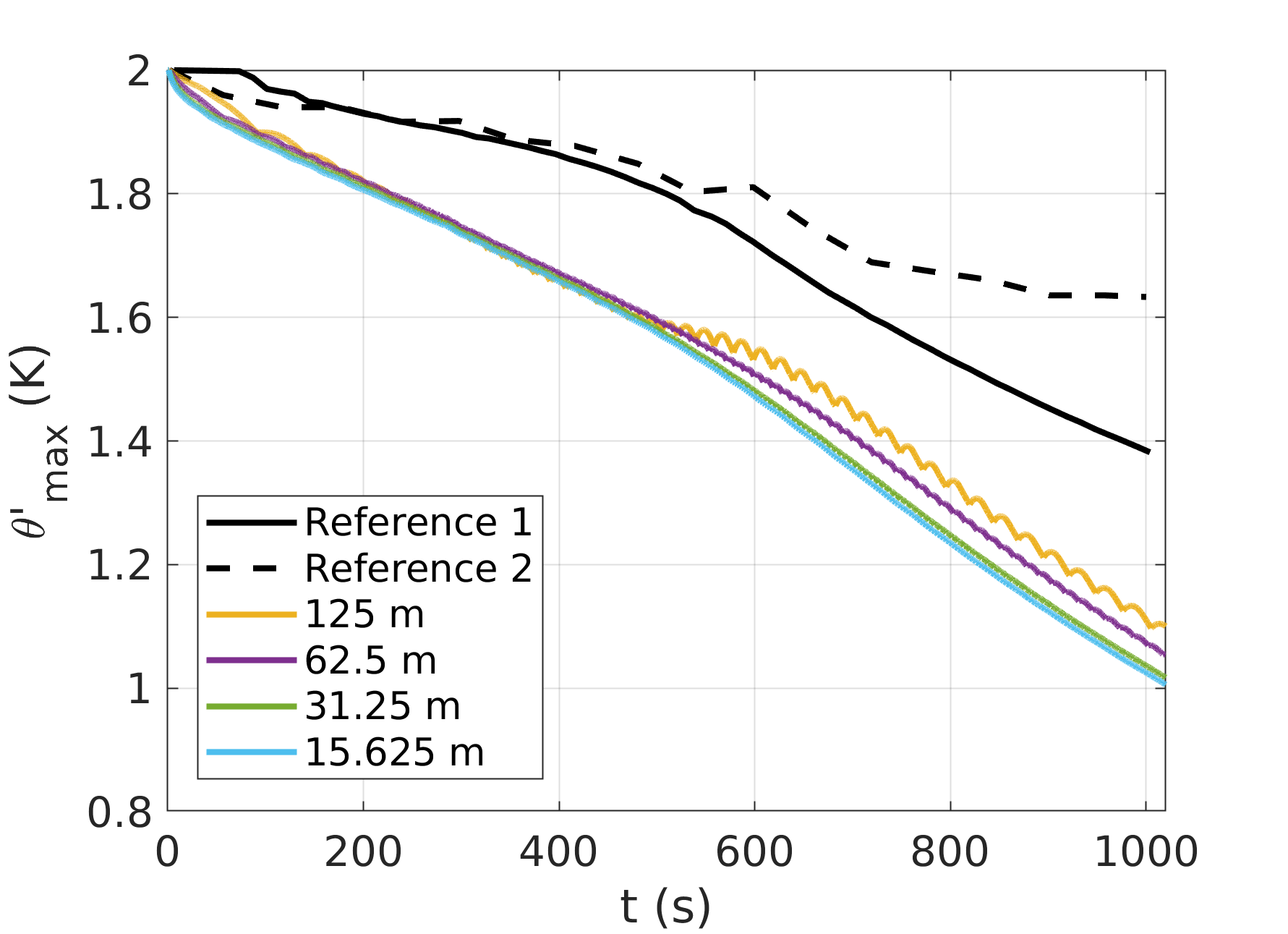}
        \put(90,73){CE1 model}
      \end{overpic}
 \begin{overpic}[width=0.45\textwidth]{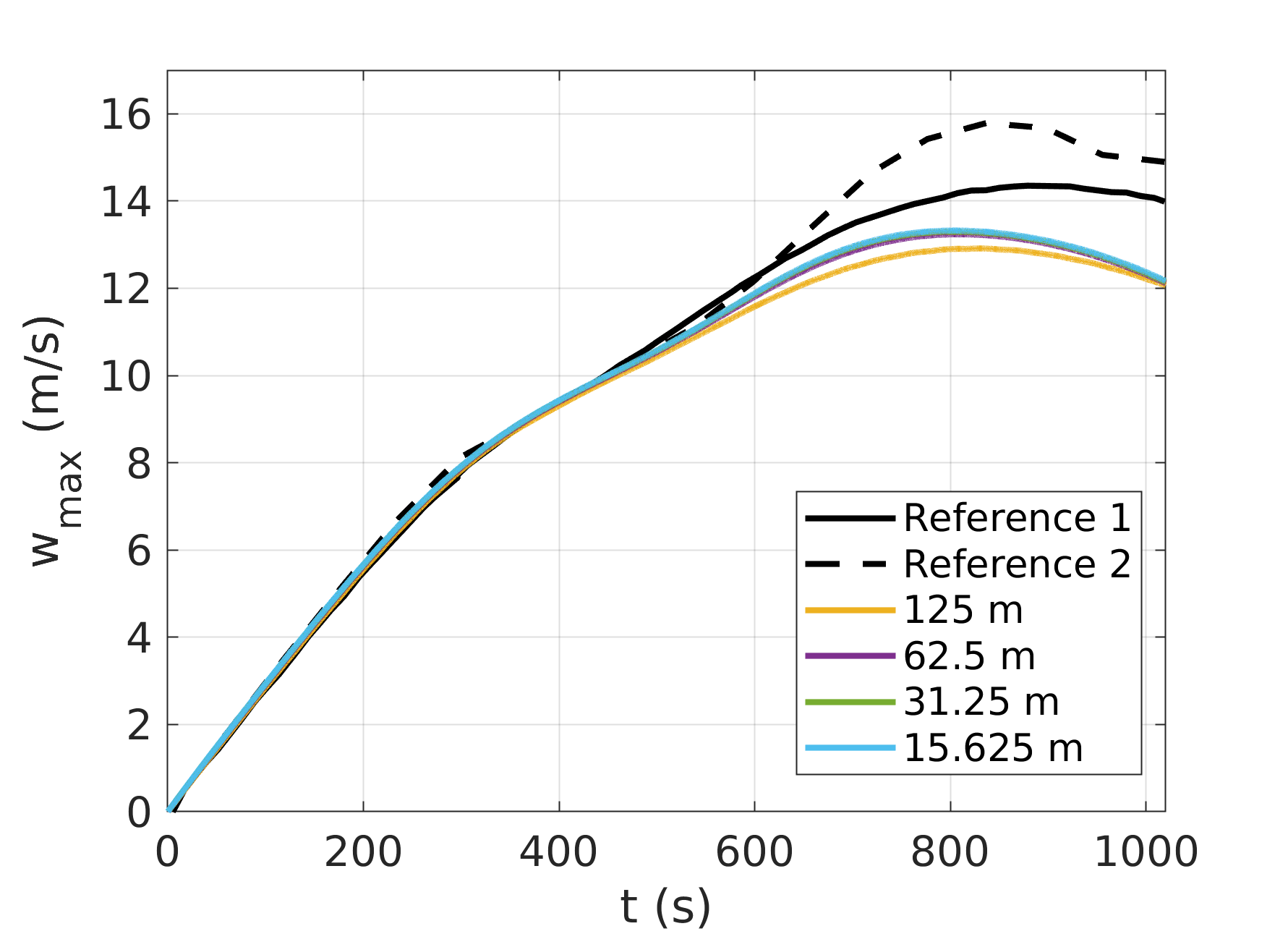}
      \end{overpic}\\
      \vskip .2cm
\begin{overpic}[width=0.45\textwidth]{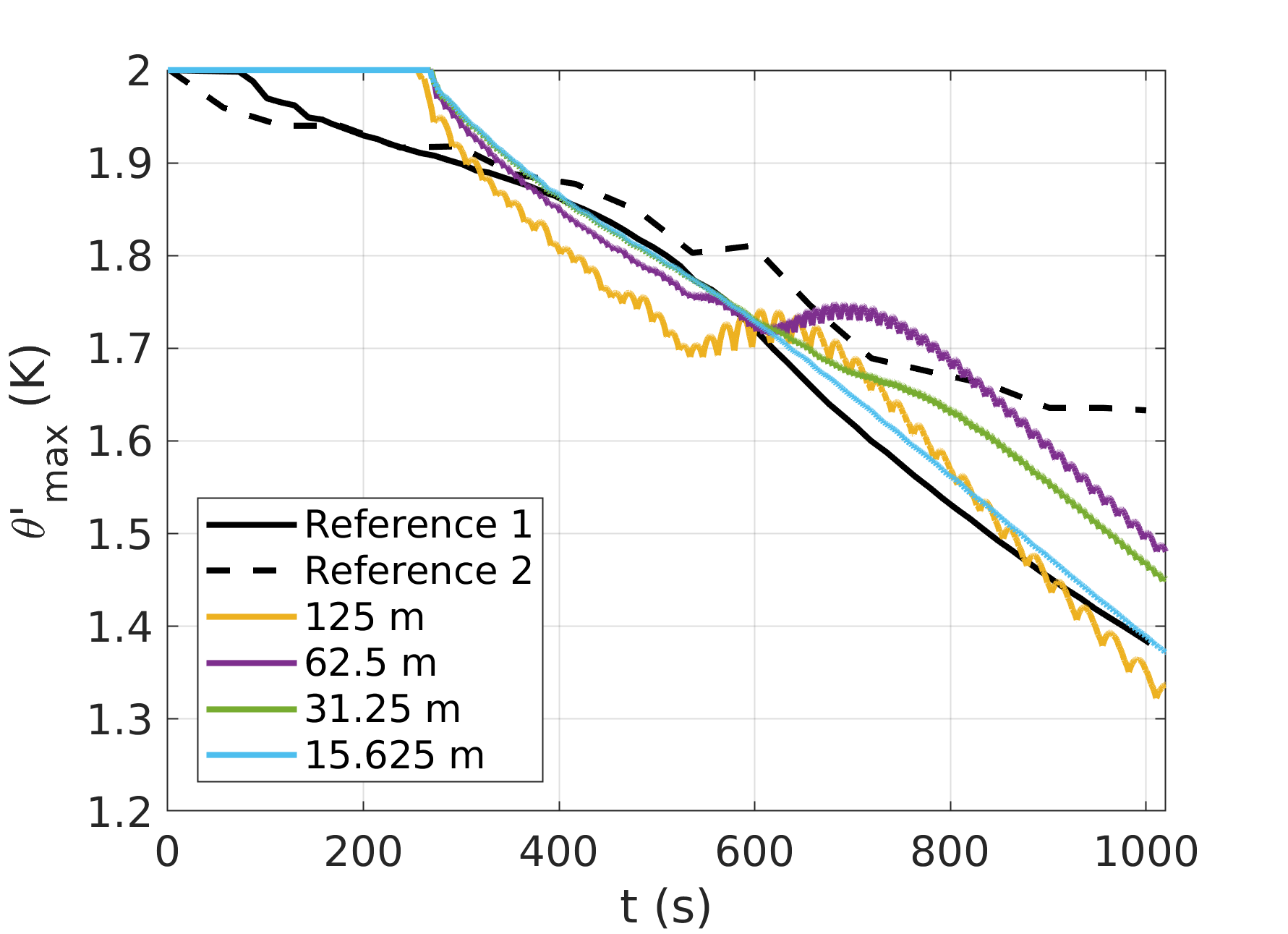}
        \put(90,73){CE2 model}
      \end{overpic}
      \begin{overpic}[width=0.45\textwidth]{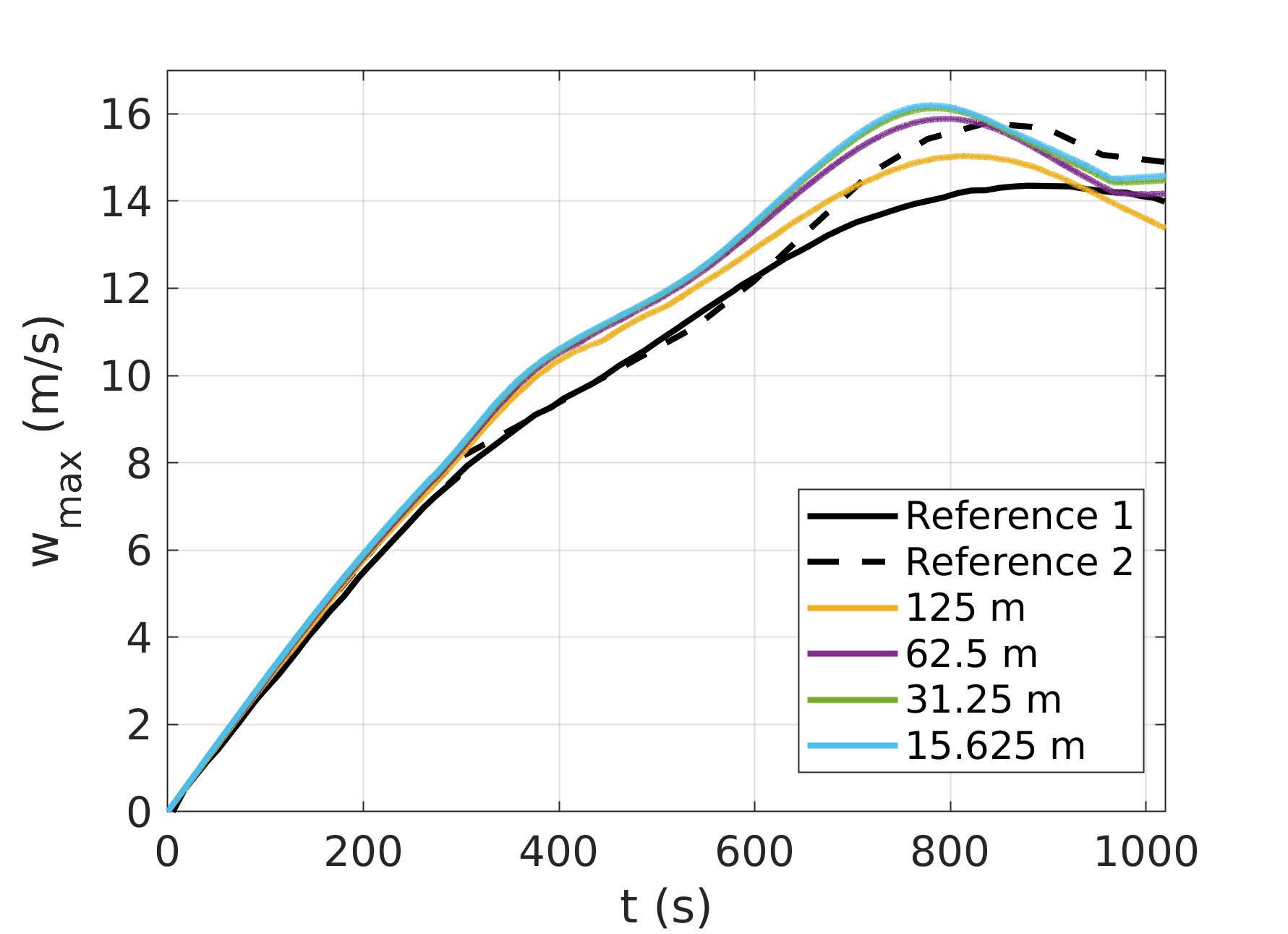}
      \end{overpic}\\
      \vskip .2cm
            \begin{overpic}[width=0.45\textwidth]{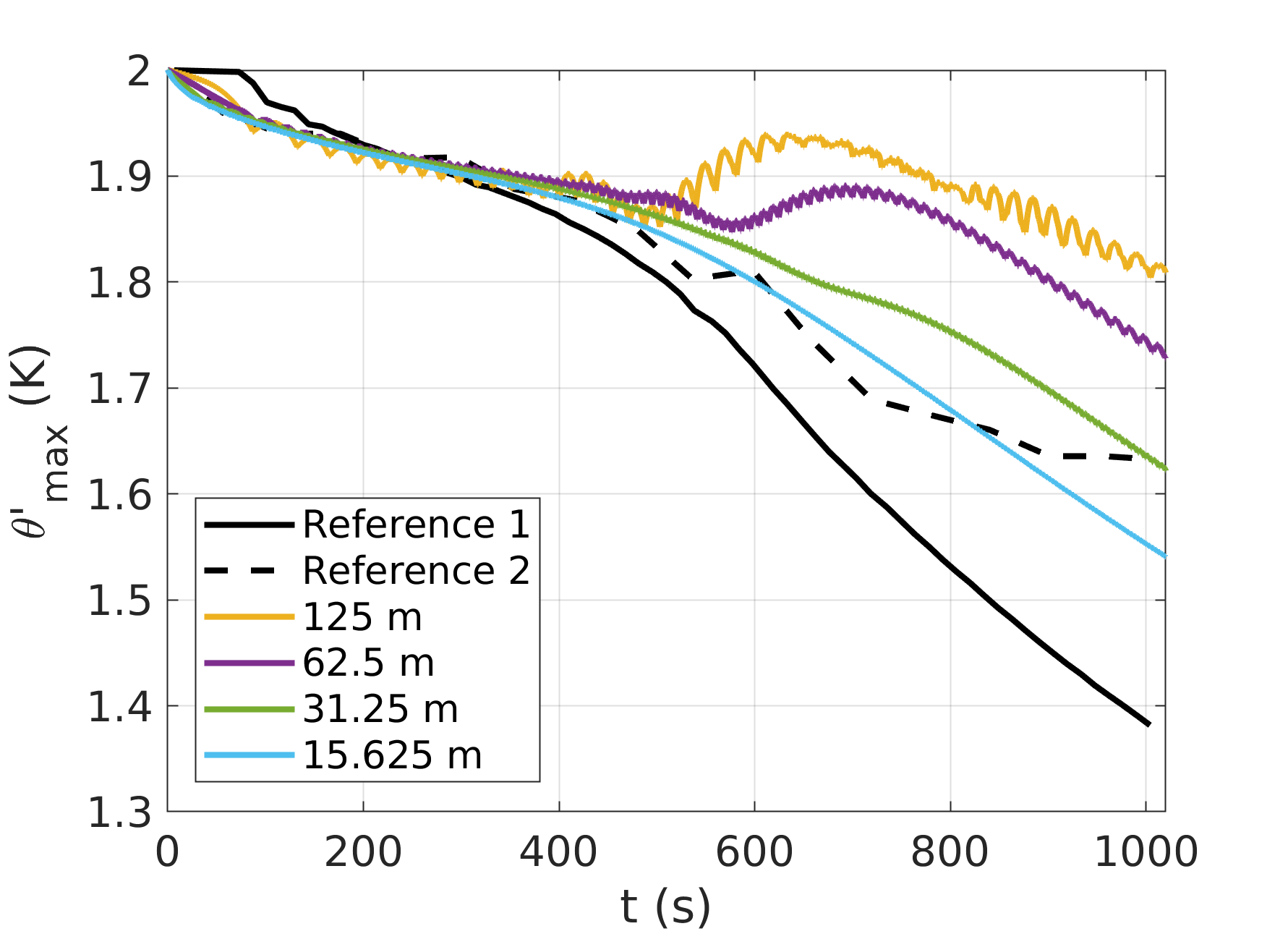}
        \put(90,73){C3 model}
      \end{overpic}
                  \begin{overpic}[width=0.45\textwidth]{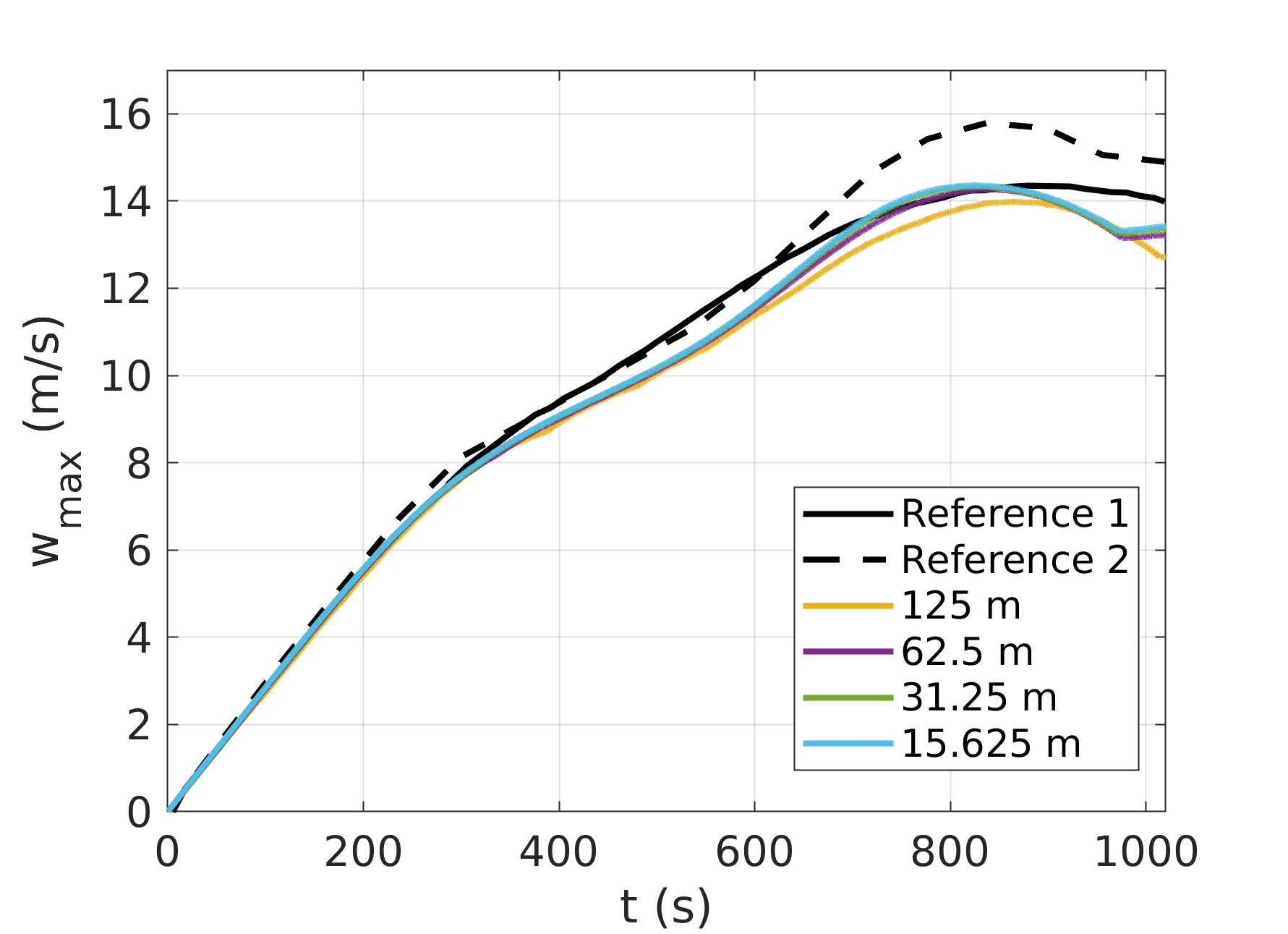}
      \end{overpic}
\caption{Rising thermal bubble: time evolution of the maximum
perturbation of potential temperature $\theta'_{max}$ (first column) and the maximum vertical component of the velocity $w_{max}$ (second column) computed with the CE1 model (first row), CE2 model (second row), and the C3 model (third row)
and all the meshes under consideration. The time step is set
to $\Delta t = 0.1$ s.
The reference curves are taken from \cite{ahmadLindeman2007} and refer to resolution 125 m.}
\label{fig:RTB1}
\end{figure}

Fig.~\ref{fig:RTB1} shows
the time evolution of the maximum perturbation of potential temperature $\theta'_{max}$ and maximum vertical component of the velocity $w_{max}$ 
computed by the CE1, CE2 and CE3 models with all the meshes, together with 
the two references curves from \cite{ahmadLindeman2007}
associated to mesh resolution of 125 m. 
The first reference curve, denoted with Reference 1, 
is obtained with a density-based approach developed from a Godunov-type scheme that employs flux-based wave decompositions for the solution of the Riemann problem. The second reference curve, 
denoted with Reference 2, is computed by the WRF model using a 
fifth-order centred finite difference scheme for 
horizontal advection and a third-order scheme for vertical advection (generally recommended by the developers of WRF).
The authors of \cite{ahmadLindeman2007} start from 
the  CE3 form. 

Let us start from the top row in 
Fig.~\ref{fig:RTB1}, which shows the results for CE1. We observe that the evolution of $\theta'_{max}$ computed with mesh $h = 125$ m is affected by spurious oscillations and do not affect the evolution of $w_{max}$. In addition, 
we see that $\theta'_{max}$ and $w_{max}$ 
computed with meshes $h = 31.25$ m and $h = 15.625$ m are practically overlapped for the entire time interval, indicating that we are close to convergence. The ``converged'' $w_{max}$ overlaps with the values of Reference 1 till about $t = 500$ s and it remains close to them till about $t = 800$ s. 
However, the agreement of the ``converged'' $\theta'_{max}$ with the results from \cite{ahmadLindeman2007} is poor.
These trends are confirmed from
Table \ref{tab:2}  which reports the extrema for the vertical velocity $w$ and potential temperature perturbation $\theta'$ at $t = 1020$ s, together with the values extracted from the figures in \cite{ahmadLindeman2007}. 
The reasons for this mismatch are discussed in \cite{GQR_OF_clima}.


\begin{table}[htb]
\begin{tabular}{|c|c|c|c|c|c|} \hline
Model & Resolution [m] & $w_{min}$ (m/s) & $w_{max}$ (m/s) & $\theta'_{min}$ (K) & $\theta'_{max}$ (K) \\
 \hline
 CE1 & 15.625 & -10.02 & 12.17 & -0.038 & 1\\
 CE2 & 15.625  & -11.28 & 14.58 & -0.084 & 1.37\\
 CE3 & 15.625  &  -10 & 13.14 & 0 & 1.54\\  Ref.~1 & 125 & -7.75 & 13.95 & -0.01 & 1.4 \\
 Ref.~2 & 125 & N/A & N/A & -0.06 & 1.65 \\
 \hline  
\end{tabular}
\caption{Rising thermal bubble: minimum and maximum vertical velocity $w$ and potential temperature perturbation $\theta'$ at $t = 1020$ s for all the models under consideration and for References 1 and 2 from \cite{ahmadLindeman2007}.
}\label{tab:2}
\end{table}

Next, we discuss the results for CE2 shown in the second row of Fig.~\ref{fig:RTB1}. Like in the case of the CE1 model, the evolution of $\theta'_{max}$ computed with mesh $h = 125$ m is affected by spurious oscillations that do not affect the evolution of $w_{max}$. Oscillations are visible for $\theta'_{max}$ computed with mesh $h = 62.5$ m as well, but they disappear at higher resolutions. 
The $\theta'_{max}$ computed with meshes $h = 31.25$ m and $h = 15.625$ m overlap up to $t = 600$ s, after which the curves grow apart up to about 0.1 K. Instead, the curves for $w_{max}$ computed with meshes $h = 31.25$ m and $h = 15.625$ m overlap for the entire time interval. 
The “converged” $w_{max}$ overlaps with the reference curves till about $t = 200$ s and it remains close to either Ref.~1 or Ref.~2 for the rest of the time interval.
All the computed $\theta_{max}$
fall close to either Ref.~1 or Ref.~2 for $0 \leq t \leq 80$ and $t \geq 350$. The lack of agreement for $80 \leq t \leq 350$ is due to the fact that the computed $\theta_{max}$ holds steady at 2K up to $t \approx 280$ s for all the meshes. 
Overall, the results given by 
CE2 model compare with the data from \cite{ahmadLindeman2007}
better the CE1 results.
See also Table \ref{tab:2}. 

Let us now look at the third row in Fig.~\ref{fig:RTB1}, which shows the results for CE3.
The evolution of $\theta'_{max}$ computed with
mesh $h = 125$ m and $h = 62.5$ m is affected by spurious oscillations, like in the case of the CE2 model. 
The $\theta'_{max}$ computed with meshes $h = 31.25$ m and $h = 15.625$ m overlap
up to $t = 400$ s and then they grow apart, similarly to what observed for the CE2 model.
Once again, the evolution of $w_{max}$ does not show any oscillation for any mesh. In addition, all the curves for $w_{max}$ are very close to each other and to Ref. 1 for the entire time interval, while showing the same kink at $t \approx 950$ as in Ref. 2. The curves for $\theta_{max}$ are more spread apart, with the curve corresponding to mesh $h = 15.625$ m closest to Ref.~2, in particular for $t < 400$ s. This is an improvement over the CE2 model.
From this first set of experiments, we can conclude that overall the CE3 model provides the best comparison with the data from \cite{ahmadLindeman2007}.
See also Table \ref{tab:2}.   

\begin{figure}[htb]
\centering
 \begin{overpic}[width=0.45\textwidth]{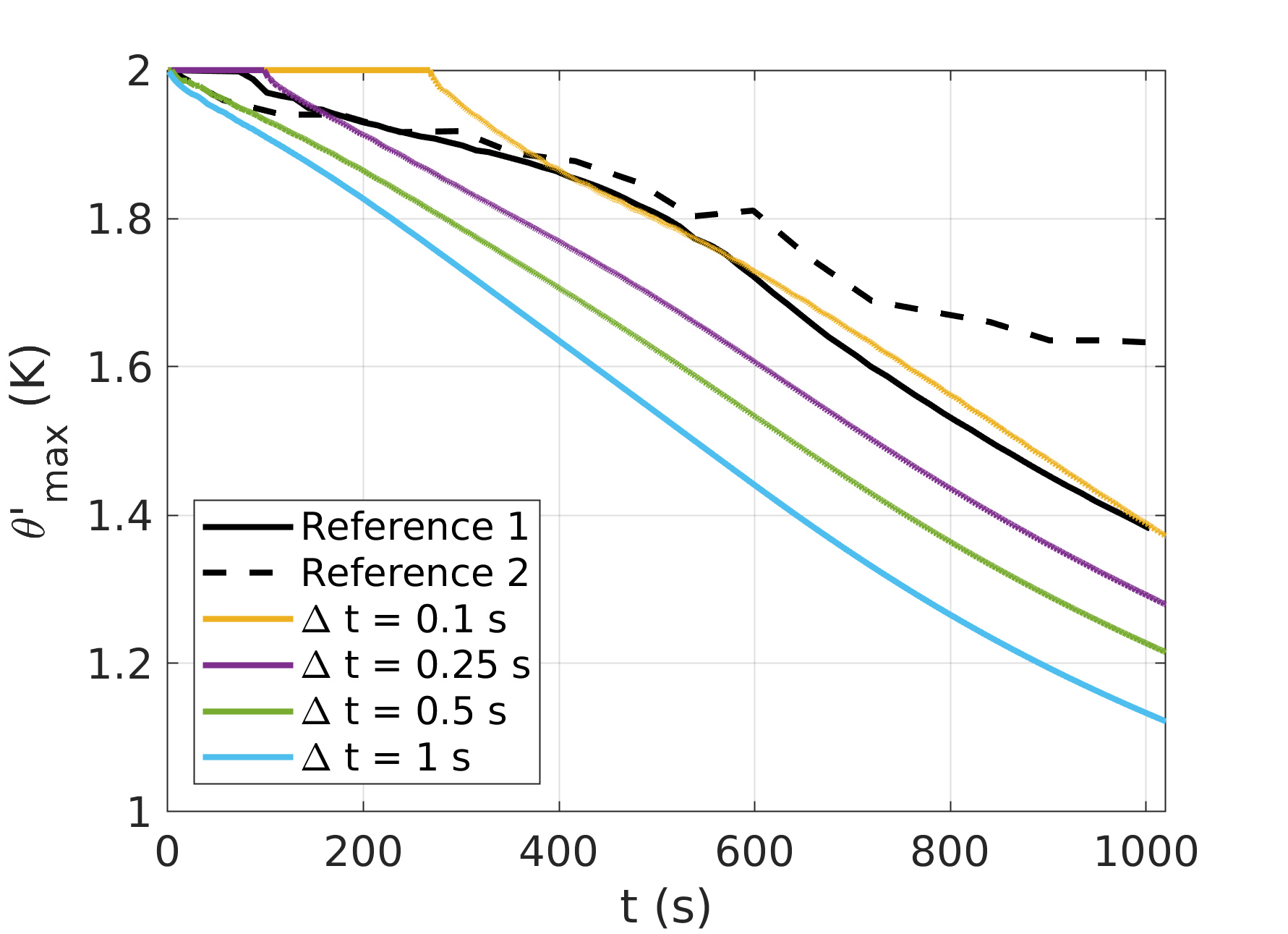}
        \put(90,72.5){CE2 model}
      \end{overpic}
 \begin{overpic}[width=0.45\textwidth]{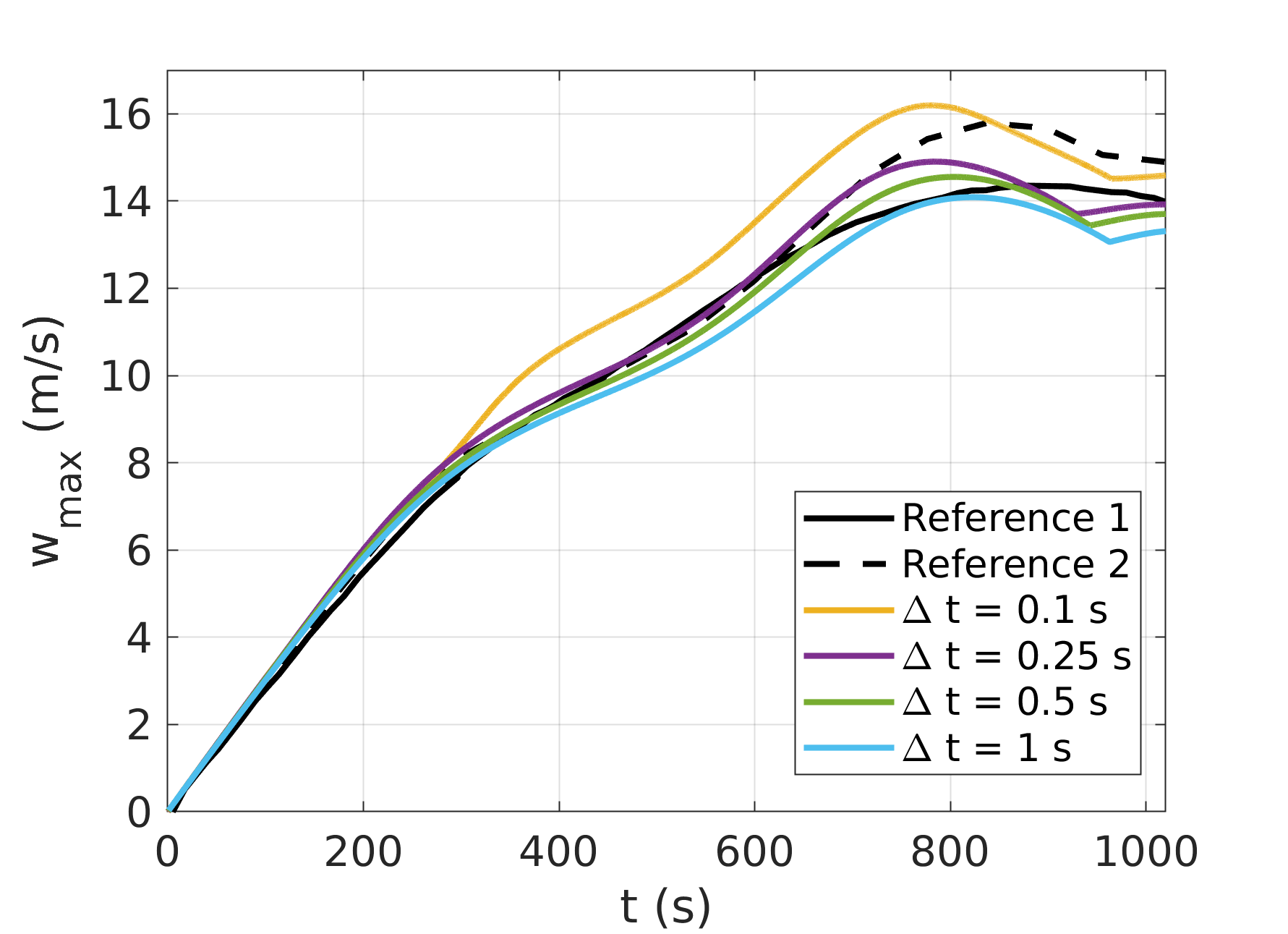}
      \end{overpic}\\
      \begin{overpic}[width=0.45\textwidth]{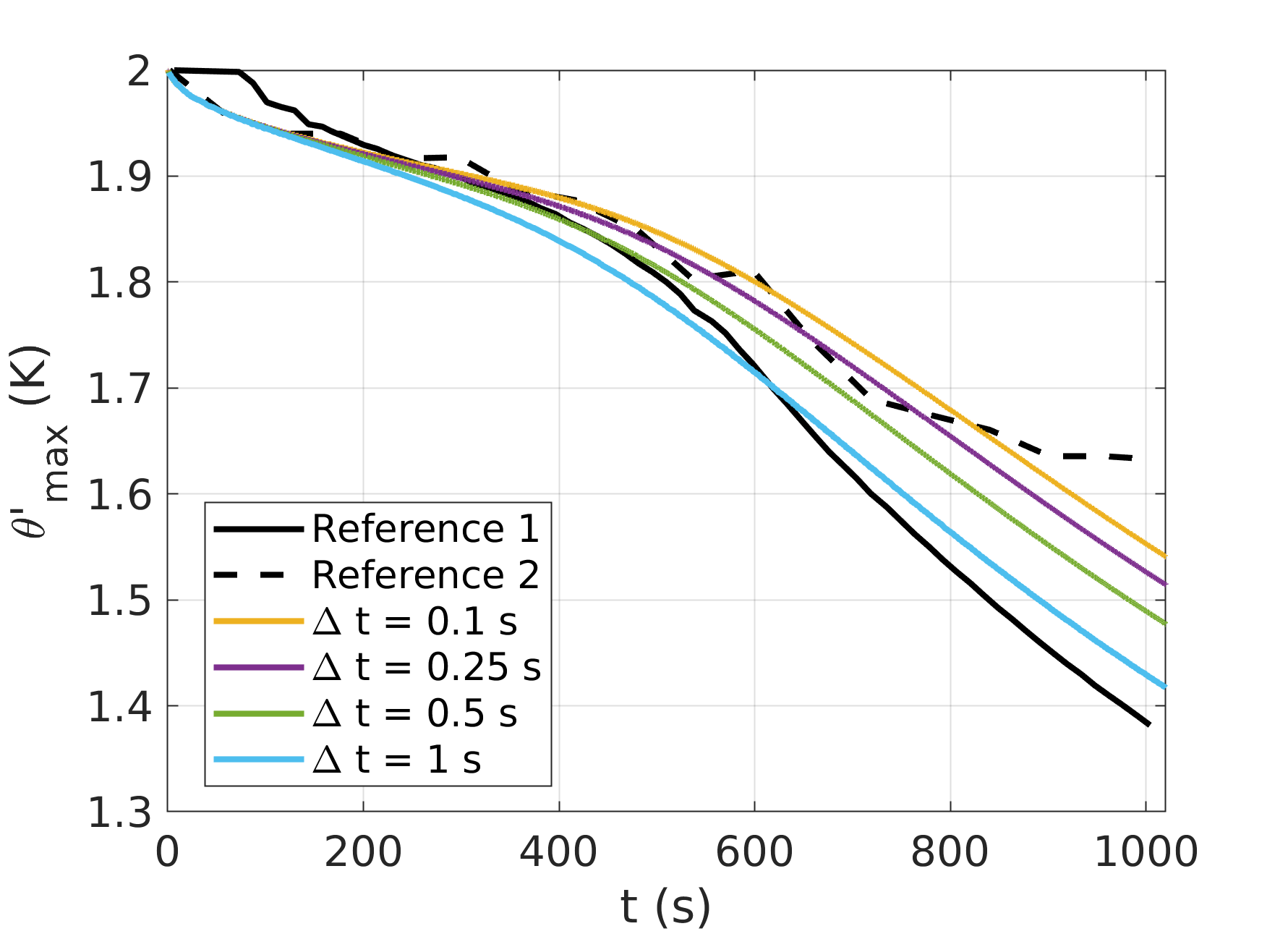}
        \put(90,72.5){CE3 model}
      \end{overpic}
 \begin{overpic}[width=0.45\textwidth]{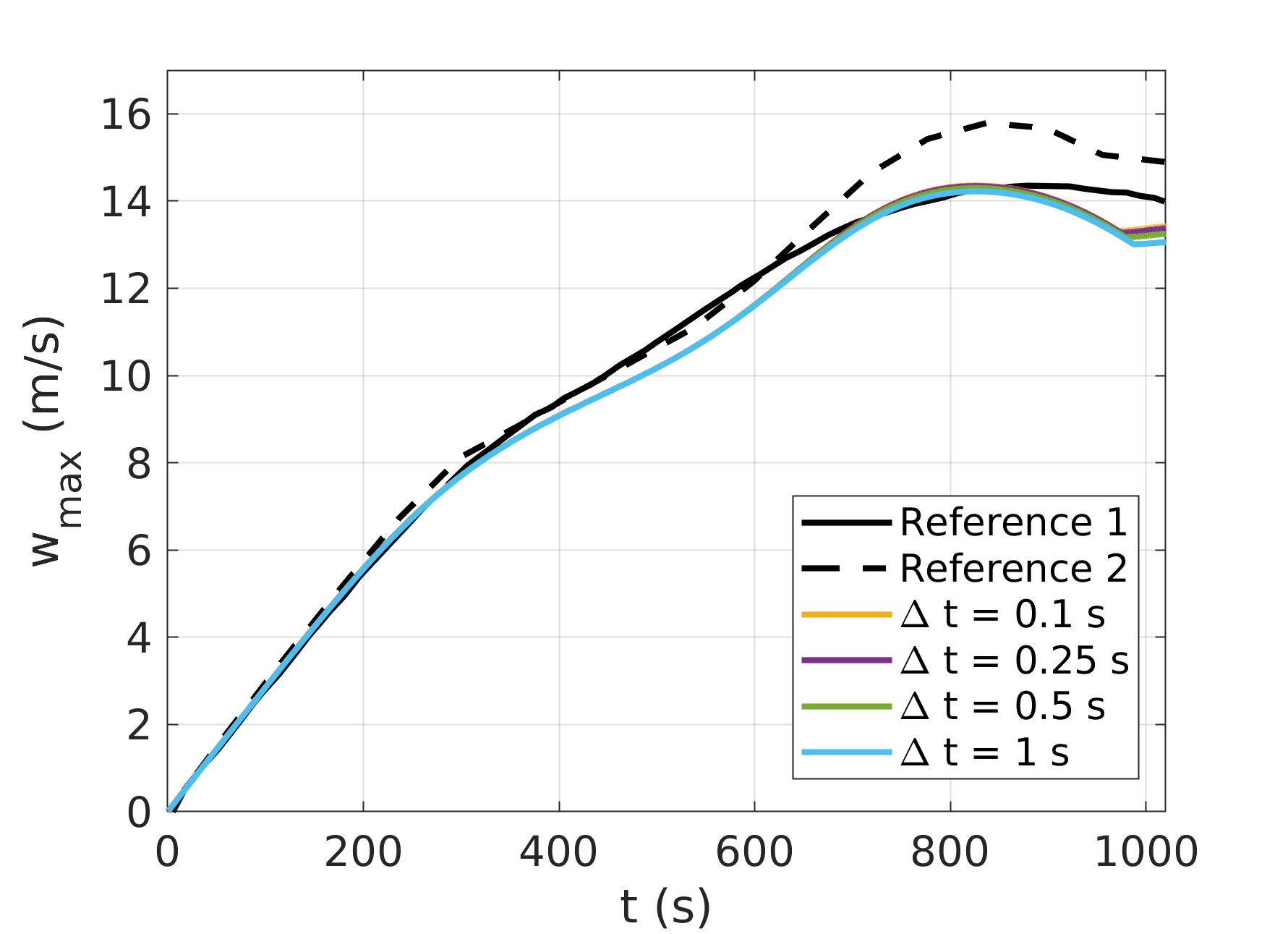}
      \end{overpic}
\caption{Rising thermal bubble: time evolution of the maximum
perturbation of potential temperature $\theta'_{max}$
(first column) and the maximum vertical component of the velocity $w_{max}$ (second column) computed with the CE2 model (top row)
and the CE3 model (bottom row) and different values of $\Delta t$.
The mesh used for the simulations is $h = 15.625$ m.
The reference values are taken from \cite{ahmadLindeman2007} and refer to resolution 125 m, with fixed CFL (0.9) for Reference 1  and fixed time step (0.25 s) for Reference 2. 
}
\label{fig:RTB4}
\end{figure}

Fig.~\ref{fig:RTB4} reports the time evolution of $\theta'_{max}$ and $w_{max}$ computed with mesh 15.625 m and $\Delta t = 0.1, 0.25, 0.5, 1$ s. The results in Fig.~\ref{fig:RTB4} refer to the CE2 and CE3 model because the CE1 solutions become unstable for $\Delta t \geq 0.25$ 
The effect of an increase in $\Delta t$ on the computed $\theta'_{max}$ is to lower its value as time progresses. In the case of the CE2 model, this means that the results move away from Ref.~1 and 2, while all the $\theta'_{max}$ computed with the CE3 model fall between the Ref.~1 and 2 curves.
The computed $w_{max}$ is less sensitive to a variation of $\Delta t$. This is true for the CE2 model when $\Delta t \geq 0.25$ s and, even more so, for the the CE3 model. 

\begin{figure}[htb]
\centering
\begin{overpic}[width=0.2\textwidth]{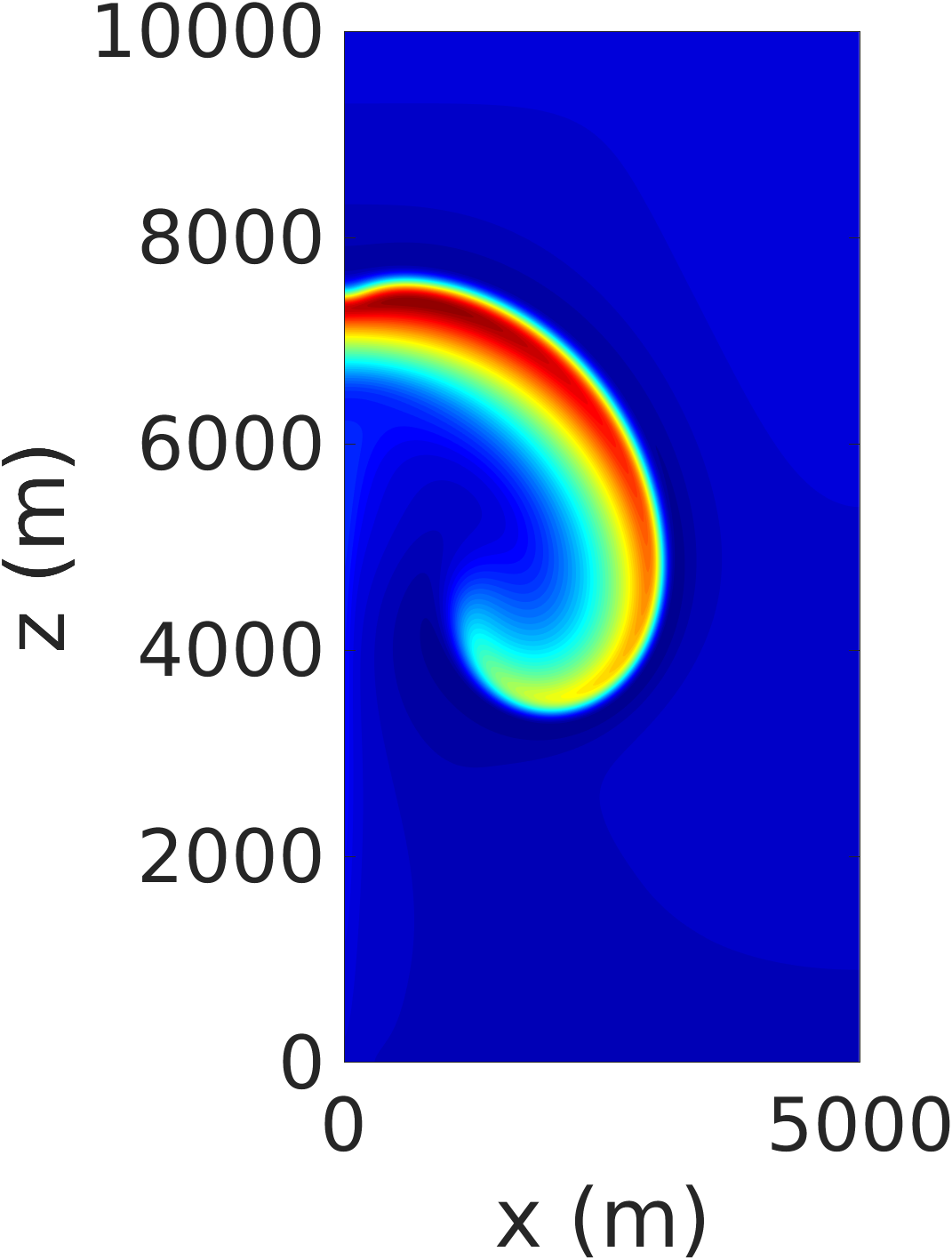}
        \put(40,87){\textcolor{white}{CE2}}
      \end{overpic}
 \begin{overpic}[width=0.235\textwidth]{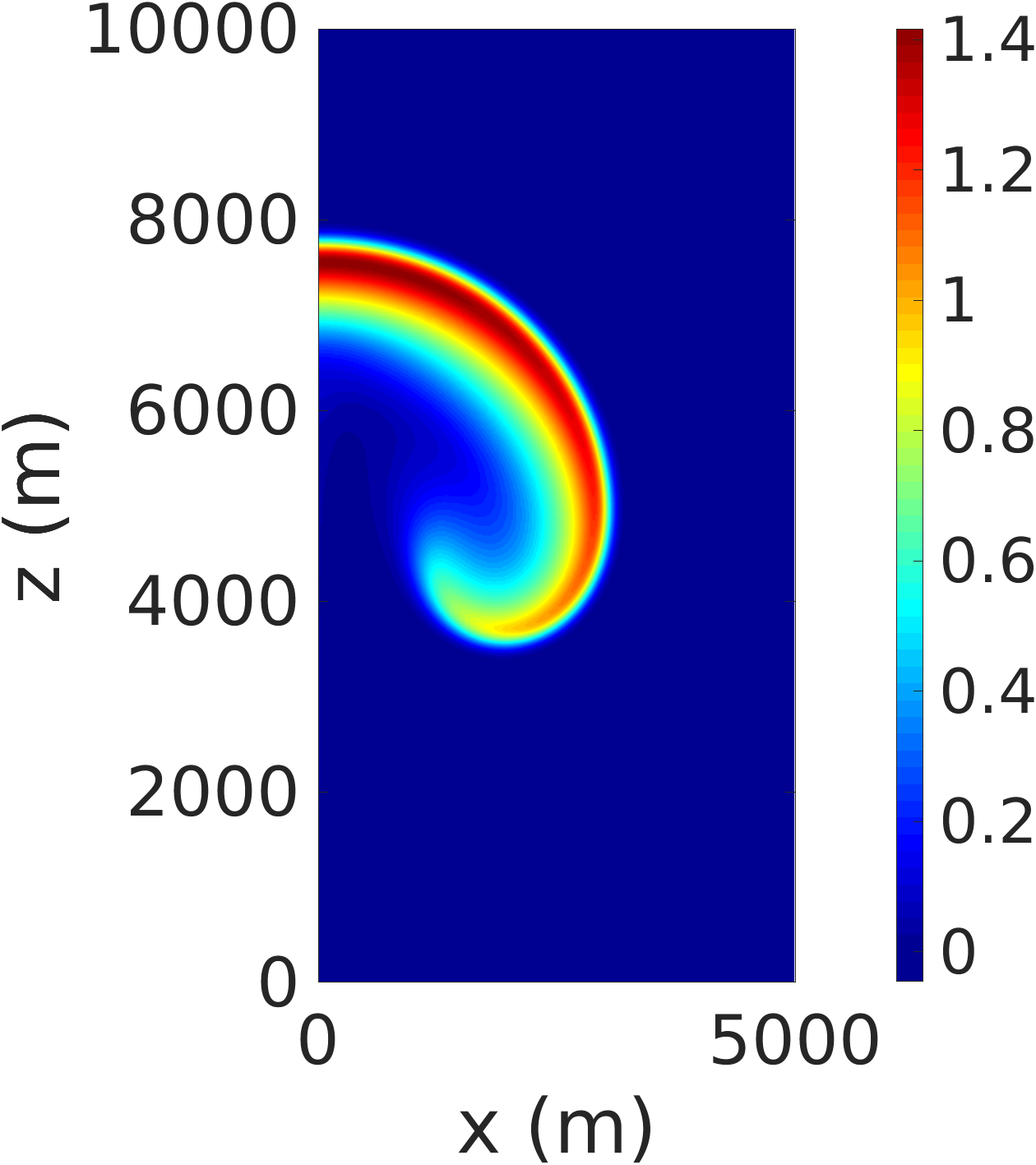}
        \put(40,87){\textcolor{white}{CE3}}
      \end{overpic}\\
\caption{Rising thermal bubble: perturbation of potential temperature at $t = 1020$ s computed with mesh $h = 15.625$ and 
time step $\Delta t = 1$ s.}
\label{fig:RTB3_dt1}
\end{figure}

We conclude this subsection with the plots of $\theta'$ at $t = 1020$ s 
computed by the CE2 and CE3 model with mesh $h = 15.625$ m and $\Delta t = 1$ s shown in Fig.~\ref{fig:RTB3_dt1}. While both simulations are stable and produce results comparable with the 
data from \cite{ahmadLindeman2007}, the CE2 solution shows an
unphysical dent at the top of the bubble. 
On the other hand, the CE3 model is able to reproduce the proper shape, showing a significantly greater robustness and reliability (compare Fig.~\ref{fig:RTB3_dt1} with Fig.~\ref{fig:RTB3}).


\subsection{Density current}\label{sec:Straka}
The computational domain in the $xz$-plane is $\Omega=[0,25600]\times[0,6400]~\mathrm{m}^2$ and the time interval of interest is $(0,900]$ s. 
Impenetrable, free-slip boundary conditions are imposed on all the walls. 
The initial temperature $T^0$ for the CE2 model 
is given by:
\begin{equation}
T^0 = 300 - \dfrac{gz}{c_p} - \frac{15}{2}\left[  1 + \cos(\pi r)\right] ~ \textrm{if $r\leq 1$},\quad T^0 = 300 - \dfrac{gz}{c_p}
~ \textrm{otherwise},
\label{dcEqn1_1}
\end{equation}
where $r = \sqrt[]{\left(\frac{x-x_{c}}{x_r}\right)^{2} + \left(\frac{z-z_{c}}{z_r}\right)^{2}}$, with $(x_r,z_r)=(4000, 2000)~{\rm m}$ and $(x_c,z_c) = (0,3000)~\mathrm{m}$.
This corresponds to the following initial potential temperature for the
CE3 model:
\begin{equation}
\theta^0 = 300 - \frac{15}{2}\left[  1 + \cos(\pi r)\right] ~ \textrm{if $r\leq 1$},\quad\theta^0 = 300
~ \textrm{otherwise}.
\label{dcEqn1}
\end{equation}
Notice that in this case the initial perturbation is a 
bubble of cold air. 
The initial specific enthalpy is given by eq.~\eqref{eq:h0}.
The initial density $\rho^0$ is computed by solving the hydrostatic balance, i.e., eq.~\eqref{eq:hydroCE1} for the CE1 model and eq.~\eqref{eq:p0_hydro} for the CE2 and CE3 models. The initial velocity field is zero everywhere.

We consider two uniform meshes with size $h = \Delta x = \Delta z = 25, 50$ m, which provide a solution close to convergence for this benchmark \cite{strakaWilhelmson1993,ahmadLindeman2007,giraldo_2008,marrasEtAl2013a,marrasNazarovGiraldo2015} when
the time step is set to $\Delta t $ = 0.1 s. 
Following \cite{strakaWilhelmson1993,ahmadLindeman2007}, we set $\mu_a = 75$ and $Pr = 1$. Like in the case of the previous benchmark, these are ah-hoc values used to stabilize the numerical simulations.

\begin{figure}[htb]
\centering
 \begin{overpic}[width=0.5\textwidth]{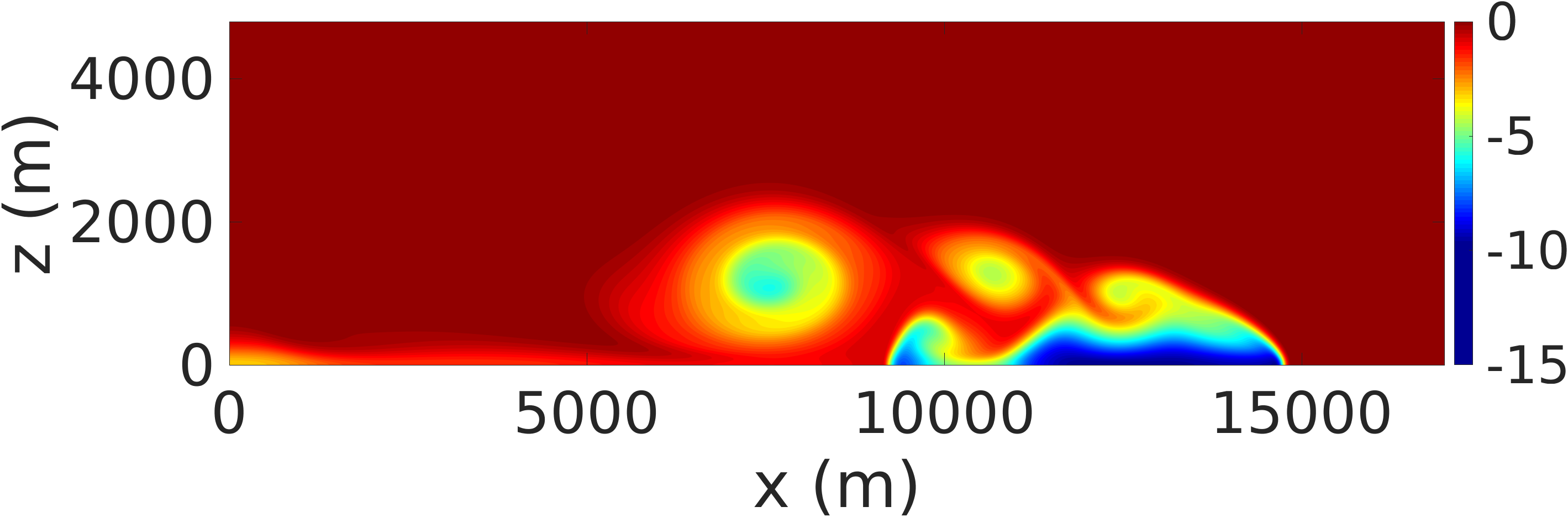}
        \put(75,25){\textcolor{white}{CE1}}
      \end{overpic}\\
 \begin{overpic}[width=0.5\textwidth]{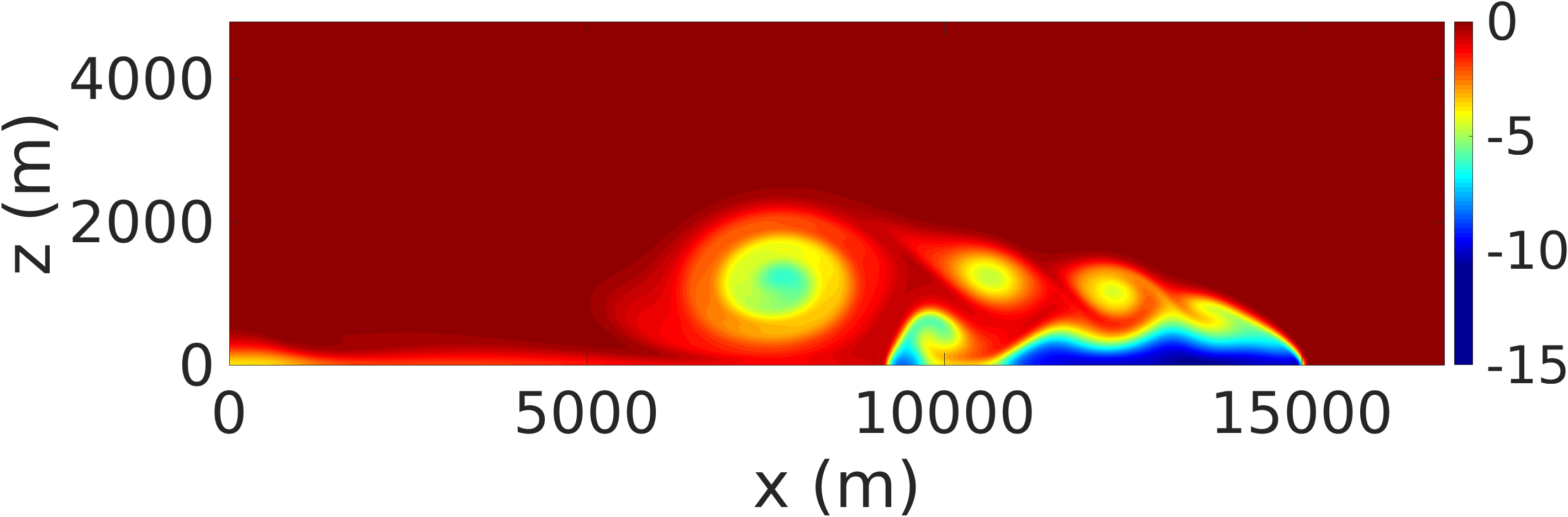}
      \put(75,25){\textcolor{white}{CE2}}
      \end{overpic}\\ 
      \begin{overpic}[width=0.5\textwidth]{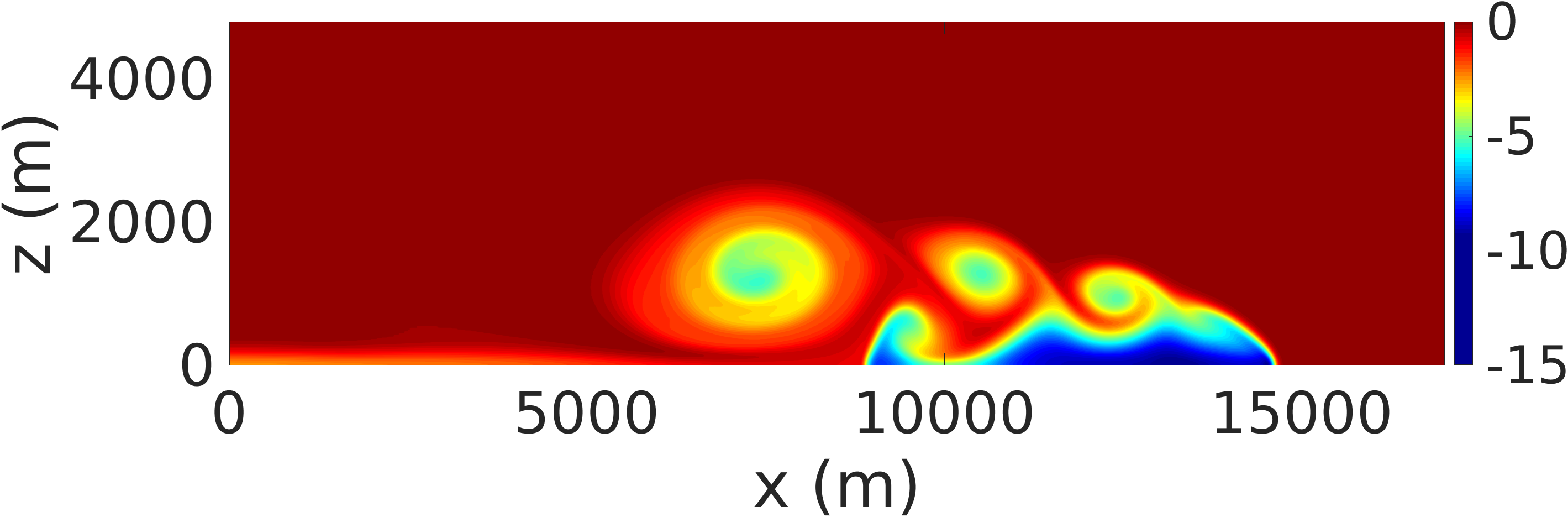}
       \put(75,25){\textcolor{white}{CE3}}
      \end{overpic}\\
\caption{Density current: perturbation of potential temperature $\theta'$ at $t$ = 900 s computed by CE1, CE2 and CE3 models with mesh $h = 25$ m and time step $\Delta t = 0.1$ s.
}
\label{fig:DC2}
\end{figure}

We illustrate in Fig.~\ref{fig:DC2} the perturbation of potential temperature $\theta'$ at $t = 900$ s
computed with CE1, CE2 and CE3 equations and mesh $h = 25$ m.  
We see some difference near the smallest (rightmost) rotor structure that results from a Kelvin-Helmholtz type instability. 
Overall, these results are in good agreement with those reported in the literature. See, e.g., \cite{strakaWilhelmson1993,ahmadLindeman2007,giraldo_2008,marrasEtAl2013a,marrasNazarovGiraldo2015}.

\begin{figure}[htb]
\centering
 \begin{overpic}[width=0.475\textwidth]{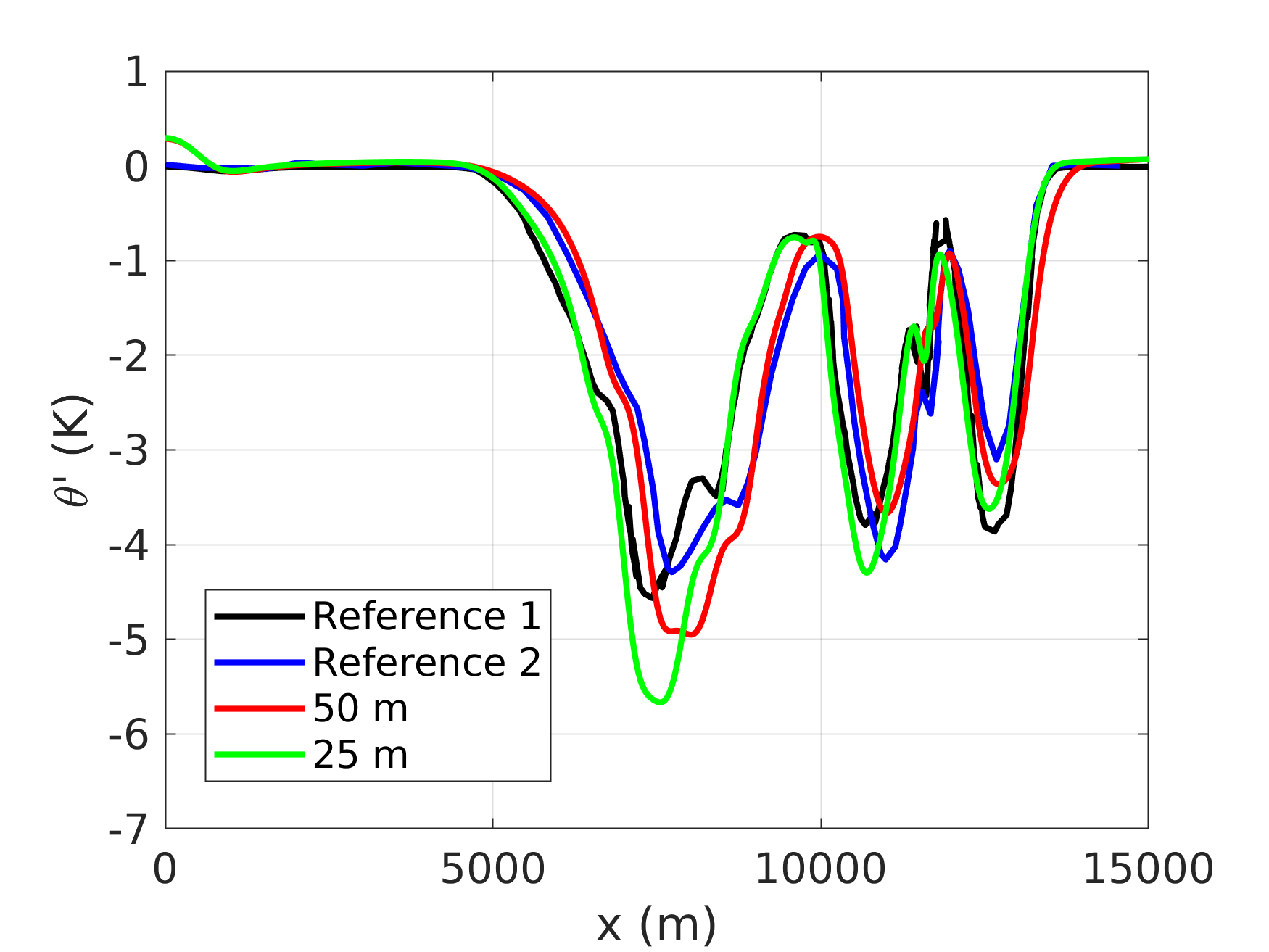}
        \put(40,73){CE1 model}
      \end{overpic}
 \begin{overpic}[width=0.49\textwidth]{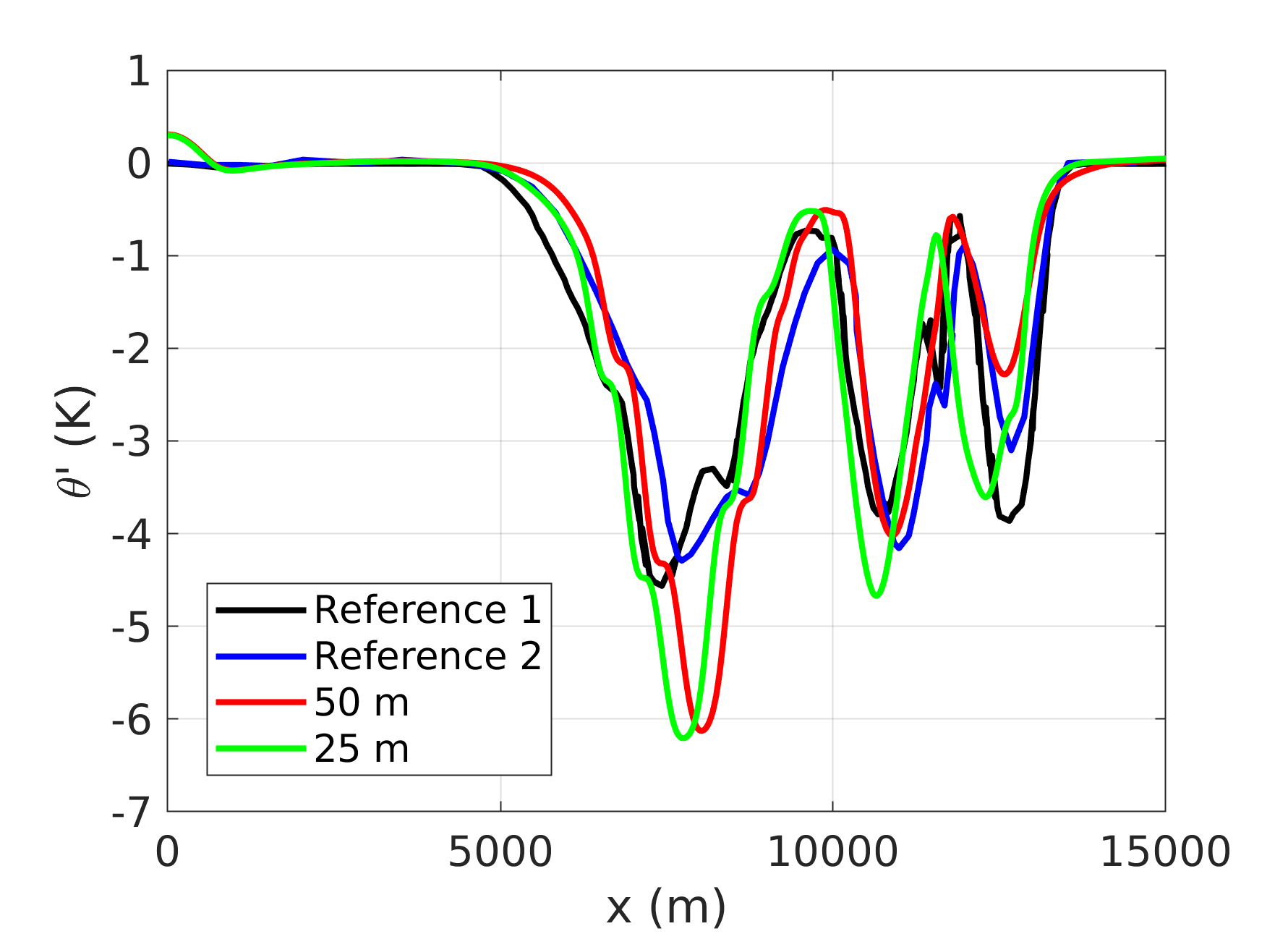}
      \put(40,70){CE2 model}
      \end{overpic}\\
 \begin{overpic}[width=0.49\textwidth]{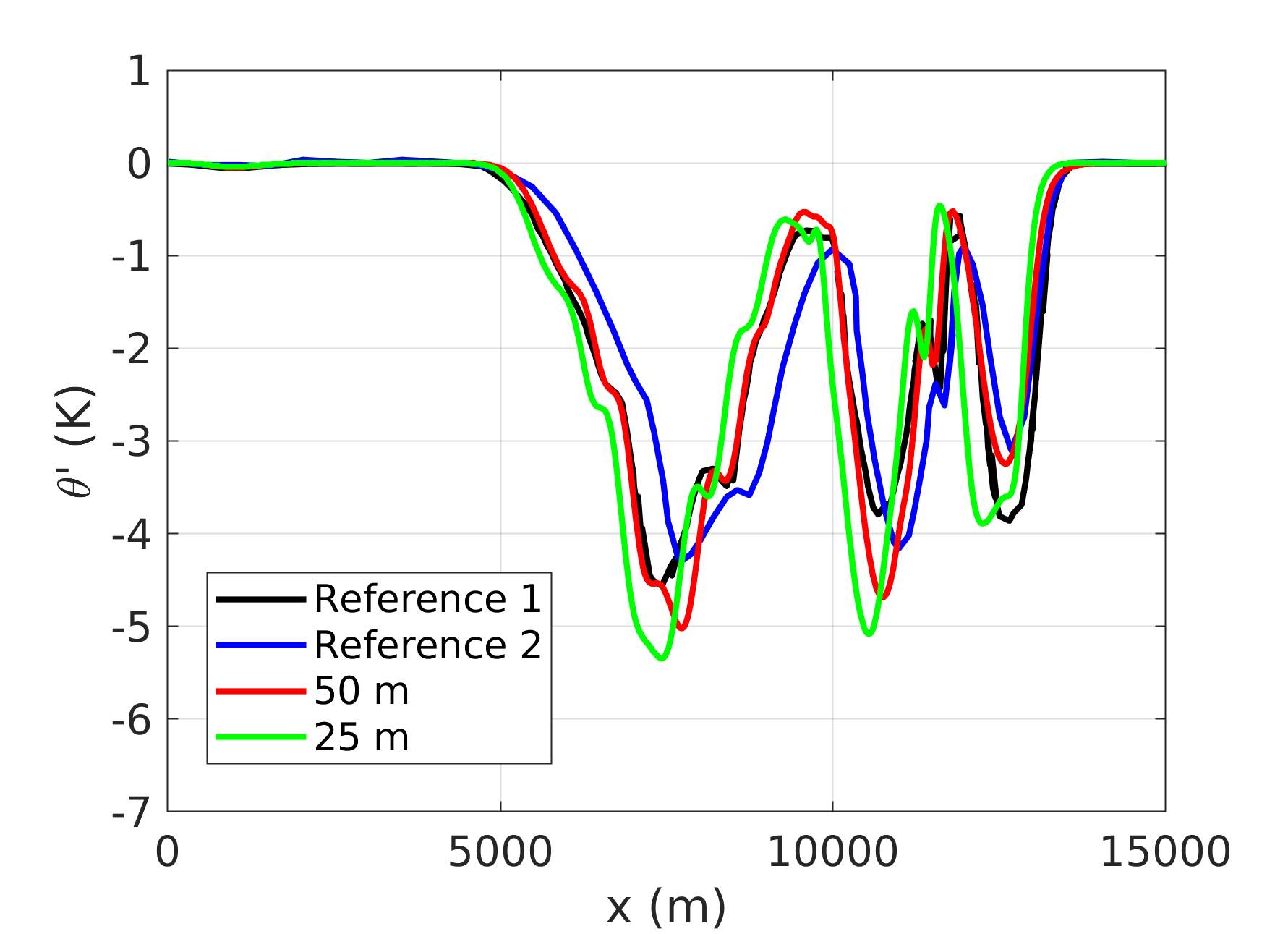}
\put(40,72){CE3 model}
      \end{overpic}
\caption{Density current: potential temperature perturbation $\theta'$ at $t = 900$ s
along the horizontal direction at a height of $z = 1200$ m given by the CE1 (top left panel), CE2 (top right panel) and CE3 (bottom panel) models with meshes $h = 50, 25$ m. The reference curves are from \cite{giraldo_2008} (denoted as “Reference 1”, mesh resolution 25 m) and
\cite{Nazari2017} (denoted as “Reference 2”, mesh resolution 50 m).}
\label{fig:DC1}
\end{figure}

For a quantitative comparison, we consider the potential temperature perturbation $\theta'$ along the horizontal direction at height $z = 1200$ m. Fig.~\ref{fig:DC1} compares 
the results given by CE1, CE2 and CE3 models with the results from \cite{giraldo_2008,Nazari2017} denoted with Reference~1 and Reference~2. 
Reference~1 \cite{giraldo_2008} was obtained with model CE3,
mesh resolution of 25 m, and a 
spectral element method that provided very similar results to a
discontinuous Galerkin method. 
Reference~2 \cite{Nazari2017} was computed with model CE3, mesh resolution of 50 m, and 
a a Godunov‐type finite‐volume solver. 
We see that, both in the reference curves and in our results, the effect of refining the mesh is to slow down the two larger recirculations, which correspond to the two main dips in the curves.  
The potential temperature fluctuation obtained with CE1 and mesh $h = 25$ m shows larger negatives peaks than the reference curves. This is true also for CE2 results, whose negative peaks are even farther from the reference curves. However, while coarsening the mesh to $h = 50$ m gets the results given by the CE1 model closer to the references, it does not help with the CE2 model. Instead, the results obtained with the CE3 model and both meshes are closer to the reference curves. 
We remark that the CE1 and CE2 results exhibit 
a bump for $x<1000$ m, which disappears when one adopts the CE3 model. 
We consider the agreement between the references and the results obtained with the CE3 model satisfactory, especially given all the differences between our approach and the approaches in \cite{giraldo_2008, Nazari2017}.

Finally, we compare the different formulations in terms of front location, defined as the location on the
ground where $\theta'$ = 1 K at $t = 900$ s.
Table \ref{tab:front_loc} reports the space interval that contains the front
location for the different models and meshes.  We see that  our results fall well within
the values reported in \cite{strakaWilhelmson1993}, which refer to 14 different approaches and meshes ranging in resolution from 25 m to 200 m. In addition,  Table \ref{tab:front_loc} demonstrates that the front becomes slower as the mesh is refined. For a given mesh, the front locations obtained with the three models are within about 400 m from each other. 

\begin{table}[htb]
\begin{tabular}{|c|c|c|c|} \hline
Model & Resolution [m] & Front Location [m] \\
 \hline
CE1 &  50 &  [15218, 15243]\\
CE1 &  25 &  [14808, 14833]\\
CE2 &  50 &  [15576, 15602]\\
CE2 &  25 &  [15064, 15090]\\
CE3 &  50 &  [15141, 15167]\\
CE3 &  25 &  [14680, 14706]\\

 Ref.~\cite{strakaWilhelmson1993} & (25, 200) & (14533,17070)
 \\
 \hline  
\end{tabular}
\caption{Density current: front location at $t = 900$ s obtained with the CE1, CE2 and CE3 models and meshes $h = 25, 50$ m compared against results reported in 
\cite{strakaWilhelmson1993}. 
For reference \cite{strakaWilhelmson1993}, we report the range of mesh sizes and front location values obtained with different methods.
} 
\label{tab:front_loc}
\end{table}

\subsection{Inertia-gravity waves}\label{sec:gravity_waves}

The computational domain in the $xz$-plane is $\Omega = [-150000, 150000] \times [0, 10000]$ m$^2$
and the time interval of interest is $(0, 3000]$ s. Impenetrable, free-slip boundary conditions are imposed
on the top bottom boundaries while 
a homogeneous Neumann conditions 
were used in the lateral boundaries. 
The initial temperature for the CE1 and CE2 models is given by: 
\begin{equation}\label{eq:IGW_1}
   T^0 = T_0 \left[1 + \left(1 - \frac{g^2}{c_p N^2 T_0}\right) \left(e^{\frac{N^2}{g} z} - 1 \right) \right] +  T_c \dfrac{\sin \left(\dfrac{\pi z}{h_c}\right)}{1 + \left(\dfrac{x - x_c}{a_c}\right)^2},
\end{equation}
where $T_0 = 300$ K, $T_c = 0.01$ K, $h_c = 10000$ m, $a_c = 5000$ m, $x_c = -50000$ m, and $N = 0.01$ is the Brunt-V\"ais\"al\"a frequency. The initial specific enthalpy is given by eq.~\eqref{eq:h0}. This is equivalent to the following initial potential temperature perturbation for the CE3 model: 
\begin{equation}\label{eq:IGW_2}
   \theta^0 = \theta_0 e^{\frac{N^2}{g} z} + \theta_c \dfrac{\sin \left(\dfrac{\pi z}{h_c}\right)}{1 + \left(\dfrac{x - x_c}{a_c}\right)^2},
\end{equation}
where $\theta_0 = 300$ K and $\theta_c = 0.01$ K.
The initial velocity field is $\mathbf{u}^0 = (20, 0)$ m/s, while the initial density $\rho^0$ is computed by solving the hydrostatic balance.
 %

\begin{figure}[htb]
\centering
 \begin{overpic}[width=0.49\textwidth]{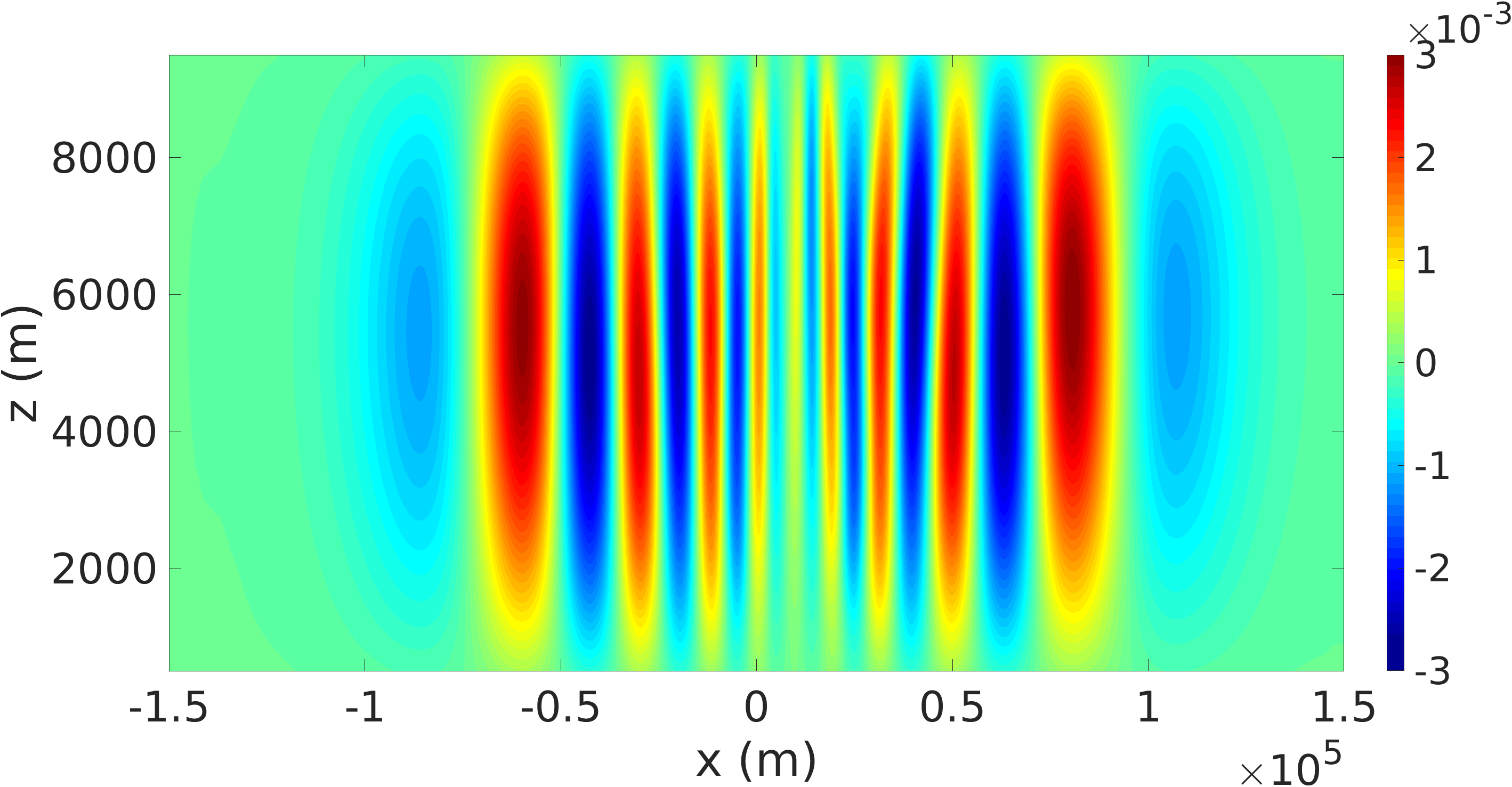}
        \put(12,43){CE2}
      \end{overpic}
       \begin{overpic}[width=0.49\textwidth]{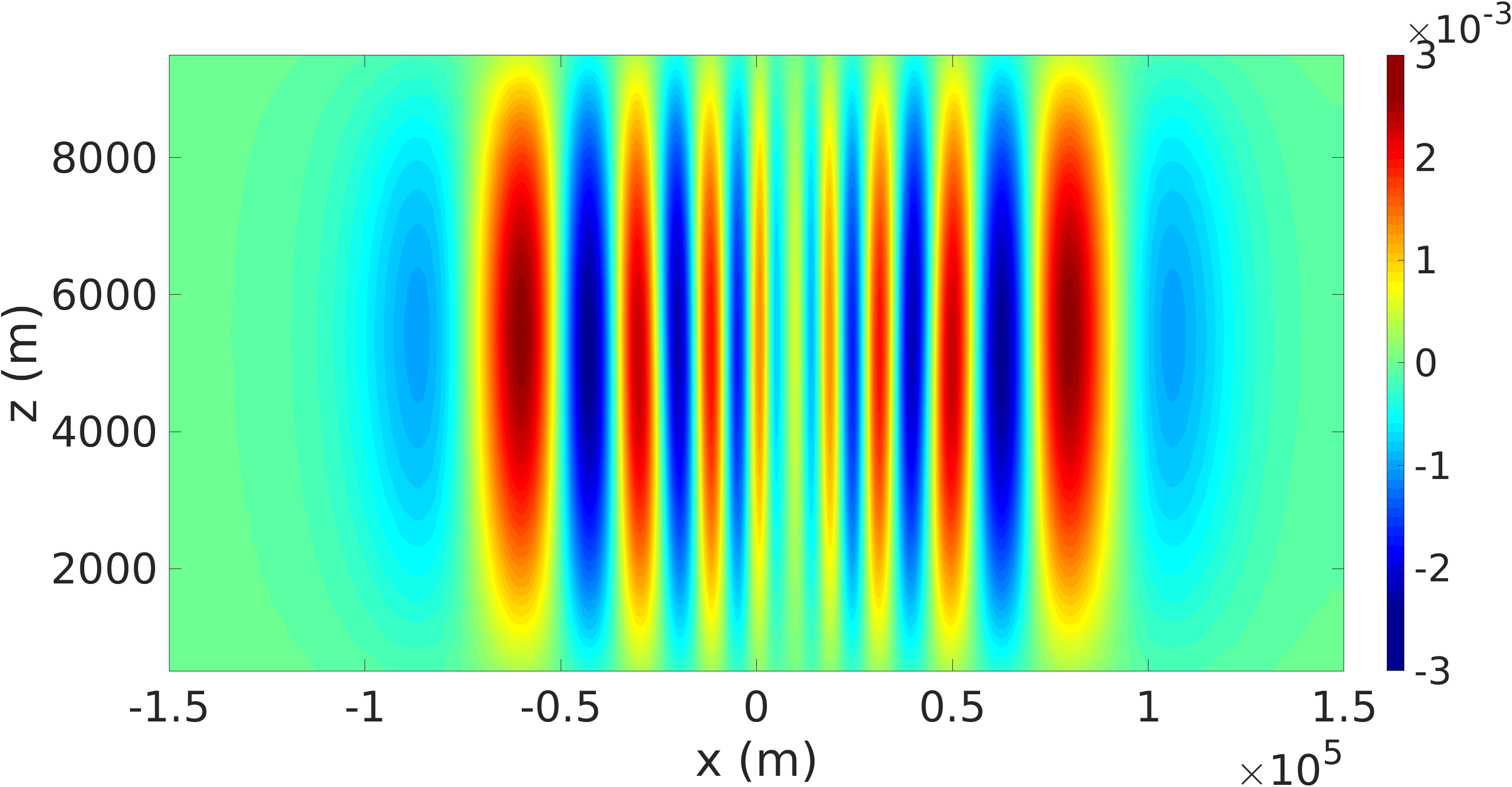}
        \put(12,43){CE3}
      \end{overpic}\\
\caption{
Inertia-gravity waves: vertical velocity $w$ at $t = 3000$ s given by the CE2 (left) and
CE3 (right) models.} 
\label{fig:GW2}
\end{figure}

We consider a uniform mesh with size $h = \Delta x = \Delta z = 250$ m \cite{giraldo_2008, bonaventura2023} and set the time step to $\Delta t$ = 1 s. For this test, we set $\mu_a$ = 0, i.e, no stabilization is introduced. 
The CE1 solution becomes unstable and thus it is not reported, while the spatial distribution of the vertical velocity $w$ at $t = 3000$ s given by CE2 and CE3 models is shown in Fig.~\ref{fig:GW2}. We see that these solutions are perfectly symmetric about the position $x=10000$ m, as expected. In fact, the initial perturbation radiates to the left and right symmetrically, but because of the mean horizontal flow, it does not remain centered about the initial position $x = 0$ m.
The plots in Fig.~\ref{fig:GW2} agree well with the corresponding figures in \cite{ahmadLindeman2007}.

\begin{figure}[htb]
\centering
 \begin{overpic}[width=0.6\textwidth]{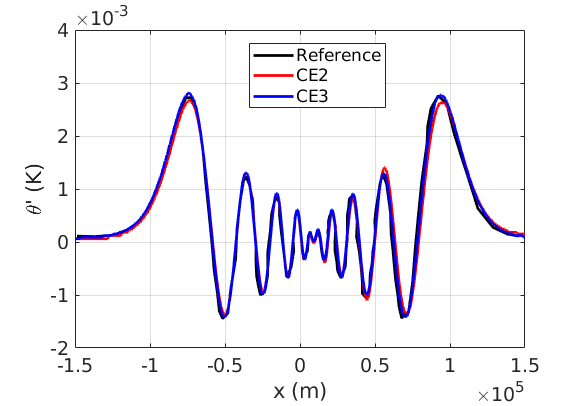}
      \end{overpic}\\
\caption{Inertia-gravity waves: potential temperature perturbation $\theta'$ at $t = 3000$ s along the horizontal direction at a height of $z = 5000$ m given by the CE2 and 
CE3 models for mesh $h = \Delta x = \Delta z = 250$ m. 
 The reference data are from \cite{giraldo_2008} and refer to the same resolution.}
\label{fig:GW1}
\end{figure}

Next, we consider the potential temperature perturbation $\theta'$ at  $t = 3000$ s along the horizontal direction at a height of $z = 5000$ m. Fig. \ref{fig:GW1} displays the results given by CE2 and CE3 models, which are very close to each other and almost superimposed to
the reference results from \cite{giraldo_2008}.  

Finally, we report in Table \ref{tab:GW} the extrema for the velocity components and the potential temperature computed with the CE2 and CE3 models, which are close to the values from \cite{giraldo_2008, ahmadLindeman2007}. 

\begin{table}[htb]
\begin{tabular}{|c|c|c|c|c|c|c|} \hline
Model &  $u_{min}$ (m/s) & $u_{max}$ (m/s)& $w_{min}$ (m/s) & $w_{max}$ (m/s) & $\theta'_{min}$ (K) & $\theta'_{max}$ (K) \\
 \hline
CE2 & -1.16e-2 & 1.32e-2 & -2.74e-3 & 3.03e-3 & -1.65e-3 & 2.81e-3\\
CE3 & -1.02e-2 & 1.03e-2 & -2.42e-3 & 2.63e-3 & -1.53e-3 & 2.83e-3\\
 Ref.~\cite{ahmadLindeman2007} & / & / & -2.8e-3 & 2.4e-3 & -1.49e-3 & 2.82e-3\\
Ref.~\cite{giraldo_2008} & -1.067e-2 & 1.069e-2 & -2.775e-3 &  2.698e-3 & -1.51e-3 & 2.78e-3\\
 \hline  
\end{tabular}
\caption{Inertia-gravity waves: our results for the extrema of the velocity components and perturbation of potential temperature at $t = 3000$ s obtained with the CE2 and CE3 models compared against results reported in \cite{ahmadLindeman2007} and \cite{giraldo_2008}. 
} 
\label{tab:GW}
\end{table}

\section{Numerical results for tests with orography}\label{sec:orography}
In this section, we consider three classical benchmarks featuring a single-peaked mountain: the hydrostatic equilibrium of an initially resting atmosphere (in Sec. \ref{sec:rest}), the steady state solution of linear hydrostatic flow (in Sec. \ref{sec:hydro_linear}), and the steady state solution of linear non-hydrostatic flow (in Sec. \ref{sec:linear_nonhydro}). In all test cases, 
the single-peaked mountain is described by the function 
\begin{align}
    h(x) = \dfrac{h_m}{1+\left(\dfrac{x-x_c}{a_c}\right)^2}, \label{eq:mountain}
\end{align}
where $h_m$ is the height of the mountain, $a_c$ is the half-width, and $x_c$ is the center. 
We use the setting from \cite{bottaKlein2004, marrasEtAl2013a, ECMWF2010}
for the equilibrium test and the setting provided in \cite{bonaventura2023, giraldo_2008} for the linear hydrostatic and non-hydrostatic mountain tests.

The linear hydrostatic and non-hydrostatic mountain tests have an analytical solution, which will be used to validate our results. 

\subsection{Hydrostatic atmosphere}\label{sec:rest}

The goal of this first test case is to verify that an initial resting atmosphere over
steep topography \eqref{eq:mountain}, with $h_m = 2000$ m, $a_c = 800$ m and $x_c = 0$ m,
remains still with reasonable accuracy for a long time interval. 
The computational domain in the $xz$-plane is $\Omega = [-8000, 8000] \times [0, 8000]$ m$^2$. We let the hydrostatic atmosphere, initially at rest
in this domain, evolve for $25$ days \cite{bottaKlein2004}. Impenetrable,
free-slip boundary conditions are imposed at all the boundaries. 

The initial temperature for the CE1 and CE2 models is given by

\begin{align*}
   T^0 =   \begin{cases}
T_0 \left[1 + \left(1 - \frac{g^2}{c_p N_0^2 T_0}\right) \left(e^{\frac{N_0^2}{g} z} - 1 \right) \right] & \text{if} \quad z \leq z_0 \\
T(z_0) \left[1 + \left(1 - \frac{g^2}{c_p \left(N_0 + \Delta N\right)^2 T(z_0) }\right) \left(e^{\frac{\left(N_0 + \Delta N\right)^2}{g} \left(z - z_0\right)} - 1 \right) \right], \qquad \qquad & \text{if} \quad z_0 < z \leq z_1 \\
T(z_1) \left[1 + \left(1 - \frac{g^2}{c_p N_0^2 T(z_1)}\right) \left(e^{\frac{N_0^2}{g} \left(z - z_1\right)} - 1 \right) \right], \qquad \qquad & \text{if} \quad z > z_1
\end{cases} 
\end{align*}
where $T_0= 300$ K, $z_0 = 750$ m and $z_1 = 1200$ m. This is equivalent to the following initial potential temperature
perturbation
\begin{align*}
   \theta^0 =   \begin{cases}
    \theta_0 e^{\frac{N_0^2}{g} z} & \text{if} \quad z \leq z_0 \\
\theta(z_0) e^{\frac{\left(N_0 + \Delta N\right)^2}{g} \left(z - z_0\right)}, \qquad \qquad & \text{if} \quad z_0 < z \leq z_1 \\
\theta(z_1) e^{\frac{N_0^2}{g} \left(z - z_1\right)}, \qquad \qquad & \text{if} \quad z > z_1
\end{cases} 
\end{align*}
for the CE3 model where $\theta_0 = 300$ K. 
We consider three different configurations proposed in \cite{bottaKlein2004}: i) $N_0 = 0.01$ and $\Delta N = 0$,  ii) $N_0 = 0.02$ and $\Delta N = 0$ and iii) $N_0 = 0.01$ and $\Delta N = 0.01$. Case iii) allows us to assess the performance of our approach for more realistic stratifications.
Finally, the initial density $\rho^0$ is computed by solving the
hydrostatic balance, i.e., eq.~\eqref{eq:hydroCE1} for CE1 model and eq.~\eqref{eq:p0_hydro} for CE2 and CE3 models. 

\begin{figure}[htb]
\centering
 \begin{overpic}[width=0.48\textwidth]{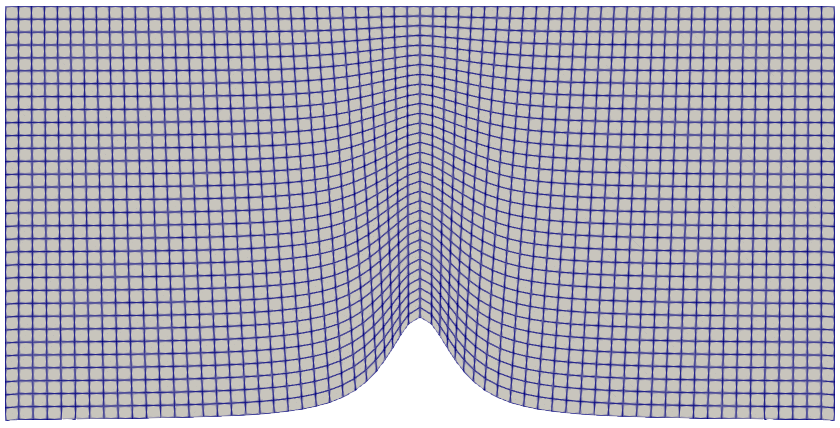}
      \end{overpic}\\
\caption{Hydrostatic atmosphere: computational grid around the mountain.}
\label{fig:mountain1}
\end{figure}

We consider a uniform mesh with
mesh size $h = \Delta x = \Delta z = 250$ m, whose maximum non-orthogonality is 55.3$^\circ$, average non-orthogonality is 13$^\circ$, and skewness is 0.76. Fig.~\ref{fig:mountain1} shows the mesh. We set the time step to $\Delta t = 1$ s.

\begin{figure}[htb]
\centering
 \begin{overpic}[width=0.48\textwidth]{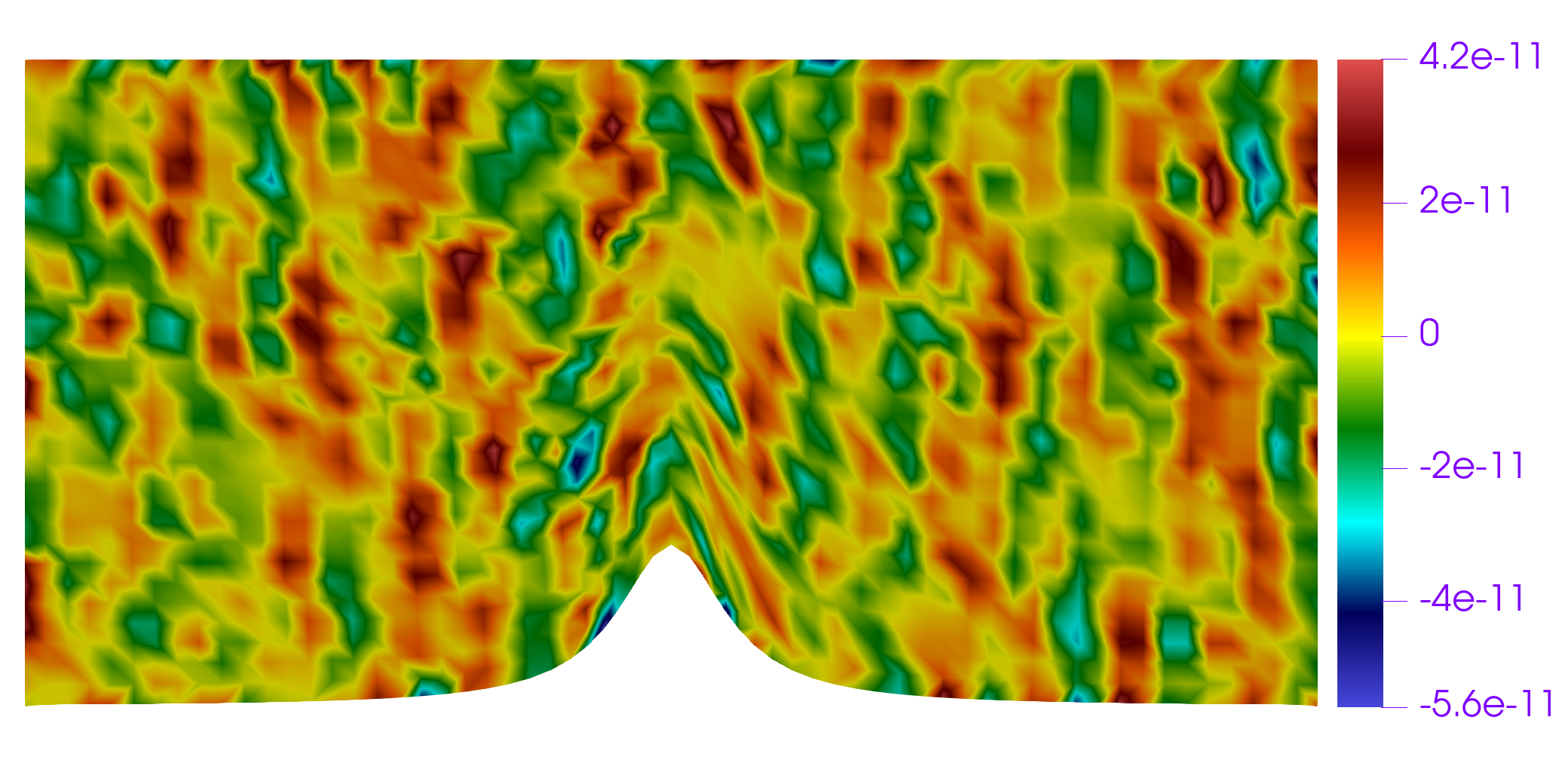}
        \put(75,50){$N_0 = 0.01$, $\Delta N = 0$}
        \put(37,47){CE2}
      \end{overpic}
 \begin{overpic}[width=0.48\textwidth]{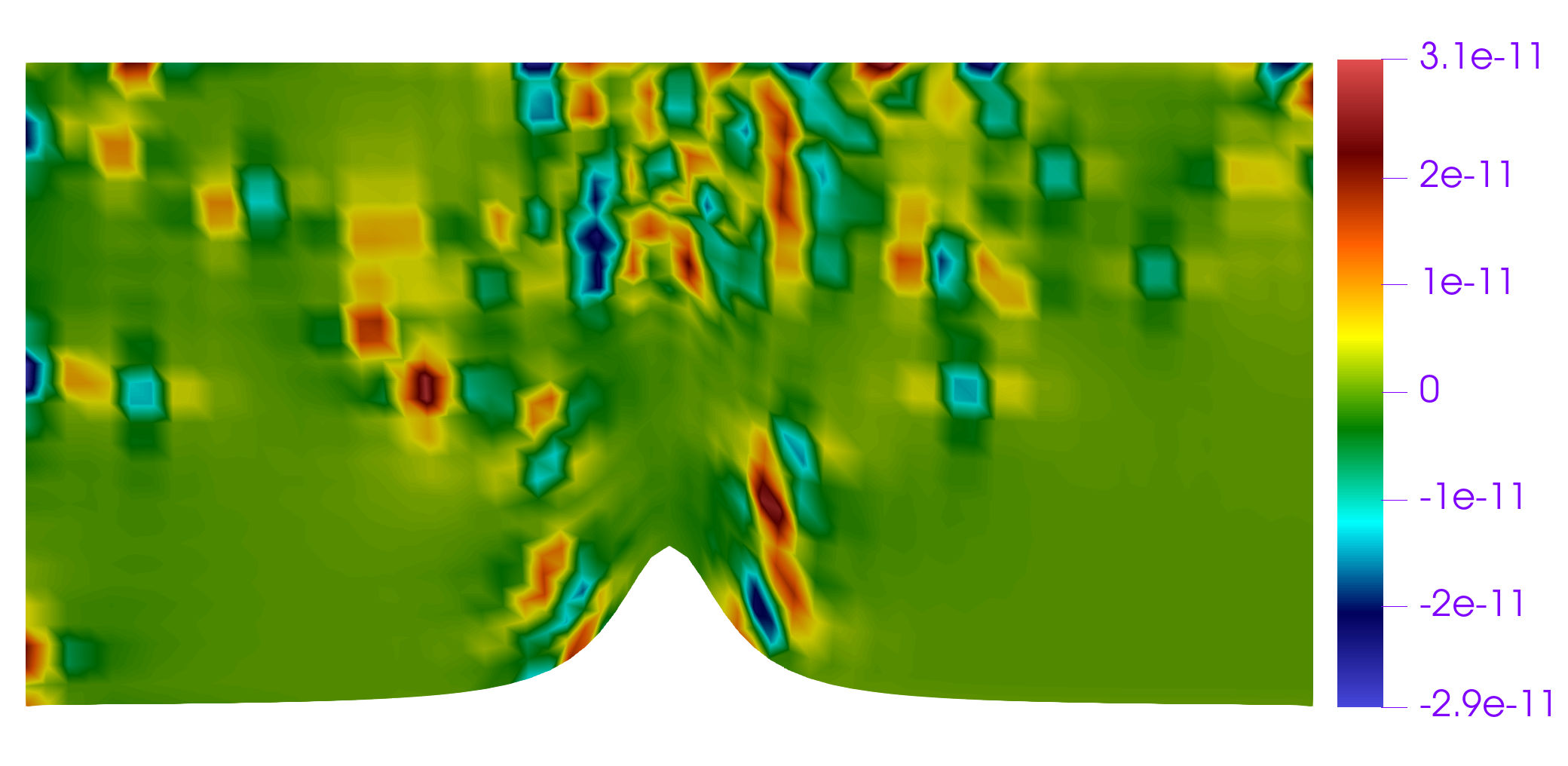}
         \put(37,47){CE3}
      \end{overpic}\\
      \vskip .4cm
       \begin{overpic}[width=0.48\textwidth]{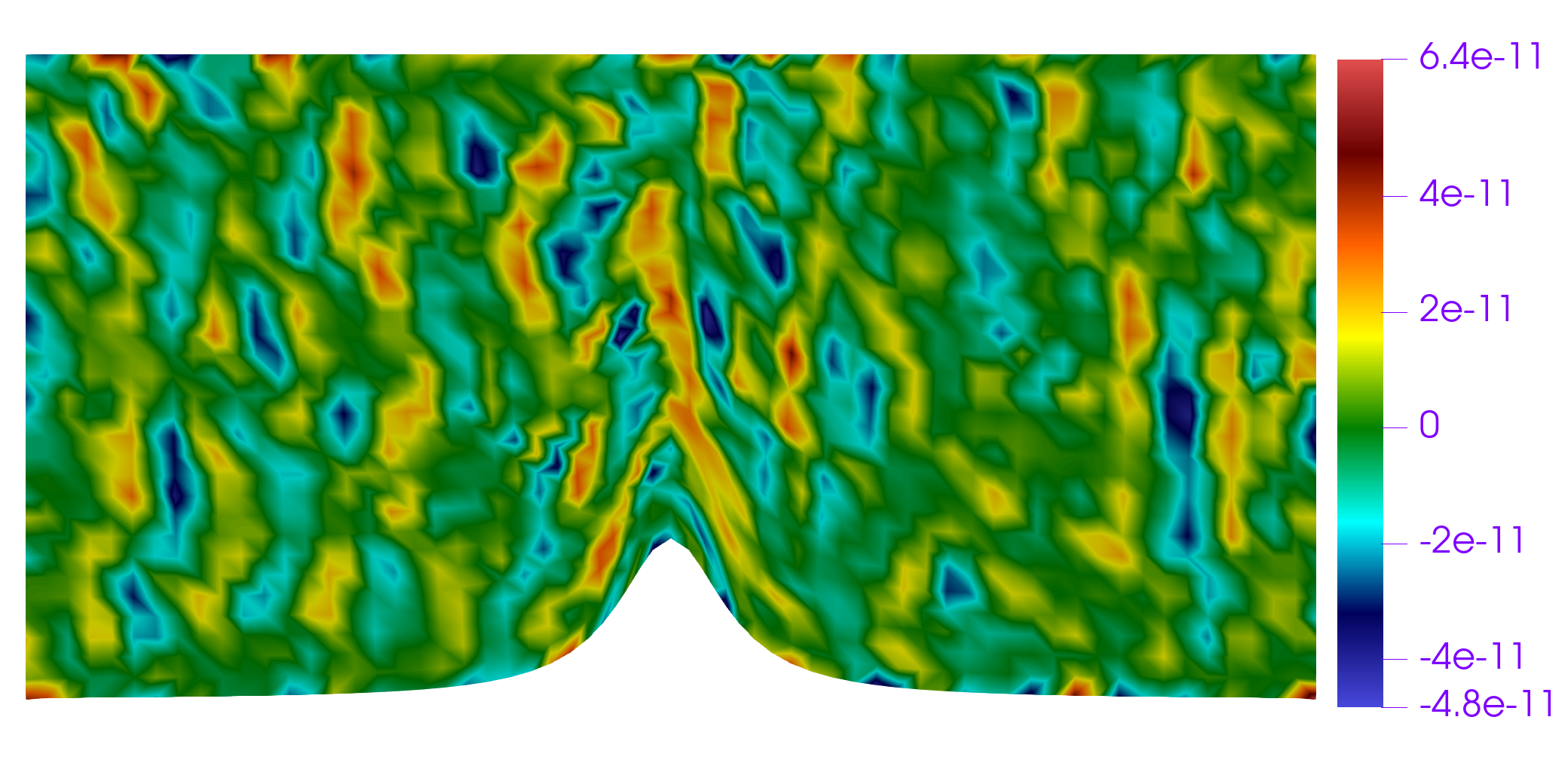}
         \put(75,50){$N_0 = 0.02$, $\Delta N = 0$}
        \put(37,47){CE2}
      \end{overpic}
       \begin{overpic}[width=0.48\textwidth]{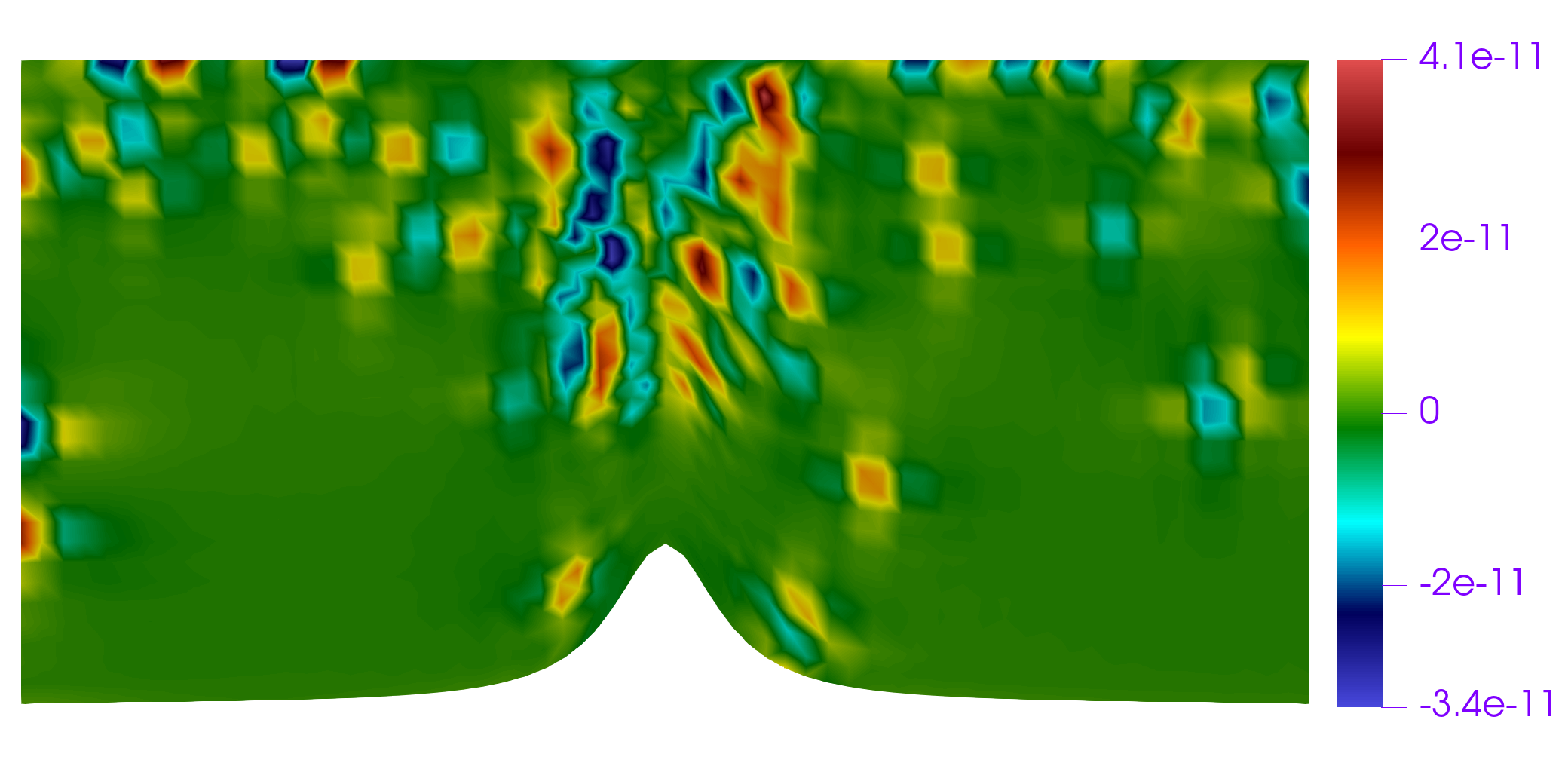}
                \put(37,47){CE3}
      \end{overpic}\\
      \vskip .4cm
             \begin{overpic}[width=0.48\textwidth]{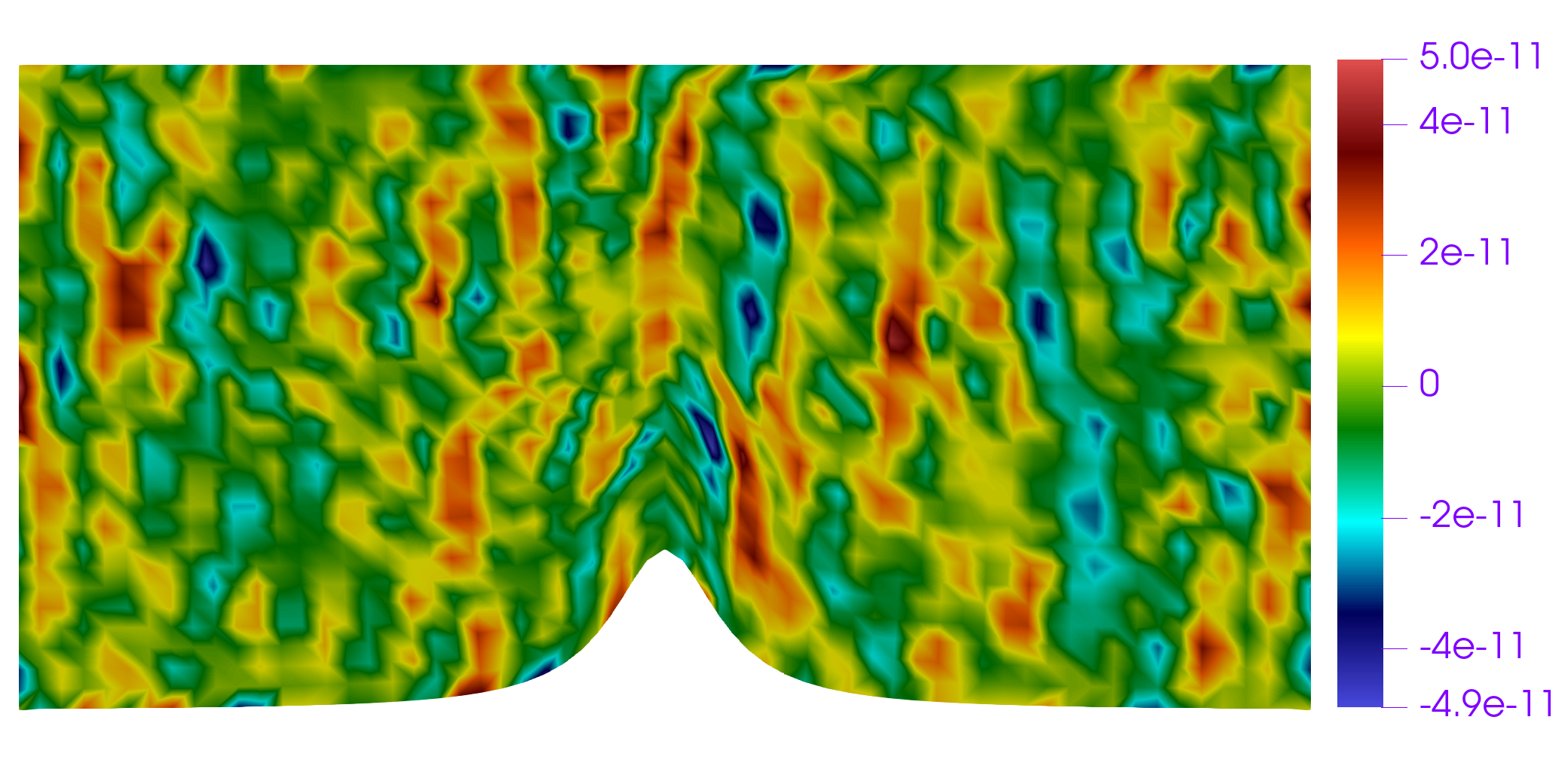}
              \put(75,50){$N_0 = 0.01$, $\Delta N = 0.01$}
              \put(37,47){CE2}
      \end{overpic}
       \begin{overpic}[width=0.48\textwidth]{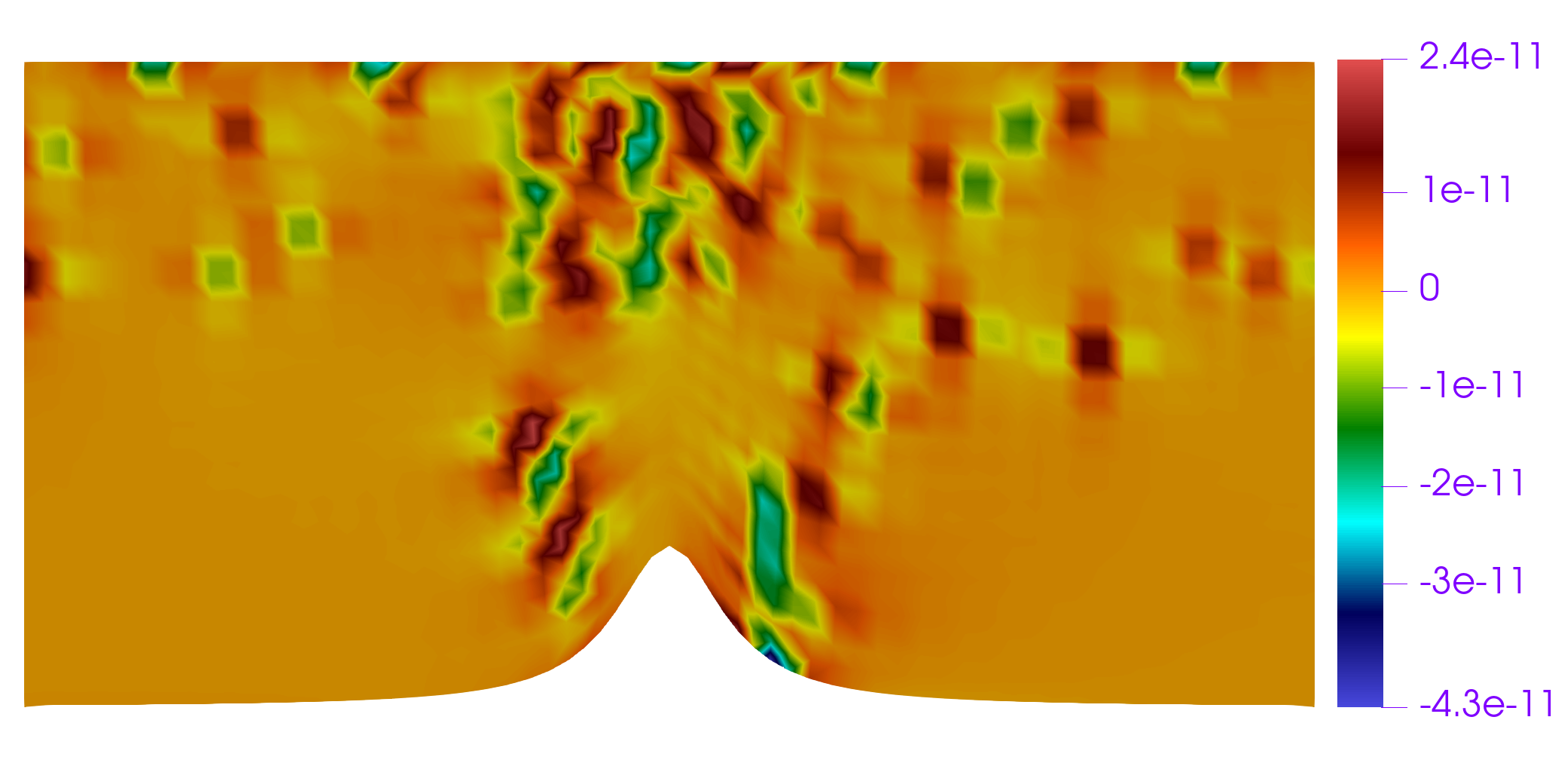}
       \put(37,47){CE3}
      \end{overpic}
\caption{Hydrostatic atmosphere: vertical velocity $w$ at $t = 60000$ s by the CE2 model (first column) and CE3 model (second column) for configuration i) (top), ii) (middle), and iii) (bottom).}
\label{fig:mountain2}
\end{figure}
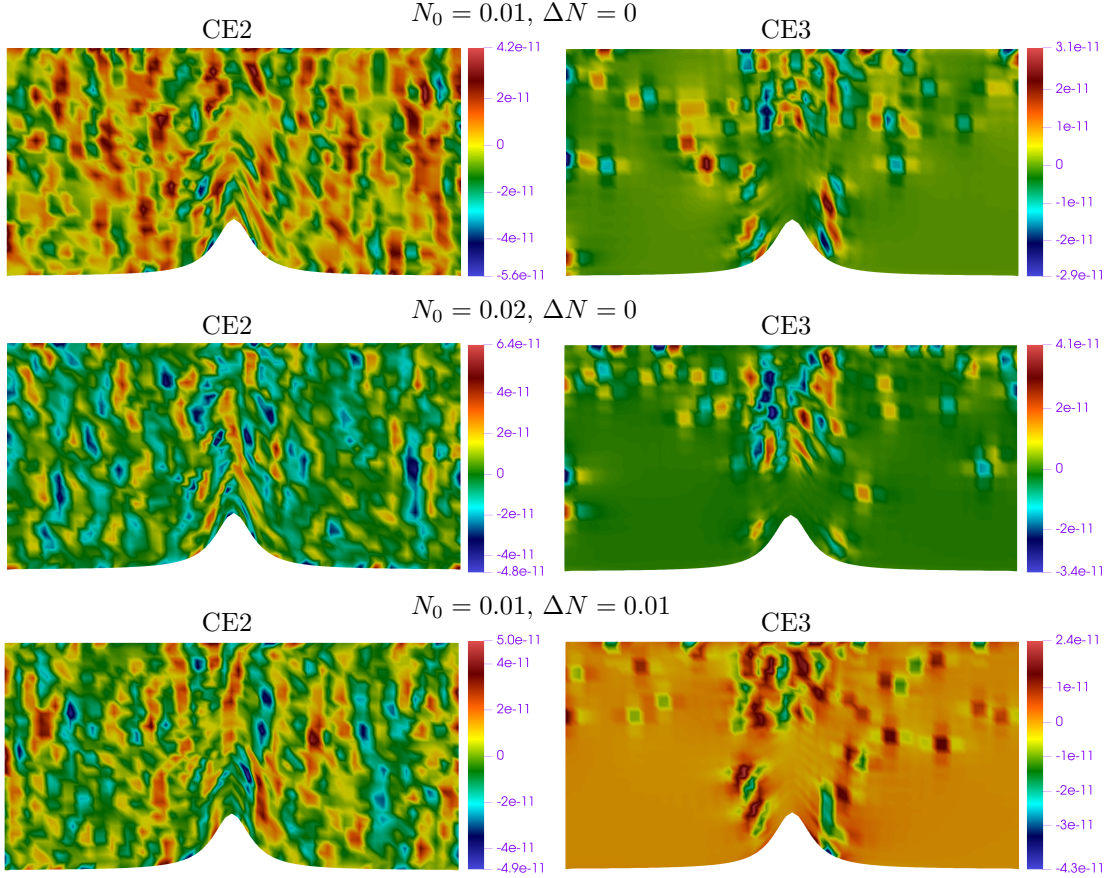

Fig.~\ref{fig:mountain2} shows the vertical velocity $w$ at $t = 60000$ s ($\approx 16.6$ h) computed by the CE2 and CE3 models for configurations i), ii), 
and iii). We do not report the solutions for CE1 because they are unstable because of spurious currents occurring in the simulation after a few seconds. 
Fig.~\ref{fig:mountain2} shows that $w$ given by CE2 is between $-5.6e-11$ m/s and $6.4e-11$ m/s, while the range is slightly smaller for $w$ given by CE3, i.e., $[-4.3e-11, 4.1e-11$] m/s.
This is comparable with the
range reported in \cite{marrasEtAl2013a},  which is $[-1e-12, 1e-11]$ m/s, and smaller than the one reported in \cite{ECMWF2010}, which is $[-2e-3, 2e-3]$ m/s for $N_0 = 0.02$ and $\Delta N = 0$.

\begin{figure}[htb]
\centering
 \begin{overpic}[width=0.56\textwidth]{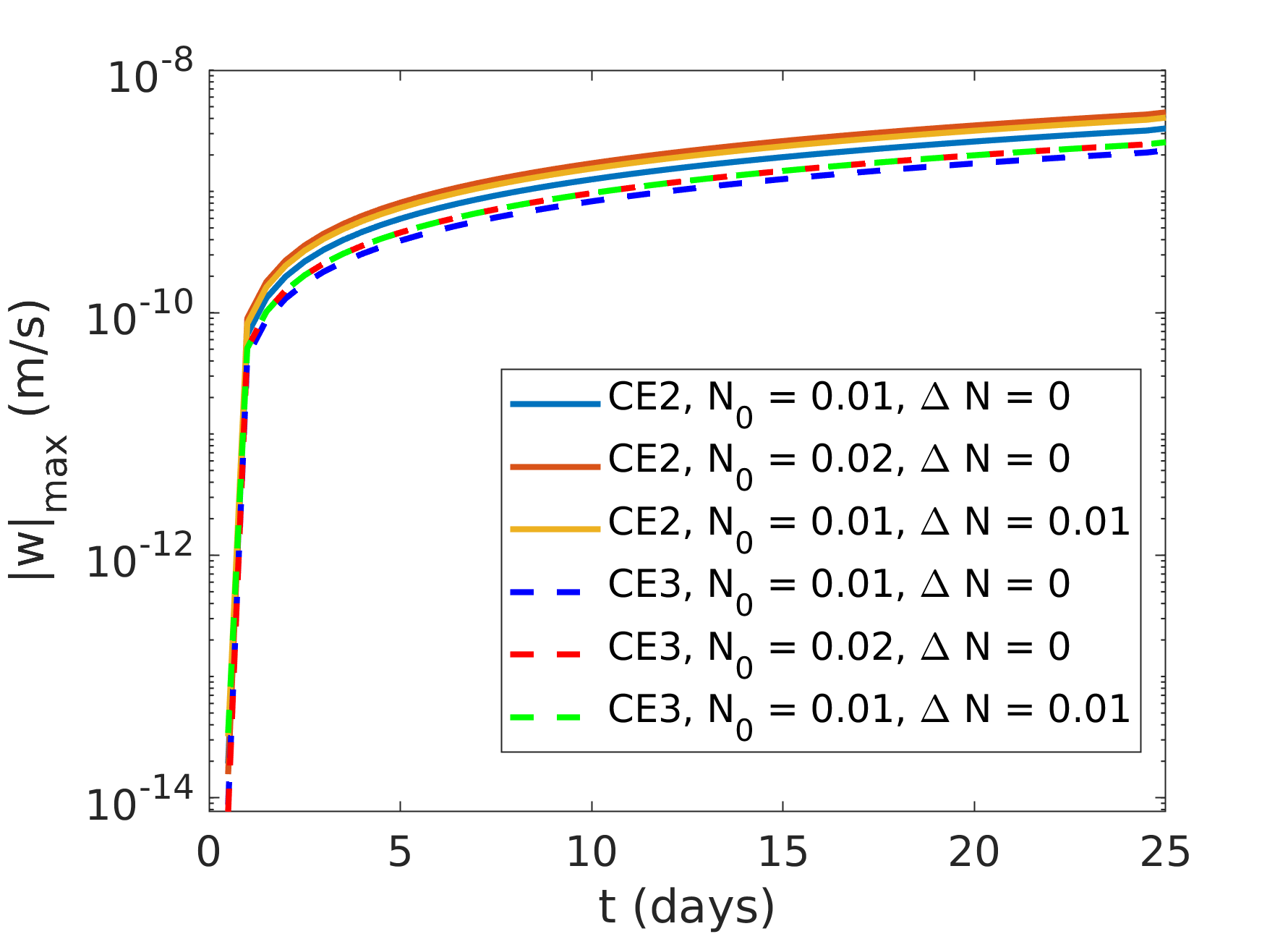}
      \end{overpic}
\caption{Hydrostatic atmosphere: evolution of the maximal absolute value of the vertical velocity $|w|_{max}$ given by the CE2 and CE3 models for all the configurations under consideration.}
\label{fig:mountain3}
\end{figure}

Fig.~\ref{fig:mountain3} reports the maximal absolute value of the vertical velocity $|w|_{max}$ over time.   
After first 24 hours, the initial steep increase in $|w|_{max}$ 
slows down and for the remaining 24 days the increase is less than 2 orders of magnitude for both models. This is a significant improvement over the results in \cite{bottaKlein2004}, which show an accuracy of $1e-8$ m/s over 25 days in the case of uniform background potential temperature, a simpler case than stratified atmosphere. In the case of stratified atmosphere, in \cite{bottaKlein2004} the authors report $|w|_{max}$ only  the first 30 minutes, showing that it does not go below $1e-3$ m/s. 

 




We conclude that the CE2 and CE3 models are able to preserve the hydrostatic equilibrium in presence of steep orography, with accuracy comparable to or higher than previous works \cite{bottaKlein2004, marrasEtAl2013a, ECMWF2010}. 
Since the results in this and the preceding sections indicate that the CE3 model is more accurate, 
Sec.~\ref{sec:hydro_linear} and \ref{sec:linear_nonhydro} will present results obtained with the CE3 model only.





\subsection{Linear hydrostatic  mountain}\label{sec:hydro_linear}
The goal of this second test is to compute the steady state solution of linear hydrostatic flow over single-peaked mountain \eqref{eq:mountain}, with
$h_m$= 100 m, $a_c$ = 10000 m, and $x_c$ = 0 m.
We let the hydrostatic atmosphere evolve until $t = 45000$ s (i.e., $12.5$ h)  in computational domain $\Omega = [-120000, 120000] \times [0, 30000]$ m$^2$.

The initial state of the atmosphere
consists of a constant mean flow with $\bm{u}^0 = (20, 0)$ m/s. 
The initial potential temperature is given by
\begin{align}
   \theta^0 = 
    \theta_0 e^{\frac{N^2}{g} z}, \label{eq:theta0_linh}
\end{align}
where $\theta_0 = 250$ K and $N = 0.02$.  The initial density $\rho^0$ is computed by solving eq.~\eqref{eq:p0_hydro}. It can be verified that $({N a_c}/{u^0}) > 1$, so that the flow is in the
hydrostatic range \cite{giraldo_2008}.

On the the bottom boundary, we impose an
impenetrable, free-slip boundary condition, while non-reflecting boundary conditions need to be enforced on the top boundary and the lateral boundaries \cite{bonaventura2023, giraldo_2008}.  To this aim, we introduce the following forcing terms in the momentum equation \eqref{eq:momeCE2} and in the potential temperature equation \eqref{eq:energyCE3}: 
\begin{align}
    \boldsymbol{f_u} &= \rho S_v \cdot \left(\bm u - \bm u^0\right) + \rho S_{h_1}  \cdot \left(\bm u - \bm u^0\right) + \rho S_{h_2}  \cdot \left(\bm u - \bm u^0\right), \label{eq:fu}\\
    f_\theta &= \rho S_v \cdot \left(\theta - \theta^0 \right) + \rho S_{h_1} \cdot \left(\theta - \theta^0 \right) + \rho S_{h_2} \cdot \left(\theta - \theta^0 \right), \label{eq:ftheta}
\end{align} 
where $S_v$, $S_{h_1}$ and $S_{h_2}$ are Raylegh damping profiles defined as \cite{giraldo_2008}:
\begin{align}
   S_v &=   \begin{cases}
0, \qquad \qquad & \text{if} \quad z < z_B \\\alpha_v \sin^2\left[\frac{\pi}{2}\left(\frac{z-z_B}{z_T - z_B}\right)\right], \qquad \qquad & \text{if} \quad z \geq z_B
\end{cases} \label{eq:Sv} \\
   S_{h_1} &=\begin{cases}
0, \qquad \qquad & \text{if} \quad x < x_{B_1} \\\alpha_{h_1} \sin^2\left[\frac{\pi}{2}\left(\frac{x-x_{B_1}}{x_{R} - x_{B_1}}\right)\right], \qquad \qquad & \text{if} \quad x \geq x_{B_1}
\end{cases} \label{eq:Sh1} \\
   S_{h_2} &=\begin{cases}
0, \qquad \qquad & \text{if} \quad x > x_{B_2} \\\alpha_{h_2} \sin^2\left[\frac{\pi}{2}\left(\frac{x-x_{B_2}}{x_{S} - x_{B_2}}\right)\right], \qquad \qquad & \text{if} \quad x \leq x_{B_2}.
\end{cases} \label{eq:Sh2}
\end{align}
We set $z_T = 30000$ m, $x_R = 120000$ m and $x_S = -120000$ m, which are related to the dimensions of the computational domain. Following \cite{bonaventura2023}, we set $z_B = 15000$ m, $x_{B_1}$ = 40000 m and $x_{B_2} = -40000$ m. This means that we apply the damping layer in
the topmost 15 km of the domain and in the first and last 40 km  along the horizontal direction.
Finally, we set $\alpha_v = \alpha_{h_1} = \alpha_{h_2} = 0.1$. 
Note that in \cite{giraldo_2008, gmd-15-6259-2022} in order to maintain stability for long-time integrations, the authors make use of spatial filters to keep under control the Gibbs oscillations. On the other hand, we obtain the same effect by simply adopting an upwind scheme for the discretization of the convective term.

The quantity of interest for this benchmark is the momentum flux \cite{Smith1979}: 
\begin{equation}
    m(z) = \int_{x_S}^{x_R}  \rho(z) u'(x,z) w'(x,z) dx,
\end{equation}
where $u' = u - u^0$ and $w = w - w^0$. We note that $w^0 = 0$ m/s. The computed momentum flux is compared with the analytic hydrostatic momentum flux known from linear theory:
\begin{equation}
   m^H = -\frac{\pi}{4}\rho_{g} u^0 N h_m^2, \label{eq:flux_hydro}
\end{equation}
where $\rho_g$ is the value of the density at ground. 
For such comparison, we define the normalized momentum flux: 
\begin{equation}
    m^\star = \frac{m(z)}{m^H}, \label{eq:mstar1}
\end{equation}
which should be as close as possible to 1. 

\begin{figure}[htb]
\centering
 \begin{overpic}[width=0.45\textwidth]{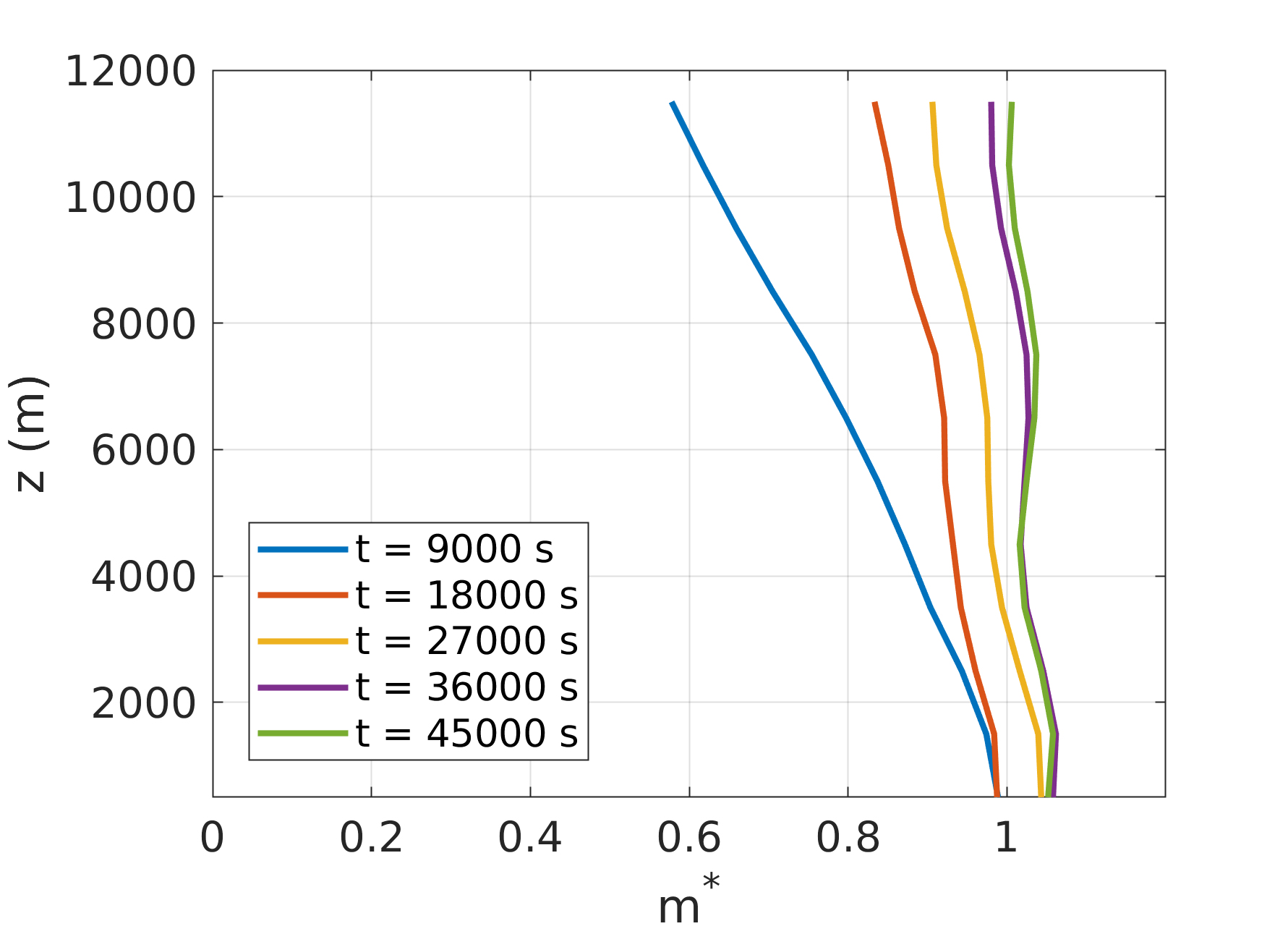}
      \end{overpic}
\caption{Linear hydrostatic mountain test: time-dependent profile of the normalized momentum flux $m^\star$ given by the CE3 model.}
\label{fig:mountain6}
\end{figure}

\begin{figure}[htb]
\centering
 \begin{overpic}[width=0.45\textwidth]{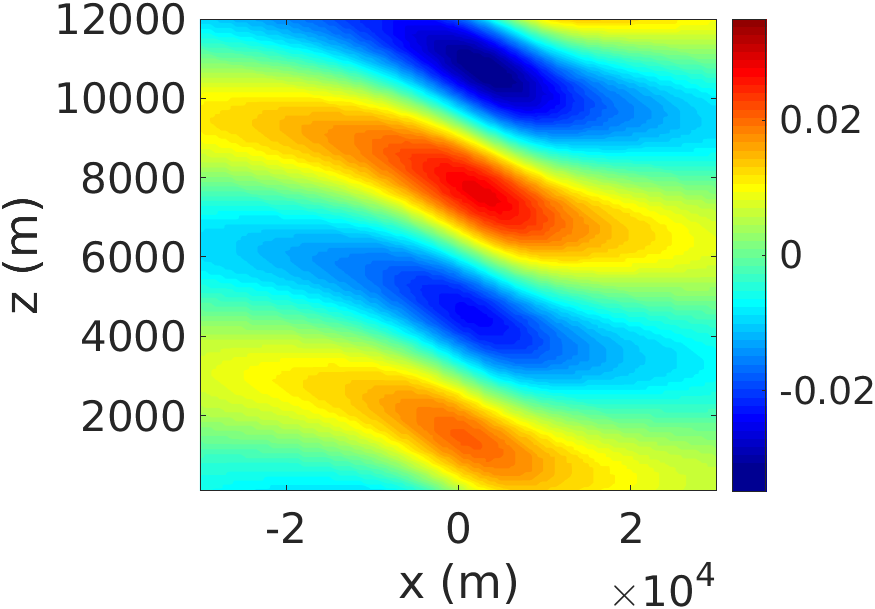}
      \end{overpic}
 \begin{overpic}[width=0.45\textwidth]{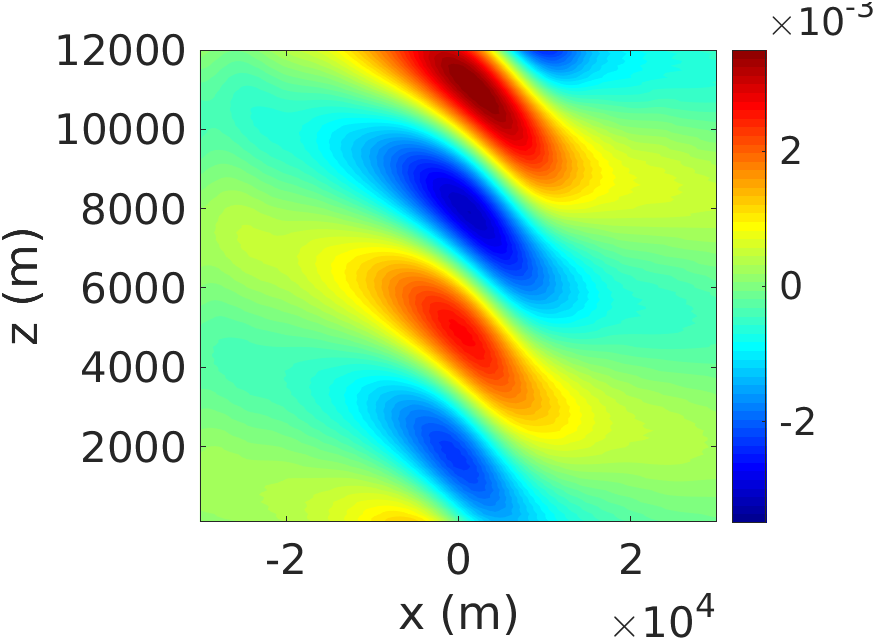}
      \end{overpic}
\caption{Linear hydrostatic mountain test: horizontal velocity $u$ (left) and vertical velocity $w$ (right) given by the CE3 model at $t = 45000$ s.}
\label{fig:mountain5}
\end{figure}

Following \cite{bonaventura2023}, we consider a mesh with sizes $\Delta x = 600$ m and $\Delta z = 100$ m.
We set $\Delta t = 0.05$ s. 
Fig.~\ref{fig:mountain6} shows the normalized momentum flux at various times.  We see that the simulation gets close to the 
steady state after
$t = 36000$ s (i.e., 10 h). This is in agreement with \cite{giraldo_2008, bonaventura2023}. The ``converged'' values of $m^\star$ in 
Fig.~\ref{fig:mountain6} range between 1 and 1.06, which is comparable to the interval in \cite{giraldo_2008}, which is [0.95, 1.01], and the interval in \cite{bonaventura2023}, which seems to be [0.99, 1.01] from Fig. 10.  
The velocity components at $t = 45000$ s, which are close to the steady state, are illustrated in Fig.~\ref{fig:mountain5}. 
These figures are comparable to the corresponding figures in \cite{bonaventura2023, giraldo_2008, gmd-15-6259-2022}.

\subsection{Linear non-hydrostatic mountain}\label{sec:linear_nonhydro}
The goal of this third test is to compute the steady state
solution of linear non-hydrostatic flow over single-peaked mountain \eqref{eq:mountain}, with $h_m = 1$ m, $a_c = 1000$ m and $x_c = 0$ m.
We let the atmosphere evolve until $t = 28800$ s (i.e., $8$ h) in computational domain $\Omega =[-72000, 72000] \times [0, 30000]$ m$^2$. 

The initial state of the atmosphere consists of a constant mean flow with $\bm{u}^0 = (10, 0)$ m/s. The initial potential temperature is given by \eqref{eq:theta0_linh}, with $\theta_0 = 280$ K and $N = 0.01$. Since in this case we obtain ${N a_c}/{u^0} = 1$, the flow is in the non-hydrostatic regime. Again, we set an
impenetrable, free-slip boundary condition 
on the bottom boundary and non-reflecting boundary conditions on the top and the lateral boundaries. For the non-reflecting boundary conditions, we define the forcing terms as in \eqref{eq:fu} and \eqref{eq:ftheta}, with 
$z_T = 30000$ m, $x_R = 72000$ m and $x_S = -72000$ m. Following \cite{bonaventura2023}, we set $z_B$ = 14000 m, $x_{B_1}$ = 32000 m and $x_{B_2} = -32000$ m. Moreover, we set $\alpha_v = \alpha_{h_1} = \alpha_{h_2} = 0.1$. 

We define the analytic non-hydrostatic momentum flux \cite{Klemp1983} as
\begin{equation}
   m^{NH} = -0.457 m^H,
\end{equation}
where $m^H$ is given by eq. \eqref{eq:flux_hydro}. Note that this is valid only for ${N a_c}/{u^0} = 1$, as is our case. 
Then, the normalized momentum flux becomes
\begin{equation}
   m^\star = \dfrac{m(z)}{m^{NH}}. \label{eq:mstar2}
\end{equation} 

\begin{figure}[htb]
\centering
 \begin{overpic}[width=0.45\textwidth]{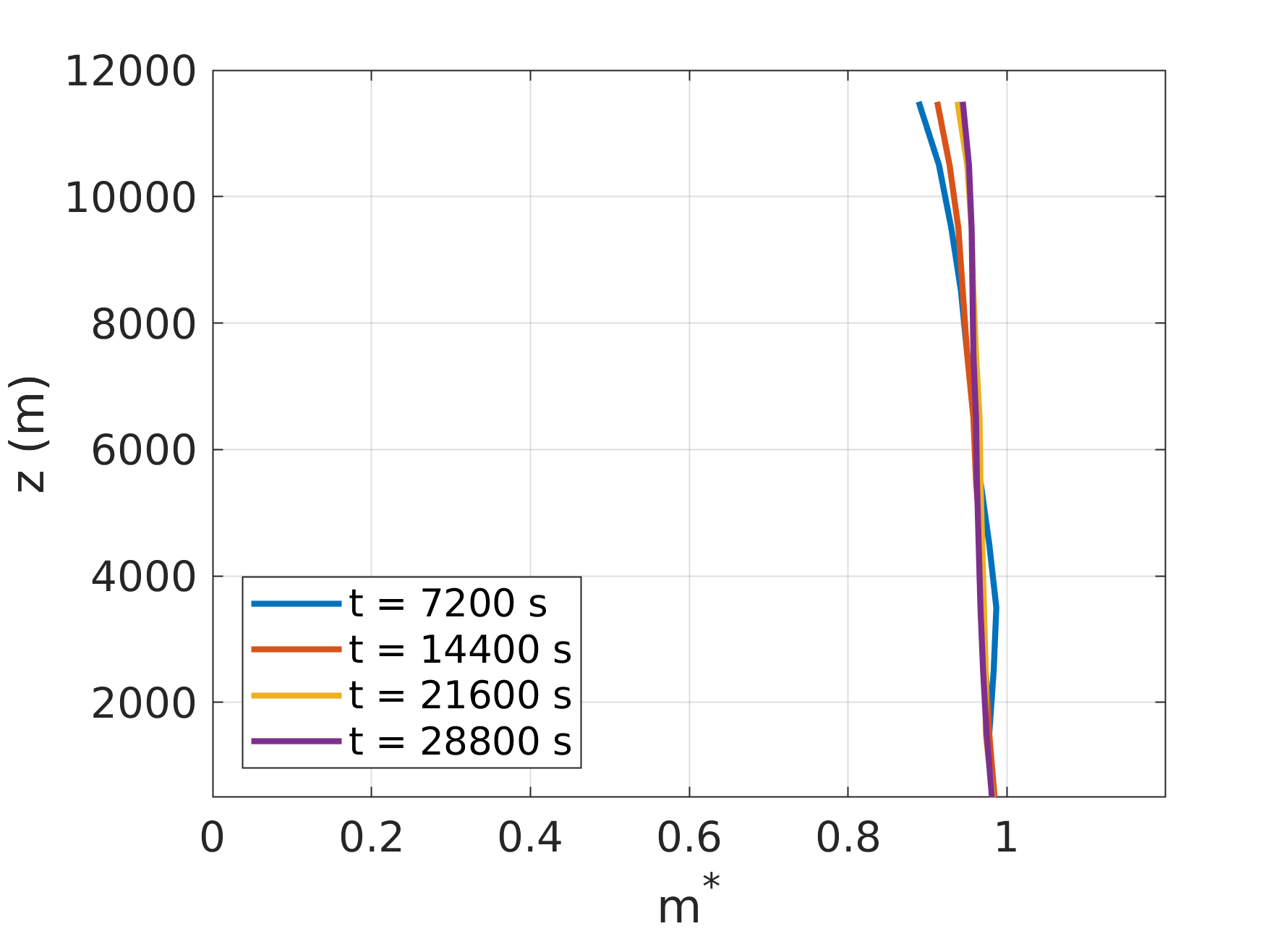}
      \end{overpic}\\
\caption{Linear non-hydrostatic mountain test: time-dependent profile of the normalized momentum flux $m^\star$ given by the CE3 model.}
\label{fig:mountain6_n}
\end{figure}

\begin{figure}[htb]
\centering
            \begin{overpic}[width=0.45\textwidth]{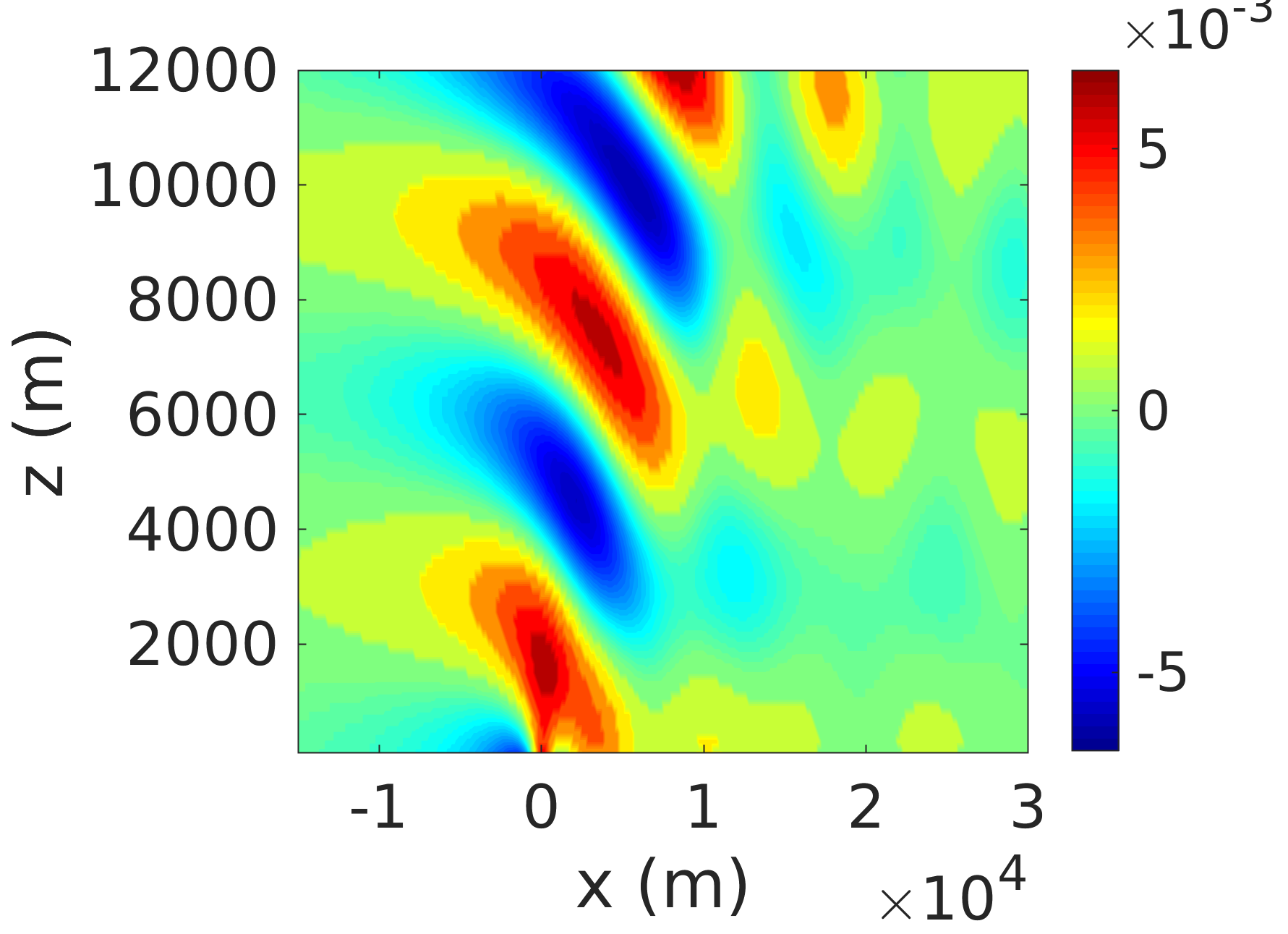}
      \end{overpic}
      \begin{overpic}[width=0.45\textwidth]{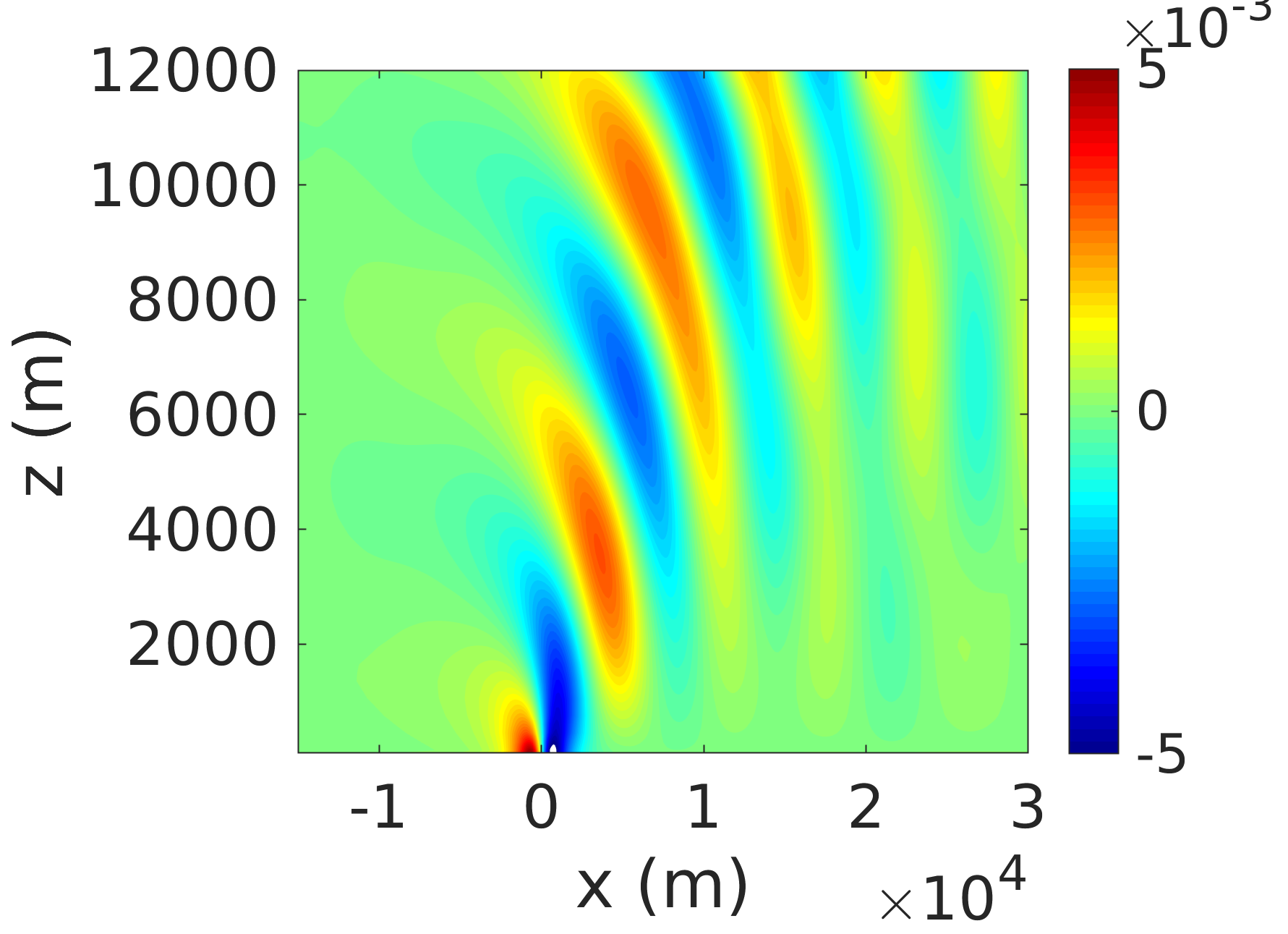}
      \end{overpic}\\
\caption{Linear non-hydrostatic mountain test: horizontal velocity $u$ (left) and vertical velocity $w$ (right) given by the CE3 model at $t = 28800$ s.}
\label{fig:mountain5_n}
\end{figure}

We consider a mesh with sizes $\Delta x = 180$ m and $\Delta z = 150 $ m and set $\Delta t$ = 0.05 s. 
Fig. \ref{fig:mountain6_n} depicts the normalized momentum flux at various times. We see that the simulations gets close to the steady state around $t = 21600$ s, i.e., after 
about 6 h, which is in agreement with \cite{giraldo_2008, bonaventura2023}. The values of $m^\star$ in Fig. \ref{fig:mountain6_n} are between 0.94 and 0.98, again in agreement with the interval found in \cite{giraldo_2008}, which is [0.95, 1], and the interval in \cite{bonaventura2023}, which seems to  be  [0.97, 1] from the Fig.~15. 
Finally, Fig.~\ref{fig:mountain5_n} shows the velocity components at $t = 28800$ s, which are close to the steady state. 
These figures compare well with the corresponding figures in \cite{bonaventura2023, giraldo_2008, gmd-15-6259-2022}.

\section{Conclusion and Perspective}\label{sec:conc}

We implemented and numerically analyzed pressure-based solvers for three different conservative forms the compressible Euler equations, called CE1, CE2, and CE3. CE1 and CE2 are written in density, momentum, and specific enthalpy and employ two different treatments of the buoyancy and the pressure gradient terms. Specifically, CE1 uses the standard pressure splitting implemented in 
open-source finite volume solvers involving a gravity field, while CE2 uses the typical splitting adopted in atmospheric studies. 
CE3 replaces specific enthaply with potential temperature and treats the buoyancy and the pressure terms like CE2. All the formulations are discretized in space with a finite volume method.
We tested the three formulations against numerical data available in the literature for classical benchmarks for mesoscale atmospheric flow on a flat terrain, i.e., the rising thermal bubble, the density current, and the inertia-gravity waves. In addition, we  considered benchmark tests featuring a single-peaked mountain, namely the hydrostatic equilibrium of an initially resting atmosphere, the steady state solution of linear hydrostatic flow, and the steady state solution of linear non-hydrostatic flow.

We found that for the rising bubble and the density current all three forms provide accurate results, although CE3 shows a significantly greater robustness and realiability. For the intertial-gravity waves, the CE1 solution becomes unstable while both  CE2 and CE3 provide results in excellent agreement with the reference results. Additionally, CE1 fails at the tests with orography. While both CE2 and CE3 models are able to preserve the hydrostatic equilibrium in presence of steep orography, the CE3 model is more accurate. 
Furthermore, the results obtained with CE3 for the linear hydrostatic and non-hydrostatic mountain tests are in very good agreement with the numerical data available in literature. Thus, also when using a pressure-based approach and space discretization by a finite volume method, the CE3 model is the most accurate, reliable, and robust for the simulation of mesoscale atmospheric flows.


\section*{Aknowledgements}
We acknowledge the support provided by the European Research Council Executive Agency by the Consolidator Grant project AROMA-CFD “Advanced Reduced Order Methods with Applications in
Computational Fluid Dynamics” - GA 681447, H2020-ERC CoG 2015 AROMA-CFD, PI G. Rozza, and INdAM-GNCS 2019–2020 projects. This work was also partially supported by the U.S. National Science Foundation through Grant No. DMS-1953535
(PI A. Quaini). 



\bibliographystyle{plain}
\bibliography{bibliography}
\end{document}

%% file: packages.tex
\usepackage{amsmath}
\usepackage{amssymb}
\usepackage{mathrsfs}
\usepackage{color}
\usepackage{graphicx}
\usepackage[percent]{overpic}
\usepackage{url}
\usepackage{subfig}
\usepackage{enumerate}
\usepackage{overpic}
\usepackage{amsthm}
\usepackage{moreverb}
\usepackage[pdftex,colorlinks,bookmarksopen,bookmarksnumbered,citecolor=red,urlcolor=red]{hyperref}

%% file: commands.tex
\def\0{\boldsymbol{0}}

\def\cl {\nonumber \\}
\def\el {\nonumber }

\newtheorem{rem}{Remark}[section]

\newcommand{\bm}[1]{\mbox{\boldmath{$#1$}}}